\documentclass[A4paper,11pt]{article}

\pdfoutput=1

\usepackage{latexsym}
\usepackage{epsfig}
\usepackage{verbatim}
\usepackage{multirow}
\usepackage{fancybox}
\usepackage{shadow}
\usepackage{cite}
\usepackage{amssymb}
\usepackage{amsmath}
\usepackage{amscd}
\usepackage{graphicx}

\newtheorem{theorem}{\sc Theorem}[section]
\newtheorem{lemma}[theorem]{\sc Lemma}

\newcommand{\eps}{\varepsilon}

\newcommand{\proofend}{{\medskip\medskip}}
\newcommand{\proof}{{\noindent\em Proof. }}

\author{
  {\sc Bernard Chazelle}
\thanks{Department of Computer Science,
       Princeton University, 
{\tt chazelle}@{\tt cs.princeton.edu }}
}

\title{
The Dynamics of Influence Systems
\thanks{This work was supported in part by NSF grants
CCF-0832797, CCF-0963825, and CCF-1016250.
Prelim. version available in Proc. 53rd FOCS 2012.
}}

\vspace{2cm}

\date{}

\begin{document} \maketitle

\begin{abstract}
{\em Influence systems} form a large class of multiagent systems
designed to model how influence, broadly defined, spreads
across a dynamic network.
We build a general analytical framework which we then use
to prove that, while sometimes chaotic, 
influence dynamics of the diffusive kind is almost always asymptotically periodic.
Besides resolving the  dynamics
of a popular family of multiagent systems, the other contribution
of this work is to introduce a new type of renormalization-based bifurcation analysis
for multiagent systems.
\end{abstract}

\vspace{1cm}

\section{Introduction}\label{introduction}

The contribution of this paper is twofold:  (i) to formulate an ``algorithmic calculus''
for continuous, discrete-time multiagent systems; and (ii) 
to resolve the behavior of a popular type of social dynamics 
that had long resisted analysis. In the process, we also
introduce a new approach to time-varying Markov chains.
{\em  Diffusive influence systems} are piecewise-linear dynamical systems
${\mathbf x}\mapsto P({\mathbf x}){\mathbf x}$,  which are specified
by a piecewise-constant function $P$ mapping any
${\mathbf x}\in {\mathbb R}^n$ to an $n$-by-$n$ stochastic matrix
$P({\mathbf x})$.
We prove that, while sometimes chaotic, such systems
are almost surely attracted to a fixed point
or a limit cycle. 

As in statistical mechanics, the difficulty of analyzing
influence systems comes from the tension between two
opposing forces: one, caused by the map's discontinuities,
is ``entropic" and leads to chaos;
the other one, related to the Lyapunov exponents, is ``energetic'' and pulls the system toward
an attracting manifold within which the dynamics is periodic.
The challenge is to show that, outside 
a vanishingly small critical region in parameter space, entropy always loses.
Because the interaction topology changes all the time
(endogenously), the proof relies heavily on an algorithmic
framework to monitor the flow of information across the system.
As a result, this work is, at its core, an algorithmic study in dynamic networks.
Influence systems include finite Markov chains as a special case but
the differences are deep and far-reaching:
whereas Markov chains have predictable dynamics, influence
systems can be chaotic even for small $n$;
whereas the convergence of a Markov chain can be checked in polynomial time,
the convergence of an influence system is undecidable.
Our main result is that this bewildering complexity 
is in fact confined to a vanishing region of parameter space.
Typically, influence systems are asymptotically periodic.

\smallskip
\paragraph{Influence and social dynamics.}

There is a context to this work and this is where we begin.
An overarching ambition of social dynamics
is to understand and predict the collective behavior of agents
influencing one another across an endogenously changing
network~\cite{castellano2009}.
{\em HK} systems have emerged in the last decade as a prototypical platform 
for such investigations~\cite{blondelHT07, blondelHT09, hegselmanK, hegselmanK2006, krause00, kurzR, lorenz03, lorenz10, 
martinezBCF07, mirtaB}. To unify
its varied strands (eg, bounded-confidence, bounded-influence, truth-seeking, 
Friedkin-Johnsen type, deliberative exchange) into a single framework
and supply closed-loop analogs to
standard consensus models~\cite{blondelHOT05,lorenz05, Moreau2005},
we introduce {\em influence systems}.
These are discrete-time dynamical systems
${\mathbf x} \mapsto f({\mathbf x})$ in 
$({\mathbb R}^d)^n$: each ``coordinate'' $x_i$
of the state ${\mathbf x}= (x_1,\ldots, x_n)$ 
is a $d$-tuple encoding the location of {\em agent} $i$ as a point 
in ${\mathbb R}^d$; with any 
state ${\mathbf x}$ is associated a directed graph
${\mathcal G}({\mathbf x})$ with the $n$ agents as nodes.
Each coordinate function $f_i$ of the map $f=(f_1,\ldots, f_n)$ takes
as input the neighbors of agent $i$ in
${\mathcal G}({\mathbf x})$ and outputs the new location
$f_i({\mathbf x})$ of agent~$i$ in $d$-space.
One should think of agent~$i$ as a ``computer'' and $x_i$ as its ``memory."
All influence systems in this work will be assumed to be
{\em diffusive}, meaning that at each step an agent may move only within
the convex hull of its neighbors.\footnote{\, This is a standard assumption 
meant to ensure that consensus is a fixed point.}
Note that the system ${\mathbf x}\mapsto P({\mathbf x}){\mathbf x}$
in the opening paragraph corresponds to the one-dimensional case.

Influence systems arise in processes as diverse as chemotaxis, synchronization, 
opinion dynamics, flocking,
swarming, and rational social learning.\footnote{\, 
The states of an influence system can be:
opinions~\cite{blondelHT09, dittmer01, hegselmanK, krause00, kurzR, lorenz05, lorenz10},
Bayesian beliefs~\cite{acemogluDLO},
neuronal spiking sequences~\cite{cessac08},
animal herd locations~\cite{ConradtR},
consensus values~\cite{caoSM, Moreau2005},
swarming trajectories~\cite{gaziP},
cell populations~\cite{pikovskyEtal},
schooling fish velocities~\cite{OkuboL, ParrishH},
sensor networks data~\cite{bulloBk},
synchronization phases~\cite{earlS, papacJ05, scardoviSS},
heart pacemaker cell signals~\cite{strogatz00, winfree67},
cricket chirpings~\cite{walker},
firefly flashings~\cite{mirolloS},
yeast cell suspensions~\cite{pikovskyEtal},
microwave oscillator frequencies~\cite{strogatz00}, or
flocking headings~\cite{balleriniCCCCG, chazFlockPaper,
CuckerSmale1, HendrickxB, jadbabaieLM03, vicsekCBCS95}.}
Typically, a natural algorithm directs $n$ autonomous agents to
obey two sets of rules: (i) one of them determines, on the basis of the system's current 
state ${\mathbf x}$, which agent communicates with which one;
(ii) the other one specifies how an agent 
updates its state by processing the 
information it receives from its neighbors.
Diffusive influence systems are central to social dynamics
insofar as they extend the fundamental concept of  
{\em diffusion} to autonomous agents operating within
dynamic, heterogeneous environments.\footnote{\,
For a fanciful but illustrative example, imagine
$n$ insects on the ground ($d=2$), each one moving
toward the mass center of its neighbors.
Each one gets to ``choose'' who is its neighbor:
this cricket picks the five ants closest to it
within its cone of vision;
that spider goes for the ladybugs within two feet;
these ants select the 10 furthest termites; etc.
Once the insects have determined their neighbors, they move to their mass center
(or a weighted version of it). This is repeated forever.}
This stands in sharp contrast
with the classic brand of diffusion found in physics and chemistry,
which assumes passive particles subject to identical laws.
Naturally, influence systems are ``downward-compatible''
and can model standard (discrete) diffusion as well. They also allow
exogeneities (eg, diffusion-reaction) via the addition of special-purpose agents.
Autonomy and heterogeneity are the defining features of
influence systems: they grant agents the freedom to have
their own, distinct decision procedures to choose their neighbors as they please
and act on the information collected from them.
This explains their ubiquity among natural algorithms.


\smallskip
\paragraph{The model.}

In a diffusive influence system,
$f({\mathbf x}) = (P({\mathbf x})\otimes \text{Id} )\,{\mathbf x}$,
where $P({\mathbf x})$ is a stochastic matrix whose positive entries 
correspond to the edges of ${\mathcal G}({\mathbf x})$ and are rationals
larger than some arbitrarily small $\rho>0$.  We form the Kronecker product with
the $d$-by-$d$ identity matrix to perform the averaging along each coordinate axis.
We grant the agents a measure
of self-confidence by adding a self-loop to each node of ${\mathcal G}({\mathbf x})$.
Agent~$i$ computes the $i$-th row of $P({\mathbf x})$ by means of its own algebraic
decision tree; that is, on the basis of the signs of
a finite number of $dn$-variate polynomials
evaluated at the coordinates of ${\mathbf x}$.
This high level of generality allows ${\mathcal G}({\mathbf x})$ 
to be specified by any first-order sentence
over the reals:\footnote{\, 
This is the language of geometry and algebra with statements
specified by any number of quantifiers and polynomial (in)equalities.
It was shown to be decidable by Tarski and amenable to quantifier 
elimination and algebraic cell decomposition by Collins~\cite{collins75}.}
In a recent bird flocking model~\cite{balleriniCCCCG}, for instance,
the communication graph joins every agent to its 7 nearest neighbors.
We show below how to reduce the dimension to $d=1$ and linearize the system so that 
$P({\mathbf x})=P_c$, 
for any ${\mathbf x}\in c$, where $c$ is any {\em atom}
(open $n$-cell) of an arrangement of hyperplanes in ${\mathbb R}^n$,
called the {\em switching partition} ({\em SP\,}).
An influence system is called {\em bidirectional} if
${\mathcal G}_{ij}\equiv {\mathcal G}_{ji}$ (with 
${\mathcal G}= ({\mathcal G}_{ij})$), which
implies that ${\mathcal G}({\mathbf x})$ is undirected.
Such a system is further called {\em metrical} if 
${\mathcal G}_{ij}$ is solely a function of
$|x_i-x_j|$.  Homogeneous {\em HK} systems~\cite{hegselmanK, hegselmanK2006, krause00}
constitute the canonical example of a metrical system.
We assume that all the
relevant parameters (matrix entries, number and coefficients of hyperplanes, $\rho$, etc)
can be encoded as rationals over $O(\log n)$ bits: this
assumption can be freely relaxed---in fact, the bit lengths can be arbitrarily large
as a function of $n$---and is only made to simplify the notation.

\smallskip
\paragraph{Past work and present contribution.}

Beginning with their introduction by Sontag~\cite{sontag81},
piecewise-linear systems have become the subject of
an abundant literature, which we do not attempt to review here. Restricting
ourselves to influence systems, we note that the
bidirectional kind are known to be attracted to a fixed point while 
expending a {\em total $s$-energy} at most exponential in the number 
of agents and polynomial in the reversible case~\cite{chazelle-total,dittmer01, HendrickxB,lorenz05,Moreau2005}.
Convergence times are known only in the simplest cases~\cite{chazelle-total, martinezBCF07, bulloBk}.
In the nonbidirectional case, most convergence results are conditional~\cite{catsigerasB10, cessac08,mirtaB, caoSM, jadbabaieLM03,  
Moreau2005, nedicOOT, olshevskyT-06, tsitsiklisBA}.\footnote{\,
As they should be, since convergence is not assured. 
An exception is {\em truth-seeking HK systems},
which have been shown to converge unconditionally~\cite{chazelle-total}.}
The standard assumption is that some form of joint connectivity 
property should hold in perpetuity.
To check such a property is in general undecidable (see why below),
so these convergence results are somewhat of a heuristic nature. 
A significant recent advance was Bruin and Deane's unconditional resolution 
of planar piecewise contractions, which are special kinds of 
influence systems with a single mobile agent~\cite{bruinD09}.
Our main result can be interpreted as a grand generalization of theirs.

\begin{theorem}\label{general-case}
$\!\!\! .\,\,$
Given any initial state, the orbit of an influence system is attracted 
exponentially fast to a limit cycle almost surely
under an arbitrarily small perturbation.
The period and preperiod are bounded by a polynomial
in the reciprocal of the failure probability.
Without perturbation, the system can be Turing-complete.
In the bidirectional case, the system is attracted to a fixed point
in time $n^{O(n)}\log \frac{1}{\eps}$ almost surely,
where $n$ is the number of agents and $\eps$ is the distance to the fixed point.
\end{theorem}

The theorem bounds the convergence time of
bidirectional systems by a single exponential and establishes
the asymptotic periodicity of generic influence systems. 
These results are essentially optimal.  We also estimate
the attraction rate of general systems but the bounds we obtain
are probably too conservative to be useful. 
Perturbing the system means replacing each hyperplane
${\bf a}^T{\mathbf x}= a_0$ of the {\em SP} 
by ${\bf a}^T{\mathbf x}=a_0 +\delta$, for some arbitrarily small random $\delta$.
Note that neither the initial state nor the transition matrices are perturbed.\footnote{\,
This is not a noise model~\cite{braverman12}: the perturbation happens only once 
at the beginning.}
We enforce an {\em agreement rule}, which
sets ${\mathcal G}_{ij}$ to be constant
over the microscopic slab $|x_i-x_j| \leq \eps_0$, for an arbitrarily small
$\eps_0>0$.\footnote{\, Agent $i$ is free to set the function  
${\mathcal G}_{ij}$ to either 0 or 1. For notational convenience, we set
$\eps_0$ to be $n^{-O(1)}$, but smaller values would work just the same.}
Intuitively, the agreement rule stipulates that minute
fluctuations of opinion between two agents 
otherwise in full agreement should have no 
macroscopic effect on the system.\footnote{\,
Interestingly, this is precisely meant to prevent the ``narcissism of small differences,"
identified by Freud and others as a common source of social conflicts.}
We emphasize that {\em both} the perturbation and the agreement rule are
necessary: without them, the attraction claims of 
Theorem~\ref{general-case} are provably false.\footnote{\,
In the nonbidirectional case, agents are made to enforce  a timeout mechanism
to prevent edges from reappearing after an indefinite absence
of unbounded length. While probably unnecessary, this minor technical feature 
seems to simplify the proof.}
We show that finely tuned influence systems are indeed Turing-complete.

Our work resolves the long-term behavior of a fundamental
natural process which includes the 
extended family of {\em HK} systems as a special case.
The high generality of our results precludes statements
about particular restrictions which might be easier. 
A good candidate for further investigation 
is the heterogeneous {\em bounded-confidence} model, 
where each ${\mathcal G}_{ij}$ is defined by a single interval,
and which is conjectured to converge~\cite{mirtaB}. (We show below that
this is false if the averaging is not perfectly uniform.)
Such systems were not even known to be periodic,
a feature that our result implies automatically.
Generally, our work exposes a surprising gap in the expressivity of
directed and undirected dynamic networks: while the latter always
lead to stable agreement (of a consensual, polarized, or fragmented nature), 
directed graphs offer a much richer complexity landscape.

The second contribution of this work is the introduction of a new
brand of bifurcation analysis based on algorithmic renormalization.
In  a nutshell, we use a graph algorithm to decompose 
a dynamical system into a hierarchy of recursively defined
subsystems. We then develop a tensor calculus to ``compile''
the graph algorithm into a bifurcation analysis. The tension between
energy and entropy is then reduced to a question in matrix rigidity.

In the context of social dynamics, Theorem~\ref{general-case}
might be somewhat disconcerting. Influence systems model how people change opinions
over time as a result of human interaction and knowledge
acquisition. Our results show that, unless people keep varying
the modalities of their interactions, as mediated by 
trust levels, self-confidence, etc, 
{\em they will be caught forever recycling the same opinions in the same order}.
The saving grace is that the period can be exponentially long,
so the social agents might not even realize they have become
little more than a clock...

\section{The Complexity of Influence Systems}

Piecewise-linear systems are known to be 
Turing-complete~\cite{alur,blondelT00,koiranCG,siegelS}.
A typical simulation relies on the existence of Lyapunov exponents of both signs,
negative ones to move the head in one direction and positive ones to move it the other way.
Influence systems have no positive exponents and yet are Turing-complete,
as we show below. 
In dynamics, chaos is typically associated with positive topological entropy, which 
entails expansion, hence positive Lyapunov exponents. But piecewise
linearity blurs this picture and produces surprises.
For example, isometries (with only null Lyapunov exponents)
are not chaotic~\cite{buzziTopEntr}
but, paradoxically, contractions (with only negative exponents) can be~\cite{kruglikovR06}.
Influence systems, which, with only null and negative Lyapunov exponents, 
sit in the middle, can be chaotic.
The spectral lens seems to break down completely in the face of piecewise linearity!

\smallskip
\paragraph{Exponential periods.}

It is an easy exercise to use higher bit lengths to increase the period of
an oscillating influence system by any amount. 
More interesting is the observation that the period can be raised
to exponential with only logarithmic bit length.
We simulate a counter modulo $2$ by building a system with $n=3$:
the first two agents are fixed at $0$ and $3$ 
while the third oscillates between positions $1$ and $2$;
this is trivially achieved with a two-test linear decision tree.
Add another mobile agent oscillating between 1 and 2 like the
previous one, but which moves only when
the first oscillating agent is at position 1. (Adding a single test makes this possible.)
Iterating in this fashion produces an $n$-agent influence system
with $O(n)$ tests whose period is exactly $2^{n-2}$.

For a system where action and control are more closely mixed,
consider implementing ${\mathbb Z}/2{\mathbb Z}$ as a 3-agent influence system
by fixing the first two agents at positions 0 and 3, respectively,
and then letting the third one oscillate between $1$ and~$2$.
By adding $O(q)$ discontinuities, we extend this scheme to keep
an agent cycling through $1,\ldots,q$, and then back to 1, 
thus implementing ${\mathbb Z}/q{\mathbb Z}$.
Repeating this construction for the first $k$ primes $p_1<\cdots<p_k$
allows us to implement the system based on the direct sum 
${\mathbb Z}/p_1{\mathbb Z} \oplus \cdots \oplus {\mathbb Z}/p_k{\mathbb Z}$,
which has period of $\prod_{j\leq k} p_j$ for a total of 
$N=O(p_1+\cdots + p_k)$ agents and discontinuities. By the prime number 
theorem~\cite{nivenZM}, this gives us a period of 
length $2^{\Omega(\sqrt{N\log N}\,)}$.  
While the period of the system as a whole is huge, each agent cycles
through a short periodic orbit: this is easily remedied by
adding another agent attracted to the mass center 
of the $k$ cycling agents. By the Chinese Remainder Theorem,
that last agent has an exponential period and acts as a sluggish clock.

\smallskip
\paragraph{Chaos and perturbation.}

Perturbation is needed for several reasons, including
uniform bounds on the time to stationarity. We focus here
on the agreement rule and show why it is necessary by
designing a chaotic system that is resistant to perturbation.
We use a total of four agents. The first two agents stay on opposite
sides of 0.5, with the one further from 0.5 moving toward it:
\begin{equation*}
(x_1, x_2) \,\,
\longmapsto\,\, \hbox{$\frac{1}{2}$}
\begin{cases}
\, (\,2x_1, x_1+ x_2 \,) \hspace{1.3cm} \,\text{ if $x_1+x_2 \geq 1$} \\
\, (\, x_1+ x_2, 2x_2\,) \hspace{1.3cm} \, \text{ else.}
\end{cases}
\end{equation*}
The two agents converge toward $0.5$ but the order in which they
proceed (ie, their symbolic dynamics) is chaotic. To turn this into actual chaos, 
we introduce a third agent, which oscillates between a fourth agent fixed at $x_4=0$
and $x_1$ (which is roughly $0.5$),
depending on the order in which the first two agents move:
$x_3 \mapsto \frac{1}{3}(x_3+ 2x_1)$ if $x_1+x_2 \geq 1$ and
$x_3 \mapsto \frac{1}{3}(x_3+ 2x_4)$ else.
Assume that $x_1(0)<\frac{1}{2}<x_2(0)$
and consider the trajectory of a line $L$:
$X_2-\frac{1}{2}= u(X_1-\frac{1}{2})$, for $u<0$.
If the point $(x_1(t),x_2(t))$ is on the line, then $x_1(t)+x_2(t)\geq 1$
implies that $u\leq -1$ and $L$ is mapped to 
$X_2-\frac{1}{2}=  \frac{1}{2}(u+1)(X_1-\frac{1}{2})$; 
if $x_1(t)+x_2(t)< 1$, then $u> -1$ and $L$ becomes 
$$X_2-\hbox{$\frac{1}{2}$}= \frac{2u}{u+1}\,(X_1-\hbox{$\frac{1}{2}$}).$$
The parameter $u$ obeys the dynamics: 
$u\mapsto \frac{1}{2}(u+1)$ if $u\leq -1$ and 
$u\mapsto 2u/(u+1)$ if $-1<u\leq 0$.  Writing $u= (v+1)/(v-1)$ gives
$v\mapsto 2v+1$ if $v< 0$ and $v\mapsto 2v-1$ else.
(Geometrically, $v$ is the tangent of the angle between
$L$ and the line $X+Y=0$.) The system $v$ escapes for 
$|v(0)|>1$ and otherwise conjugates with the baker's map~\cite{devaney}
via the variable change: $v= 2w-1$.
Agent 3 is either at most 1/6 or at least 1/3
depending on which of agent 1 or 2 moves.
This implies that the system has 
positive topological entropy: to know where agent 3 is at  
time $t$ requires on the order of $t$ bits of accuracy 
in the initial state.
The cause of chaos is the first two agents' convergence
toward the {\em SP} discontinuity.  It is immediate
that no perturbation can prevent this, so 
the agreement rule is indeed needed.

\smallskip
\paragraph{Turing completeness.}

Absent perturbation and the agreement rule,
an influence system can simulate a piecewise-linear system
and hence a Turing machine. Here is how.
Given a nonzero $n$-by-$n$ real-valued matrix $A$,
let $A^+$ (resp. $A^-$) be the matrix obtained by
zeroing out the negative entries of $A$ (resp. $-A$),
so that $A= A^+ - A^-$. Define the matrices
\begin{equation*}
B= \,\, \rho 
\begin{pmatrix}
A^+ &  A^- \\  A^-  & A^+
\end{pmatrix} 
\hspace{.4cm}
\text{and}
\hspace{.4cm}
C= \,\,
\begin{pmatrix}
B&  (\text{\rm Id} - B) {\mathbf 1}  & {\mathbf 0} \\  {\mathbf 0} & 1 & 0 \\
{\mathbf 0} & 1-\rho & \rho
\end{pmatrix} ,
\end{equation*}
where $\rho$ is the reciprocal of the maximum row-sum
in the matrix derived from $A$ by taking absolute values.
It is immediate that $C$ is stochastic and conjugates with
the dynamics of $A$. Indeed, given
${\mathbf x}\in {\mathbb R}^n$, if 
$\overline{\mathbf x}$ denotes the $(2n+2)$-dimensional
column vector $({\mathbf x}, -{\mathbf x}, 0,1)$, then
$C\, \overline{\mathbf x} = \rho \, \overline{A\mathbf x}$;
hence the commutative diagram:
\begin{equation*}
\begin{CD}
{\mathbf x} @ >>> A{\mathbf x}\\
@VVV   @VVV\\
\overline{\mathbf x} @>>> \rho^{-1}C\, \overline{\mathbf x} \, .
\end{CD}
\end{equation*}
Imagine now a piecewise-linear system consisting of a number of
matrices $\{A\}$ and an {\em SP}. We add $n$ negated clones
to the existing set of $n$ agents, plus a {\em stochasticity} agent  permanently
positioned at $x_{-1}=0$ and a {\em projectivity} agent initially at $x_0$.
This allows us to form the 
vector $\overline{\mathbf x}=( {\mathbf x}, - {\mathbf x}, x_{-1},x_0)$.
The system scales down, 
so we must projectify the {\em SP} by rewriting with homogeneous 
coordinates any ${\bf a}^T{\mathbf x}=a_0$ 
as ${\bf a}^T{\mathbf x}=a_0 x_0$. 
We can use the same value of $\rho$ throughout by picking
the smallest one among all the matrices $A$ used in the
piecewise-linear system.

Koiran et al~\cite{koiranCG} have shown 
how to simulate a Turing machine with a 3-agent piecewise-linear system,
so we set $n=3$. We need an {\em output} agent to indicate whether
the system is in an accepting state: this is done by pointing to one
of two fixed agents. 
We can enlist one of the three original agents for that purpose, 
which brings the agent count up to $2n+3=9$.  Predicting
basic state properties of an influence system is therefore undecidable.
With a few more agents, we can easily encode as an undecidable question
whether the communication graphs (or their union over
bounded time windows) satisfy certain
connectivity property infinitely often.

\smallskip
\paragraph{Linearization.}\label{linearize}

We show how to linearize an influence system by tensor powering (for any $d$).
Let ${\tt d}$ be the maximum total degree of the polynomial tests used in the algebraic 
decision trees (recall that each agent comes equipped with its own). We can always assume the existence
of an agent confined to position 1 with no in/out-link: we use it 
to homogeneize the test polynomials, so that every monomial has degree exactly~${\tt d}$.
Given ${\mathbf x}= (x_1,\ldots, x_n)\in {\mathbb R}^n$,
we define the monomial $y_{k_1,\ldots, k_{\tt d}}= \prod_{i=1}^{\tt d} x_{k_i}$
($1\leq k_1,\ldots, k_{\tt d}\leq n$) and, listing them in lexicographic order,
form ${\mathbf y}= (y_{k_1,\ldots, k_{\tt d}}) \in {\mathbb R}^N$, 
where $N= n^{\tt d}$; note that
${\mathbf y}$ lies on a (real) algebraic variety ${\mathcal V}$ smoothly
parametrized injectively by~${\mathbf x}$.
The map ${\mathbf x}\mapsto f({\mathbf x})$ induces
the lifted map ${\mathbf y}\mapsto g({\mathbf y})$, where
$g({\mathbf y})= P({\mathbf x})^{\otimes \, {\tt d}}\,{\mathbf y}$ and
\begin{equation*}
P({\mathbf x})^{\otimes \, {\tt d}} 
\,\, =  \,\,\,
\overset{{\tt d}}
    {\overbrace{\, P({\mathbf x})\otimes\cdots\otimes P({\mathbf x})
\,}} .
\end{equation*}
Being the Kronecker product of stochastic matrices,
$P({\mathbf x})^{\otimes \, {\tt d}}$ is stochastic: its diagonal is positive
and its nonzero entries all exceed $\rho^{\tt d}$.
Its associated graph, whose edges map out its nonzero entries,
is the tensor graph product ${\mathcal G}({\mathbf x})^{\otimes \, {\tt d}}$.
We use the term {\em ground agents} to refer to the $n$ agents positioned at ${\mathbf x}$.
Including all the test polynomials from all the ground agents' decision trees
gives us as many hyperplanes in ${\mathbb R}^N$ and the sign conditions
of a cell $c$ specify a unique stochastic matrix $Q_c$.
This matrix is always a tensor power $P^{\otimes \, {\tt d}}$
but it is guaranteed to be of the form
$P({\mathbf x})^{\otimes \, {\tt d}}$ only
if $c$ contains a point ${\mathbf y}$ of $V$ parametrized by ${\mathbf x}$.

Whereas a random shift produces affine forms $a_1y_1+\cdots+ a_Ny_N+\delta$,
the agreement rule acts in a more subtle way. 
While the whole point of the lifting is to forget about the variety ${\mathcal V}$,
the tensor structure of the matrices $Q_c$ brings benefits we can exploit.
Given $K\subseteq \{1,\ldots, n\}$,
the {\em cluster} $C_K$ refers to
the subset of $|K|^{\tt d}$ agents with labels in $K^{\tt d}$.
If all the agents of a cluster fit within a tiny interval then so do
their ground agents; to see why, just expand $(x_i-x_j)^{\tt d}$.
By the agreement rule, therefore, the induced subgraph of
the cluster cannot change until it is pulled apart by outside agents.
Assume now that $d>1$. We write 
$$ {\mathbf x} = (x_{1,1},\ldots, x_{1,d},\ldots, x_{n,1},\ldots, x_{n,d}),$$
with the homogeneizing agent $1$ permanently
positioned at $( x_{1,1},\ldots, x_{1,d}) = {\mathbf 1}_d$. 
Next, we define 
${\mathbf y} = ({\mathbf y}_1,\ldots, {\mathbf y}_{N})$,
where $N=(dn)^{\tt d}$ and ${\mathbf y}_{l} =  \prod_{i=1}^{\tt d} x_{k_i, j_i}$
with $l$ denoting the lexicographic rank of the string
$(k_1,j_1,\ldots, k_{\tt d},j_{\tt d})$ for $k_i\in \{1,\ldots, n\}$ and $j_i\in\{1,\ldots, d\}$.
The matrix $Q_c$ associated with 
cell $c$ is of the form $(P\otimes {\mathbf I}_{d})^{\otimes \, {\tt d}}$;
furthermore, $P= P({\mathbf x})$ whenever 
${\mathbf y}$ satisfies the $N$ conditions 
${\mathbf y}_{l} =  \prod_{i=1}^{\tt d} x_{k_i, j_i}$ for some ${\mathbf x}\in {\mathbb R}^{dn}$.
The cluster $C_K$  consists now of $(d |K|)^{\tt d}$ agents.

\smallskip
\paragraph{Nonconvergent {\em HK} systems.}

\begin{figure}[htb]
\begin{center}
\hspace{.2cm}
\includegraphics[width=7cm]{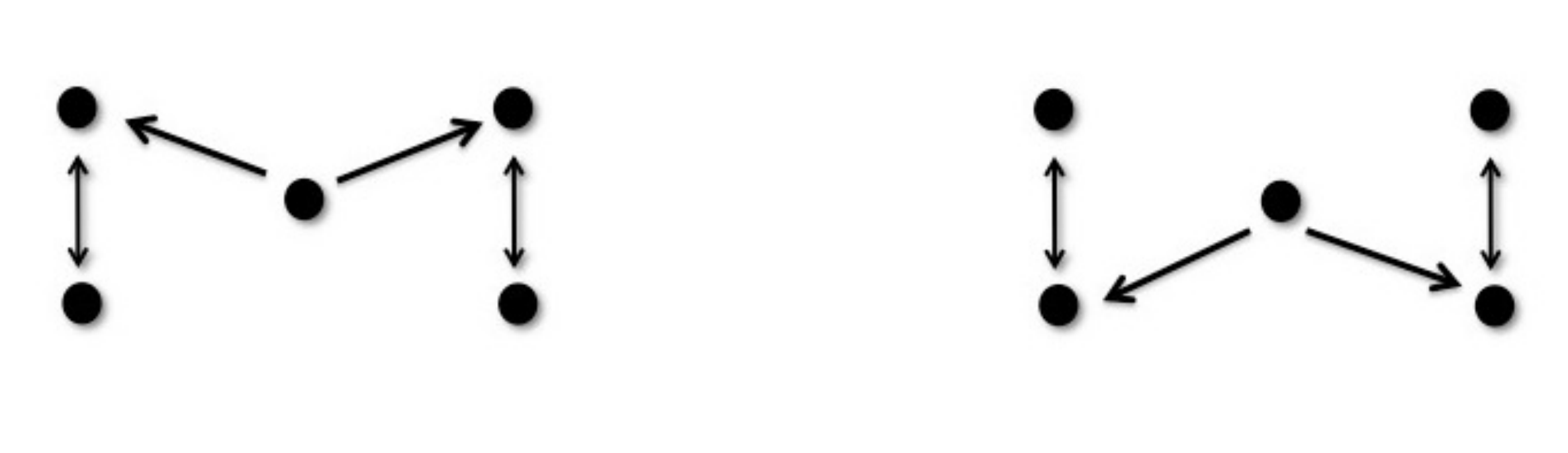}
\end{center}
\vspace{-.2cm}
\caption{\small  The communication graph of the {\em HK} system alternates between these two configurations.
\label{fig-hk-periodic}}
\end{figure}

Heterogeneous {\em HK} systems~\cite{hegselmanK, hegselmanK2006}
are influence systems where each agent $i$ is associated with 
a confidence value $r_i$ and the communication graph links $i$
to any agent $j$ such that $|x_i-x_j|\leq r_i$.
We design a periodic 5-agent system with period 2.
We start with a 2-agent system with $r_1=r_2=2$.
Instead of uniform averaging, we decrease the self-confidence weight
so that, if the two agents are linked, then
$x_1\mapsto \frac{1}{3}(x_1+ 2x_2)$ and $x_2\mapsto \frac{1}{3}(2x_1+ x_2)$.
Any self-confidence weight less than 0.5 would work, too; of course, a weight
of zero makes the problem trivial and uninteresting.
If the agents are initially positioned at $-1$ and $1$, 
they oscillate around 0 with $x_i= (-1)^{i+t} \, 3^{-t}$ at time $t$.
Now place a copy of this system with its center at $X=2$ and a mirror-image copy at $X=-2$;
then place a fifth agent at 0 and link it to any agent at distance at most $2$.
As indicated in Fig.~\ref{fig-hk-periodic}, even though the agents
themselves converge, their communication graph does not.

\vspace{.5cm}
\vspace{.5cm}

\section{An Overview of the Proof}

Given the challenge of presenting the multiple threads of the argument
in digestible form, we begin with a bird's eye view of the proof.
The standard way to establish the convergence of
an algorithm or a dynamical system is
to focus on a single unknown input and track the rate at which 
the system expends a certain resource on its way toward equilibrium:  
a potential function in algorithms; a free energy in 
statistical physics; or a Lyapunov function in dynamics.
This approach cannot work here. Instead,
we need to study the system's action on all inputs at once.
This is probably the single most important feature distinguishing
natural algorithms from the classical kind: because they run
forever, qualitative statements about their behavior will sometimes
require a global view of the algorithm's actions with respect
to {\em all} of its inputs. For this, we need a language that allows us
to model the evolution of phase space as a single geometric object.
This is our next topic. As explained earlier, we may assume that $d=1$.

\smallskip
\paragraph{The coding tree.}

\begin{figure}[htb]
\begin{center}
\hspace{.2cm}
\includegraphics[width=7cm]{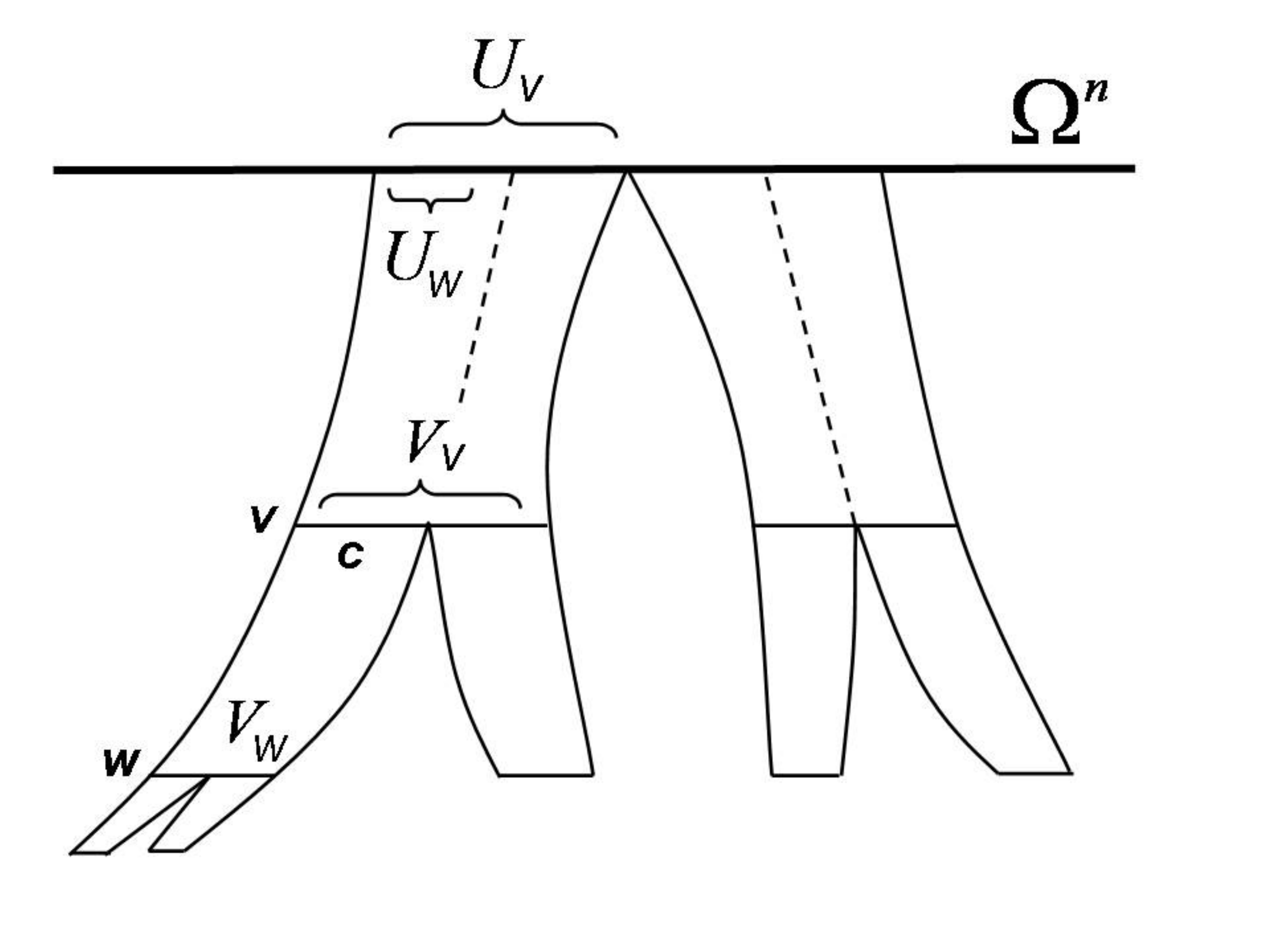}
\end{center}
\vspace{-.5cm}
\caption{\small Node $w$ at depth $t_w=2$ in the coding tree,
with its phase tube $(U_w,V_w)$ (curved for aesthetic reasons).
The cell $U_w$ lies within a single atom whereas $V_w$
splits over two of them. Time points downwards.
We have represented only two of the possibly many
children of the root.
\label{fig-coding-tree}}
\end{figure}

This infinite rooted tree encodes into one geometric
object the set of all orbits and the full symbolic dynamics.
It is the system's ``Rosetta stone,'' from which 
everything of interest can be read off.
The coding tree ${\mathcal T}$ is embedded in 
$\Omega^n\times {\mathbb R}$, where
$\Omega= (0,1)$ and the last dimension 
represents time.\footnote{\, 
By convexity, we can restrict the 
phase space to $\Omega^n$.}
Each child $v$ of the root is associated
with an atom $U_v$, while the root itself stands for
the phase space $\Omega^n$.
The {\em phase tube} $(U_v,V_v)$
of each child $v$ is the ``time cylinder'' whose cross-sections
at times $0$ and $1$ are $U_v$ and $V_v= f(U_v)$, respectively.
In general, a phase tube is a discontinuity-avoiding 
sequence of iterated images of a given cell in phase space.
The tree is built recursively
by subdividing $V_v$ into the cells $c$ formed by its
intersection with the atoms, and  
attaching a new child $w$ for each $c$: we set
$V_w=f(c)$ and $U_w= U_v\cap f^{-t_v}(c)$,
where $t_v$ is the depth of $v$ (Fig.~\ref{fig-coding-tree}).
The phase tube $(U_v,V_v)$ consists of all the cylinders
whose cross-sections at $t=0,\ldots, t_v$ are, respectively,
$U_v,f(U_v),\ldots, f^{t_v}(U_v)=V_v$. Intuitively,
${\mathcal T}$ divides up the phase space into maximal
regions over which the iterated map is linear.

The coding tree has three structural parameters
that require investigation.
One of them is combinatorial.
Label each node $w$ of the tree by the unique atom 
that contains the cell $c$ defined above. 
This allows us to interpret any path as a word of atom labels
and define the language $L({\mathcal T})$ of all such words:
the word-length growth of $L({\mathcal T})$ plays a central role,
which we capture with the {\em word-entropy} (formal definitions below).
The two other parameters are geometric: 
the {\em thinning rate} tells us how fast the  
tree's branches thin out; the {\em attraction rate} tells us 
how close to ``periodic'' the branches become. Whereas
the latter concerns the behavior of single orbits, the thinning rate
indicates how quickly a ball in the space of orbits contracts with time,
or equivalently how quickly the distribution of agent positions
loses entropy.

\vspace{0.0cm}
\vspace{.5cm}

\begin{figure}[htb]
\begin{center}
\includegraphics[width=5cm]{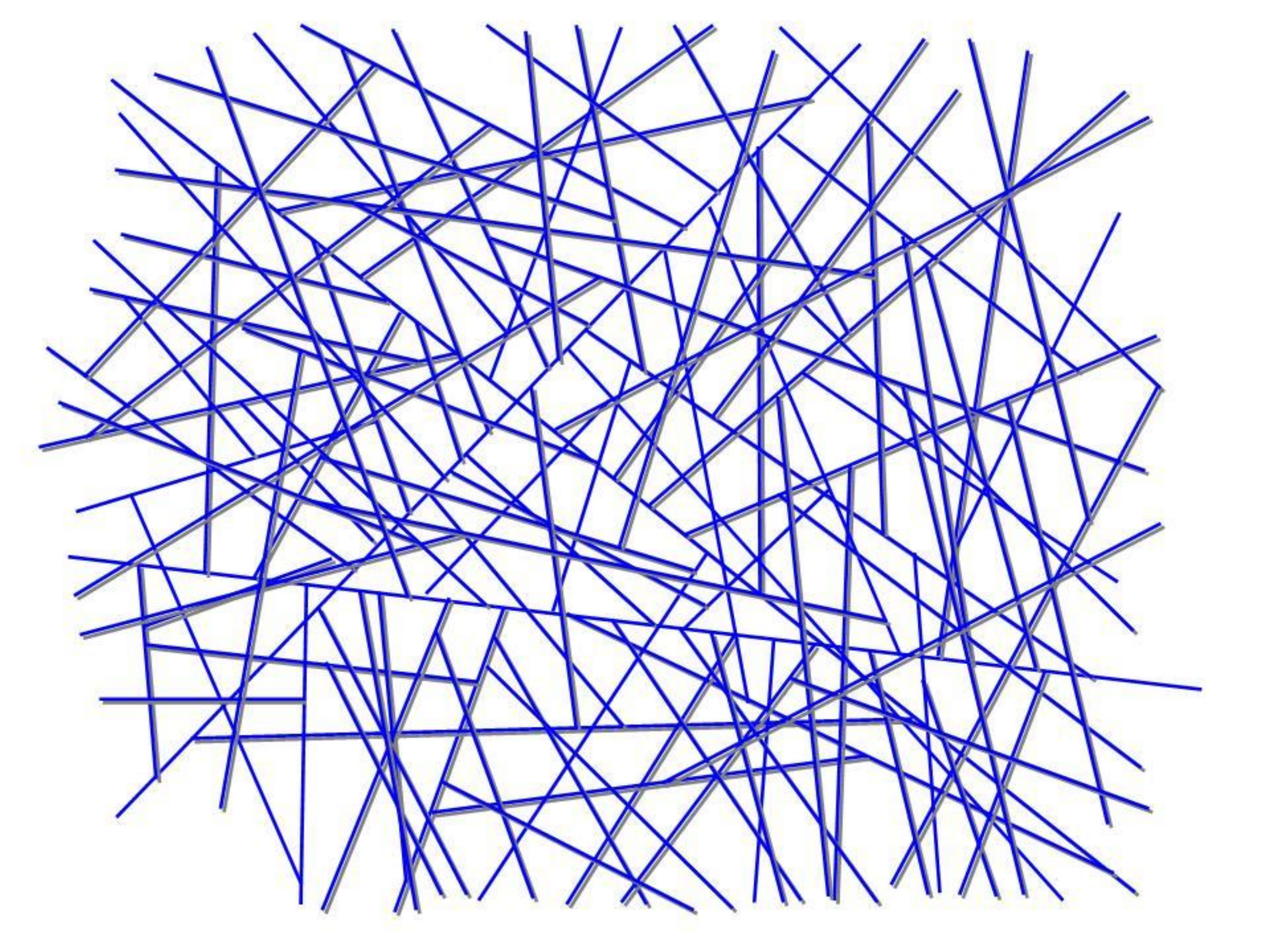}
\hspace{1cm}
\includegraphics[width=5cm]{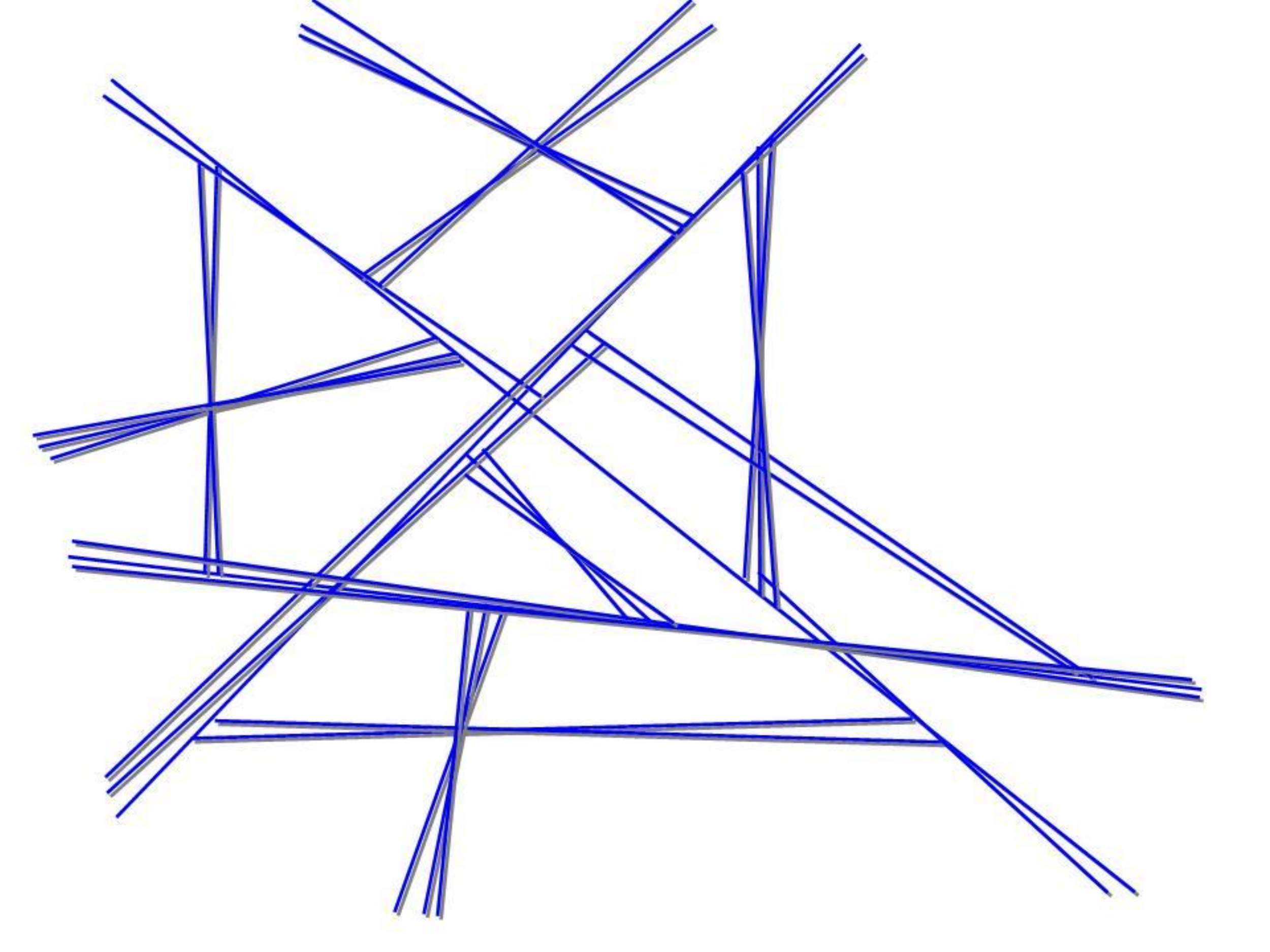}
\end{center}
\caption{\footnotesize 
Two scenarios for the iterated preimages of
the {\em SP} discontinuities ${\mathcal R}$: the set 
$ {\mathcal R} \cup f^{-1}({\mathcal R})
          \cup \cdots \cup f^{-t}({\mathcal R})$ depicted
on the left seems to spread everywhere in phase space
so as to cover all of it eventually, a symptom of chaos;
the set on the right tends to fall into clusters or escape outside of $\Omega^n$,
a sign of periodicity.
}
\label{fig-chaos}
\end{figure}

How do we read periodicity off from the coding tree?
Intuitively, one would expect that, at some time $\nu$
called the {\em nesting time},
for every $v$ of depth $t_v= \nu$,
there exists $w$ at the same depth with
$V_v\subseteq U_w$. In other words the bottom sections
of the phase tubes will, suitably permuted,
fit snugly within the top sections. 
This is not always true, however, and 
to find necessary conditions for it
necessitates a delicate bifurcation analysis.
Fig.~\ref{fig-chaos} 
suggests a visual rule-of-thumb to guide our intuition
in distinguishing between chaos and periodicity:
the set ${\mathcal R}$ consists of 
the points in phase space where the map $f$ is
not continuous.

\vspace{1cm}
\vspace{0cm}

\paragraph{The algorithmic pipeline.}

We assemble the coding tree by glueing together
smaller coding trees defined recursively. We entrust this
task to the {\em arborator}, an algorithm
expressed in a language for ``lego-like'' assembly.
The arborator needs two (infinite) sets of parameters
to do its job, the {\em coupling times} and 
the {\em renormalization scales}. To produce
these numbers, we use the {\em flow tracker},
an algorithm that, in the bidirectional case, 
works roughly like this:
(i) declare agent 1 {\em wet};
(ii) any dry agent becomes wet as soon as it links to a wet one;
(iii) if all agents ever become wet, dry them all
and go back to (i).
The instants $t_k$ at which wetness propagates
constitute the coupling times; the renormalization
scales are given by the number $w_k$ of wet agents at time $t_k$.
The key idea is that, between two coupling times $t_k$ and $t_{k+1}$,
the system breaks up into two subsystems with interaction
between them going only in one direction: from wet to dry.\footnote{\, 
This does not mean that the dynamics within the dry agents
is not influenced by the wet ones: only that dry agents do not
include wet ones in the averaging. The standout exception is
the case of a metrical system, where the dry agents act
entirely independently of the wet ones between $t_k$ and $t_{k+1}$.}
We denote by ${\mathcal A}(p\rightarrow q)$ an influence
system that consists of two groups 
of size $p$ and $q$, with none of the $q$ agents ever
linking up to any of the $p$ agents. This allows us
a recursive decomposition of the overall system:

\bigskip\bigskip
{\small
\par
\renewcommand{\sboxsep}{0.7cm}
\renewcommand{\sdim}{0.8\fboxsep}
\hspace{1.5cm}
\shabox{\parbox{9cm}{
\vspace{-0.6cm}
\begin{center}${\mathcal A}(n\rightarrow 0)$\end{center}
\smallskip
For $k=1,2,\ldots$
\begin{itemize}
\item[]
Run ${\mathcal A}(w_k\rightarrow n-w_k)$
and ${\mathcal A}(n-w_k\rightarrow 0)$
concurrently between times $t_k$ and $t_{k+1}$.
\end{itemize}
}}
\par\bigskip
}

\bigskip
\noindent
This formulation is of interest only
if we can bound $t_{k+1}-t_k$.
This is done implicitly by recursively monitoring the
long-term behavior of the two subsystems and 
inferring from it the possibility of further wetness propagation.
The flow tracker is a {\em syntactical} device because it merely 
monitors the exchange of information among agents with no regard
for what is done with it. By contrast, the arborator
models the agents' {\em interpretation} of that information
into a course of action.
The arborator is assembled as a recursive arithmetic
expression over four operations:
$\oplus$, $\otimes$, {\tt absorb}, and {\tt renorm}
(Fig.~\ref{fig-tensor}).
It comes with a dictionary that spells out
the effect of each term on the coding tree's structural parameters.
Here is a quick overview:

\vspace{0.5cm}
\begin{figure}[htb]
\begin{center}
\hspace{.2cm}
\includegraphics[width=8cm]{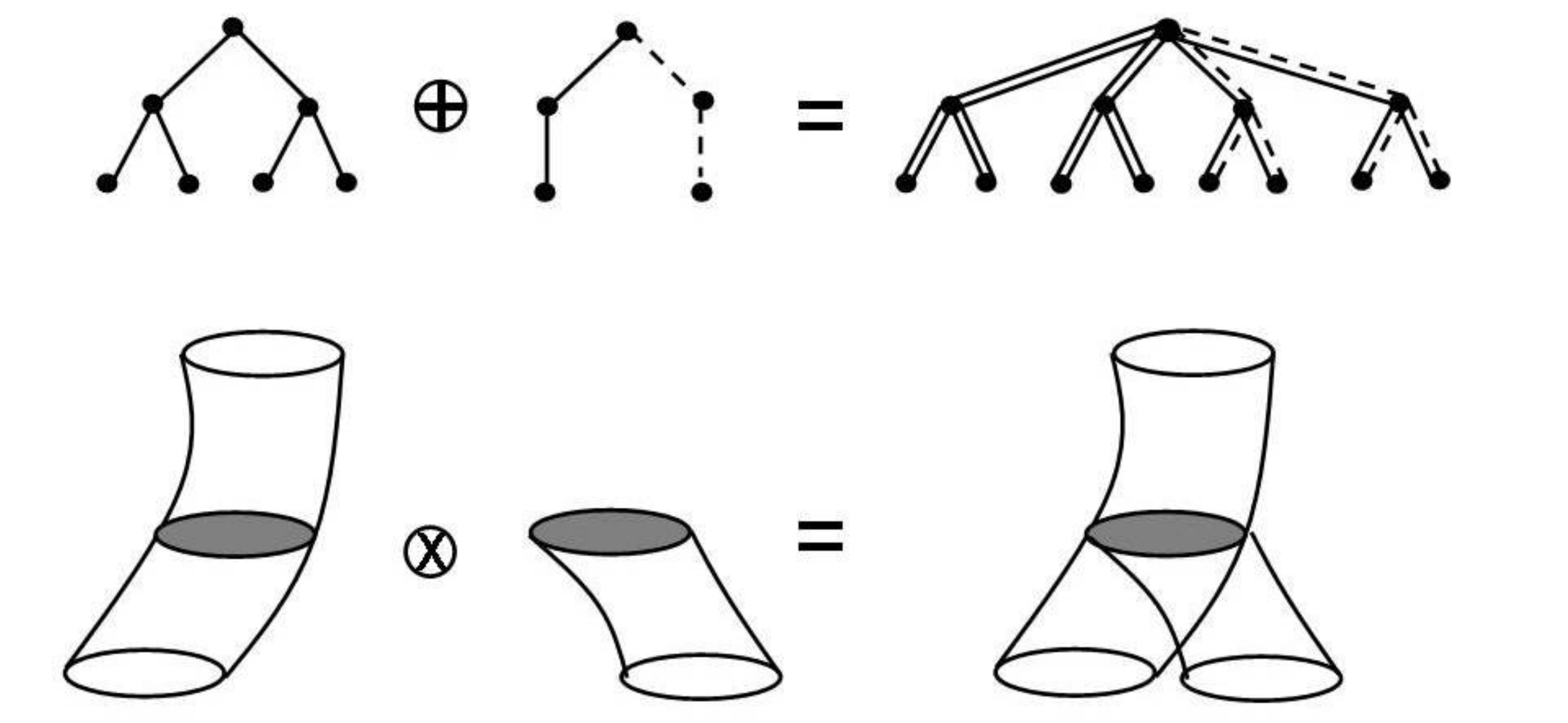}
\end{center}
\vspace{0.0cm}
\caption{\small The two tensor products.
\label{fig-tensor}}
\end{figure}

\begin{itemize}
\item
The direct sum $\oplus$ models the parallel execution
of two independent subsystems. 
Think of two agents, Bob and Alice, interacting 
with each other in one corner of the room while Carol and David 
are chatting on the other side. The coding tree of
the whole is the (pathwise) Cartesian product of both
two-agent coding trees.

\item
The direct product $\otimes$
performs tree surgery. It calls upon
another primitive, {\tt absorb}, to prune
the trees and prepare their phase tubes for ``glueing.''
Imagine Alice suddenly turning to Carol and addressing her.  
The flow tracker records that the two groups,
Bob-Alice and Carol-David, are no longer isolated.
Since this might not have happened had Alice been
at a slightly different location, 
the phase tube leading to this event may well split
into two parts: one bearing witness to the new interaction;
and the other extending the direct sum unchanged. 
By analogy with the addition of an absorbing state to a Markov chain,
the first operation is called {\tt absorb}.\footnote{\,
The dynamics multiplies transition matrices to the left.
Looking at it dually, the rightward products model
a random walk over a time-varying graph. The operation
{\tt absorb} involves adding a new leaf, which is similar
to adding an absorbing state; the direct product glues
the root of another coding tree at that~leaf.}

\item
The primitive {\tt renorm}, so named
for its kinship with the renormalization group of statistical 
physics, uses the renormalization
scales to compress subtrees into single nodes so as to
produce (nonuniform) time rescaling.
\end{itemize}

\vspace{0.0cm}
\begin{figure}[htb]
\begin{center}
\hspace{.2cm}
\includegraphics[width=7.5cm]{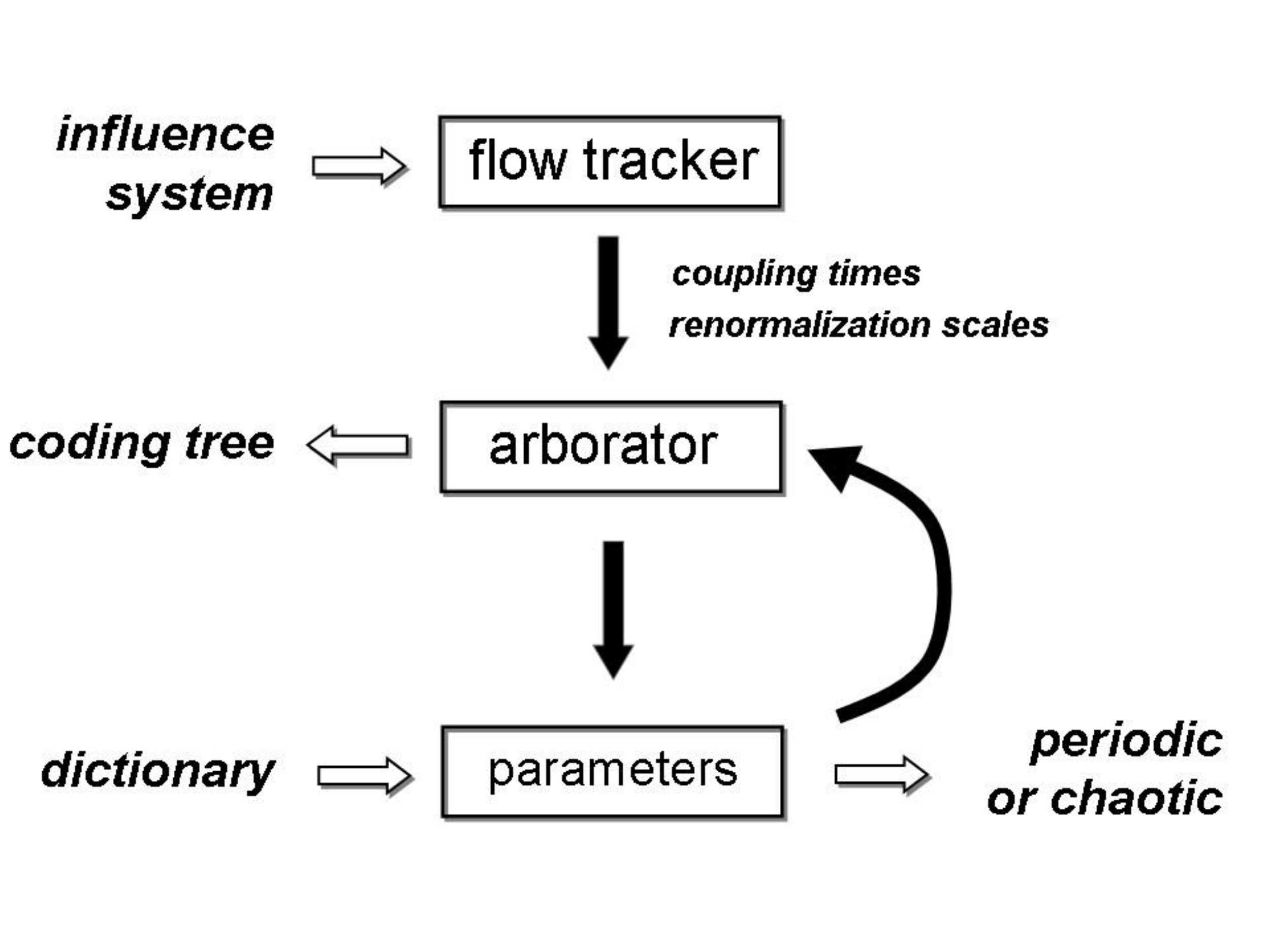}
\end{center}
\vspace{0.0cm}
\caption{\small Given the specification
of the natural algorithm, the {\em flow tracker}
computes the coupling times and the renormalization scales,
which are needed by the arborator to assemble the coding tree.
A dictionary allows us to bound the coding tree's
structural parameters by examining the arborator 
one component at a time. 
Renormalization makes this a recursive process, hence
the loop between the arborator and the parameters box.
\label{fig-pipeline}}
\end{figure}

\medskip

\smallskip
\paragraph{Attraction and chaos.}

The occurrence of chaos is mediated
by the tension between two forces: 
dissipation causes the phase tubes to become thinner,
which favors periodicity; phase tube splitting
produces a form of expansion conducive to chaos. Two arbitrarily close
orbits can indeed diverge wildly once they fall on both sides
of a discontinuity. The phase tubes snake around phase space
while getting thinner at an exponential rate, so hitting
{\em SP} discontinuities should become increasinly
rare over time. The problem is that branching
multiplies the chances of hitting discontinuities.
For dissipation to overcome branching, 
the average node degree should be small.
To show this is indeed the case requires a 
fairly technical rank argument about
the linear constraints implied by the splitting
of a phase tube.

The thinning rate is about contraction, not attraction.
To see why, consider a trivial system
with only self-loops: it is stuck at a fixed point,
yet the agents' marginal distributions suffer no loss of entropy.\footnote{\,
This is not to be confused with the word-entropy or the topological entropy.}
The information-theoretic interpretation of thinning
is illuminating. As agents are attracted to
a limit cycle, they lose memory of where 
they came from, something that would not happen in a chaotic system.
Paradoxically, interaction can then act as a memory recovery device
and thus delay the onset of periodicity.

Say the group Alice-Bob-Carol is isolated from David,
until the latter decides to interact with Alice, thus
taking in a fixed fraction of her entropy.
Fast-forward. Alice is now caught in a limit cycle
with Bob and Carol,
while David has yet to interact with anyone since
his earlier contact with Alice. His isolation means 
that he has had no chance to shed any of Alice's entropy.
Although later caught in a periodic orbit, Alice might
still be subject to tiny fluctuations, leading
to a sudden interaction with David. When this happens,
she will recapture part of the entropy she had lost: she will recover
her memory! Happy as the news might be to her, this 
only delays the inevitable, which is being caught yet again
in a limit cycle. Memory recovery cannot recur forever
because David loses some of his own memory every time.
In the end, because of dissipation, all the agents' memory will be lost.

\section{Algorithmic Dynamics}

We flesh out the ideas above, beginning with
a simple local characterization 
of periodicity.
We then proceed to define 
the coding tree~(\S\ref{sec:codingtreeT}),
the arborator~(\S\ref{sec:arborator}),
and the flow tracker~(\S\ref{flowtracker}).

\smallskip
\subsection{Conditions for asymptotic periodicity} 

It is convenient to thicken the discontinuities.
This does not change the dynamics of the system 
and is used only as an analytical device.
Fix a small parameter $\eps>0$ 
once and for all, and, for any $t\geq 0$, define
the {\em margin} ${\mathcal R}_{\eps}$, where
\begin{equation}\label{Rteps-def}
{\mathcal R}_{\eps} = \bigcup_{\text{\em SP}} \, \Bigl\{\, 
{\mathbf x}= (x_1,\ldots, x_n)\in {\mathbb R}^n 
  \,:\,|\, a_0+a_1x_1+\cdots + a_nx_n + \delta | \leq \eps \,\Bigr\},
\end{equation}
where the union extends over all the {\em SP} discontinuities.
The margin is made of $n^{O(1)}$ closed slabs of 
width at least $\eps n^{-O(1)}$. 
It is useful to classify the initial states by how long 
it takes their orbits to hit ${\mathcal R}_\eps$, if ever.
With $f^0= {\rm Id}$ and $\min \emptyset = \infty$,
we define the label of  
${\mathbf x}\in \Omega^n$ as
$$
\ell({\mathbf x}) = \min\, \Bigl\{\, t\geq 0 \, |\, f^t({\mathbf x}) \in 
      {\mathcal R}_{\eps} \,\Bigr\} .
$$
The point ${\mathbf x}$ is said to {\em vanish} at time 
$\ell({\mathbf x})$ if its label is finite.
As we shall see, the analysis needs to focus only
on the nonvanishing points. Write
${\mathcal S}_t = \{\, {\mathbf x}\in \Omega^n= (0,1)^n\, | \, 
                         \ell({\mathbf x}) \geq t\,\}$
for the set of points that do not vanish before time $t$:
${\mathcal S}_0$ is $\Omega^n$; and, for $t>0$, 
$$ {\mathcal S}_t = \Omega^n \setminus \bigcup_{k=0}^{t-1} 
 f^{-k}( {\mathcal R}_{\eps}) \, .$$
Each of its connected components is specified by 
a set of strict linear inequalities in ${\mathbb R}^n$,
so ${\mathcal S}_t$ is a union of disjoint open $n$-cells, 
whose number we denote by $\#{\mathcal S}_t$.
We redefine an atom to be a cell of ${\mathcal S}_1$
and restrict the domain of $f$ to these new atoms.
Each cell of ${\mathcal S}_{t+1}$ lies within a cell of ${\mathcal S}_t$.
The limit set ${\mathcal S}_\infty= \bigcap_{\,t\geq 0} {\mathcal S}_t$ 
collects the points that never vanish.
Unlike those of ${\mathcal S}_t$, its cells may 
not be open or full-dimensional. 


\smallskip
\paragraph{Periodic sofic shifts.}

Any cell $c$ of ${\mathcal S}_\infty \subseteq {\mathcal S}_1$
lies within a single atom, so we can define $f_{|c}$ 
as the linear map corresponding to the transition matrix $P_c$. 
Since ${\mathcal S}_\infty$ is an invariant set, 
the image $f(c)$ must, by continuity, 
lie entirely within a cell of ${\mathcal S}_\infty$.
Suppose that $\# {\mathcal S}_\infty < \infty$, a fact we will prove shortly.
We define a directed graph $F$, with
each node labeled by a cell $c$ of ${\mathcal S}_\infty$
and with an edge $(c,c')$, labeled by $f_{|c}$,
joining $c$ to the unique cell $c'$ of ${\mathcal S}_\infty$
that contains $f(c)$. The system forms a
{\em sofic shift} (ie, a regular language over the edge labels).
Furthermore, $F$ is {\em functional}, meaning that
each node has exactly one outgoing edge 
(possibly a self-loop), so any infinite path ends up in a cycle.
The {\em trajectory} of a point ${\mathbf x}$ is the string 
$s({\mathbf x})= c_0 c_1\cdots$ of atoms that it visits:
$f^t({\mathbf x})\in c_t$ for all $0\leq t < \ell({\mathbf x})$.
It is infinite if and only if ${\mathbf x}$ does not vanish,
so all infinite trajectories
are eventually periodic. The weakness of this result is
that it might be a statement about the empty set. 
To strengthen it, we declare the system to be 
{\em nesting at} $t$ if no cell $c$ of ${\mathcal S}_{t}$
contains more than one cell of ${\mathcal S}_{t+1}$.
(This does not mean that $f(c)$ lies inside an atom.)
The minimum value of $t$ is called 
the {\em nesting time} $\nu$ of the system.
Observe that $\# {\mathcal S}_\nu \geq \# {\mathcal S}_t$,
for any $t\geq \nu$.
We bound the nesting time and then 
proceed with an alternative characterization of nesting.

\begin{lemma}\label{nestingtime-bound}
$\!\!\! .\,\,$
Both the nesting time $\nu$ and the number of cells in
${\mathcal S}_t$ are bounded by $(n/\eps)^{O(n)}$,  
for $t=0,1,\ldots,\infty$.
\end{lemma}

\proof
We begin with the second claim.
If, in~(\ref{Rteps-def}), we replace ${\mathbf x}$ by
$P {\mathbf x}$, for a stochastic matrix $P$,
the coefficients of the affine form remain 
polynomially bounded, so the cells of 
${\mathcal S}_t$ are separated
from one another by slabs of thickness at least 
$\eps n^{-O(1)}$. A simple volume argument implies an upper bound of
$(n/\eps)^{O(n)}$ on the number of such cells.
To bound the nesting time, consider this procedure: 
suppose that we have placed
a special point (called a {\em witness}) in each cell $c$ of
${\mathcal S}_t$. If $c$ contains only one cell of ${\mathcal S}_{t+1}$,
we move its witness to that unique cell; if it contains more than one
cell, then we move the witness to one of them and 
create new witnesses to supply
the others; if $c$ contains no cell of ${\mathcal S}_{t+1}$, 
we leave its witness in place.
We carry out this process for $t=0,1,\ldots$, beginning with a single
witness in ${\mathcal S}_0$.
Witnesses may move around but never disappear;
furthermore, by the previous argument, any two
of them are separated by at least $\eps n^{-O(1)}$,
so their number is bounded by $(n/\eps)^{O(n)}$.
Any time $t$ at which the system fails to be nesting 
sees the creation of at least one new witness, 
and the first claim follows.
\hfill $\Box$
\proofend

\begin{lemma}\label{geometric-facts}
$\!\!\! .\,\,$
Given any cell $c$ of ${\mathcal S}_t$ 
and $k\leq t$, the function $f_{|c}^k$ is linear.
Given any cell $b\subseteq \Omega^n$ and any linear function $g$, 
if $g(b)\setminus {\mathcal R}_{\eps}$ is
connected then so is $b\setminus g^{-1}({\mathcal R}_{\eps})$.
\end{lemma}
\proof
To call $f_{|c}^k$ linear is to say that $f^k$ is
described by a single stochastic matrix over all of $c$.
We may assume that $t>0$.
Given a cell $c\subseteq {\mathcal S}_t$, 
none of the cells $c, f(c),\ldots, f^{t-1}(c)$
intersect ${\mathcal R}_{\eps}$, hence each one falls
squarely within a single atom and
$f_{|c}^k$ is linear for any $k\leq t$. 
For the second claim, 
note that, if the cell $b$ intersects more than 
one connected component of 
$\Omega^n\setminus g^{-1}({\mathcal R}_{\eps})$, 
then it contains a segment $pq$ and a point $r\in pq$ such
that $g$ maps $p$ and $q$ outside of ${\mathcal R}_{\eps}$
and $r$ inside of it. By linearity, 
$g(r)$ lies on the segment $g(p)g(q)$;
therefore $g(b)\setminus {\mathcal R}_{\eps}$ is nonconvex
hence disconnected. 
\hfill $\Box$
\proofend

\begin{lemma}\label{nesting-charact}
$\!\!\! .\,\,$
The nesting time $\nu$ is the minimum $t$ such that
$f^t(c)\setminus {\mathcal R}_{\eps}$ is connected
for each cell $c$ of ${\mathcal S}_t$;
as a corollary, if $c$ is a cell of ${\mathcal S}_\nu$,
then $f(c)$ intersects at most one cell of ${\mathcal S}_\nu$.
\end{lemma}
\proof
The claims are trivial if $\nu=0$, so assume that $\nu>0$.
For the first claim, it
suffices to show that the system is nesting at time $t>0$ 
if and only if $f^t(c)\setminus {\mathcal R}_{\eps}$ is connected
for each cell $c$ of ${\mathcal S}_t$.
For the ``only'' part, we show why $f^t(c)\setminus {\mathcal R}_{\eps}$
must be connected.
By Lemma~\ref{geometric-facts},
$f^t_{|c}$ is linear; therefore, since
$c'= c\setminus f^{-t}({\mathcal R}_{\eps})$
is connected so is $f^t(c')= f^t(c)\setminus {\mathcal R}_{\eps}$.
Conversely, assuming that each set 
$f^t(c)\setminus {\mathcal R}_{\eps}$ is connected,
then we identify the function $g$ in Lemma~\ref{geometric-facts}
with $f^t_{|c}$ (in its linear extension)
and conclude that $c\setminus g^{-1}({\mathcal R}_{\eps})=c'$ 
is connected, hence constitutes
the sole cell of ${\mathcal S}_{t+1}$ lying within $c$.
To prove the corollary, again we turn to Lemma~\ref{geometric-facts}
to observe that $f_{|c},\ldots, f^\nu_{|c}$
are all linear, hence so is $g=f^{\nu-1}_{|b}$, for $b=f(c)$.
Our new characterization of nesting implies
that $f^\nu(c)\setminus {\mathcal R}_{\eps} =
g(b)\setminus {\mathcal R}_{\eps}$ is
connected, hence so is 
$b\setminus g^{-1}({\mathcal R}_{\eps}) =
f(c)\setminus f^{1-\nu}({\mathcal R}_{\eps})$.
Since $f(c)$ lies entirely within a cell of ${\mathcal S}_{\nu-1}$,
the labels of its points are all at least $\nu-1$. Removing
from $f(c)$ the points of label $\nu-1$ leaves the connected set
$f(c)\setminus f^{1-\nu}({\mathcal R}_{\eps})$;
therefore, $f(c)$ can intersect at most one cell of~${\mathcal S}_\nu$.
\hfill $\Box$
\proofend

We define the directed graph $F$ with one node
per cell $c$ of ${\mathcal S}_\nu$ and an edge from $c$ to
$c'$, where $c'$ is the unique cell of ${\mathcal S}_{\nu}$,
if it exists, that intersects $f(c)$. Every trajectory
corresponds to a directed path in $F$.
The main difference with the previous graph is that
the converse is not true. Not only a node may lack
an outgoing edge but, worse, nothing in this framework
keeps an orbit from going around a cycle for a while
only to vanish later. The previous lemma's
failure to ensure that $f(c)$ lies strictly within another cell
of $S_\nu$ puts periodicity in jeopardy.
Perturbation is meant to get around that difficulty.
Periods and preperiods are defined with respect to
the paths of $F$, not trajectories: since the correspondence
from paths to trajectories is not injective,
the latter may have shorter periods.

\begin{lemma}\label{nesting-periodic}
$\!\!\! .\,\,$
The system is nesting at $\nu$ and any time thereafter.
Any nonvanishing orbit is eventually periodic and 
the sum of its period and preperiod is bounded
by $\# {\mathcal S}_\nu$.
\end{lemma}

\smallskip
\paragraph{The attraction rate.}

Assume that $\nu>0$ and let $c$ be a cell of ${\mathcal S}_\nu$.
Identifying the nodes of $F$ with their cells in ${\mathcal S}_\nu$,
we denote by $\sigma_0,\sigma_1,\ldots$ the path 
from $c = \sigma_0$.
Let $j$ be the smallest index such that
$\sigma_i= \sigma_j$ for some $i<j$.
This defines the period $p=p(c)= j-i$ and the preperiod
$q=q(c)= i$, with $p+q\leq \#{\mathcal S}_\nu$.
Given any ${\mathbf x}\in c$, its trajectory 
$s({\mathbf x})= c_0 c_1\cdots c_{\ell({\mathbf x})-1}$ 
is such that $c_k$ is the atom containing the cell $\sigma_k$.
Furthermore, 
for any $q\leq t\leq \ell({\mathbf x})$,
\begin{equation}\label{x(t)M}
f^t({\mathbf x}) = M_{t-q\, (\text{\scriptsize mod}\,p)} \,
                      Q^{\lfloor (t-q)/p \rfloor} f^q({\mathbf x}),
\end{equation}
where $M_k= P_{c_{q+k-1}}\cdots P_{c_{q}}$, for
$k=0,\ldots, p-1$, and $Q=M_p$, with $M_0$ the identity matrix.\footnote{\,
Note that $f^q({\mathbf x})= P_{c_{q-1}}\cdots P_{c_0} {\mathbf x}$,
with the matrix denoting the identity if $q=0$.}
Because of the self-loops in the communication graphs, the powers of $Q$ 
are known to converge to a matrix $\widetilde Q$~\cite{seneta06}.
Given $c$ and $t\geq 0$, we define 
$$
\Pi_{\, t}= 
M_{ t-q \, (\text{\scriptsize mod}\,p)}
\, \widetilde Q\, P_{c_{q-1}}\cdots P_{c_0}.
$$
The approximation $\Pi_{t}$ is one of $p$ matrices
obtained by substituting $\widetilde Q$ for as many 
``chunks'' $Q=M_p$ we can extract from the 
matrix product $P_{c_{t-1}}\cdots P_{c_0}$ that defines $f_{|c}$.
Note that this includes the case $t<p+q$,
where no such chunk is to be found.
Given any real $\alpha> 0$, we define the
{\em attraction rate} $\theta_\alpha$ as
the maximum value of $\theta_\alpha(c)$,
over all cells $c$ of ${\mathcal S}_\nu$,
where
\begin{equation}\label{theta-alphaVP}
\begin{split}
\theta_\alpha(c) = 
\min \, \Bigl\{\, \theta \geq 0 \,:\,  
&
\| f^t({\mathbf x}) - 
   \Pi_{\,t (\text{\scriptsize mod}\,p)}\,{\mathbf x} \|_\infty
       \leq \, \alpha\, ,    \\
&  \hspace{2cm}   \text{ for all ${\mathbf x}\in c$} \,\,\,\text{and} \,\,\,
   \theta\leq t\leq \ell( {\mathbf x} ) \,  
                \Bigr\}.
\end{split}
\end{equation}

\noindent
Suppose that $Q$ can be written as
\begin{equation}\label{Qmatrix-def}
Q=  
\begin{pmatrix}
A & C \\ \emptyset  & B
\end{pmatrix}
\end{equation}
and assume the existence of a limit matrix
$\widetilde B$ such that $\|B^t- \widetilde B\|_\text{max}
\leq e^{-\gamma t}$, for some $\gamma>0$.
We tie the attraction rate to 
the maximum row-sum in $A$, which is itself related to
the thinning rate (whose formal definition we postpone).

\begin{lemma}\label{matrixproduct}
$\!\!\! .\,\,$
Given $Q$ as in~{\rm (\ref{Qmatrix-def})} 
and an upper bound $\mu$ on $\|A \, {\mathbf 1}\|_\infty$ such that
$e^{-\gamma}\leq \mu<1$, for any $0<\alpha<1-\mu$,
$$\theta_\alpha =
 O\Bigl(\, \frac{\#{\mathcal S}_\nu}{1-\mu} \,\Bigr)
                    \log\hbox{$\frac{n}{\alpha}$}\,.
$$
\end{lemma}
\proof
For any $t>0$,
$$ Q^t =
\begin{pmatrix}
A^t & C_t \\ 0 & B^t
\end{pmatrix}.
$$
The matrix $A$ is strictly substochastic ($\mu<1$), so, by
standard properties of a Markov chain's fundamental matrix,
$\sum_{k\geq 0} A^k = (I-A)^{-1}$; therefore, for $t>0$,
\begin{equation*}
C_t- (I-A^t)(I-A)^{-1}C\widetilde B 
= \sum_{k=0}^{t-1} A^{t-k-1}CB^k- \sum_{k=0}^{t-1} A^k C\widetilde B \\
= \sum_{k=0}^{t-1} A^{t-k-1}CD_k,
\end{equation*}
where $D_k= B^k - \widetilde B$. Since $C$ is substochastic,
$\|C D_k\|_{\text{max}} \leq e^{- \gamma k}$.
From $\|A^k \,{\mathbf 1}\|_\infty\leq \mu^k$, we derive
$$ \|A^{t-k-1}CD_k \|_\text{max} \leq \mu^{t-k-1}e^{- \gamma k}.$$
Since $\mu\geq e^{-\gamma}$, it follows that
$\| C_t- (I-A^t)(I-A)^{-1}C\widetilde B \|_\text{max}
\leq t\mu^{t-1}$; hence,
\begin{equation*}
\widetilde Q =
\begin{pmatrix}
0 & (I - A)^{-1}C\widetilde B \, \\
0 & \widetilde B 
\end{pmatrix},
\end{equation*}
where, by $\|A^t\|_\text{max}\leq \mu^t$
and $\|(I-A)^{-1}\|_\text{max}\leq 1/(1-\mu)$,
$\| Q^t - \widetilde Q \|_\text{max}
= O(tn \mu^{t-1}/(1-\mu))$. As a result,
by Lemma~\ref{nesting-periodic},
$\theta_\alpha\leq q+ pt\leq  (\#{\mathcal S}_\nu)t$,
if $t$ satisfies $t\mu^{t-1}<\alpha(1-\mu)n^{-b}$, for 
a large enough constant $b>0$.
\hfill $\Box$
\proofend

\smallskip
\noindent
This next result argues that, although a vanishing 
point may take arbitrarily long to do so, 
it comes close to vanishing
fairly early. This gives us a useful analytical device
to avoid summing complicated series when estimating
the probability that a point will eventually vanish
under random margin perturbation.

\begin{lemma}\label{infinite-path}
$\!\!\! .\,\,$
Given any finitely-labeled point ${\mathbf x}$ in a cell $c$
of ${\mathcal S}_\nu$, 
there exists $t< \theta_\alpha +p(c)+q(c)$ such that
$f^t({\mathbf x})\in {\mathcal R}_{2\eps}$, for
some $\alpha\geq \eps n^{-O(1)}$.
\end{lemma}

\proof 
We can obviously assume that 
$\ell ({\mathbf x}) \geq \theta_\alpha +p(c)+q(c)$.
For any $t$ such that $\theta_\alpha\leq t\leq \ell({\mathbf x})$,
$f^t({\mathbf x})$ lies in an $\ell_\infty$ ball of radius $\alpha$ centered at 
$\Pi_{\,t \, \text{\scriptsize mod}\,p(c)}\, {\mathbf x}$.
This means that, between times $\theta_\alpha +q(c)$ and $\ell({\mathbf x})$,
the orbit of ${\mathbf x}$ lies entirely in the union of $p(c)$ balls of 
radius $\alpha$ and, by periodicity, each ball is visited
before time $\theta_\alpha+p(c)+q(c)$. Since ${\mathbf x}$ vanishes at 
time $\ell({\mathbf x})$, 
one of these $p(c)$ balls must intersect ${\mathcal R}_\eps$.
Thickening all the margin slabs by a width of $4\alpha\sqrt{n}$
is enough to cover that ball entirely. If $\alpha= \eps n^{-b}$
for a large enough constant $b$,
replacing $\eps$ by $2\eps$ achieves the required thickening.
\hfill $\Box$
\proofend

\noindent
Although nesting occurs within finite time, the strict inclusion
${\mathcal S}_{k+1} \subset {\mathcal S}_k$ may occur
infinitely often. We show why:

\medskip
{\small
\begin{quote}
\label{ex1}
{\sc Example \ref{ex1}}:\ 
Vanishing can take arbitrarily long.
Consider the two-agent influence system
\begin{equation*}
\begin{pmatrix}
x_1 \\ x_2 
\end{pmatrix}
\,\stackrel{f}\longmapsto \,
 \hbox{$\frac{1}{3}$}
\begin{pmatrix}
2 & 1 \\ 1 & 2 
\end{pmatrix}
\begin{pmatrix}
x_1 \\ x_2 
\end{pmatrix},
\end{equation*}
with the {\em SP} discontinuities formed by the single slab
$${\mathcal R}_\eps= \{\, {\mathbf x}\in {\mathbb R}^3 \,:\,  
                             | x_1 - 1+\delta | \leq \eps \,\}.$$
For simplicity, assume the same linear map $f$
in the two atoms. It follows that
\begin{equation*}
\begin{pmatrix}
x_1 \\ x_2 
\end{pmatrix}
\,\stackrel{f^t}\longmapsto \,\hbox{$\frac{1}{2}$}
\begin{pmatrix}
1+3^{-t} & 1-3^{-t}  \\ 1-3^{-t} & 1+3^{-t} 
\end{pmatrix}
\begin{pmatrix}
x_1 \\ x_2 
\end{pmatrix}.
\end{equation*}
The set ${\mathcal S}_\infty$ is the complement within $\Omega^2$ 
(the effective phase space) of 
$$
\bigcup_{t=0}^\infty \,\Bigl\{\,|\, (1+3^{-t})X_1+ 
                    (1-3^{-t})X_2-2+2\delta \,|\leq 2\eps \,\Bigr\}.
$$
Note that if $\eps=0$, the number of cells in ${\mathcal S}_\infty$ 
is infinite: they are defined by an infinite number of lines
passing through $(1-\delta,1-\delta)$, with increasing slopes 
tending to $-1$. As soon as we allow thickness $\eps>0$, however,
the margin creates only $O(|\! \log \eps|)$ cells. 
Not all of them are open. To see this, consider the point 
$2(1-\delta+\eps,0)$. It never vanishes yet any neighborhood
contains points that do. Some points take arbitrarily long to vanish.
Thickening the {\em SP} discontinuities into slabs
is a ``finitizing'' device meant to keep the number of 
$\infty$-labeled cells bounded.
\end{quote}
}

\smallskip
\subsection{The coding tree}\label{sec:codingtreeT}

The richly decorated tree ${\mathcal T}$ encodes the
branching structure of the sets ${\mathcal S}_k$ as a geometric
object in ``phase space $\times$ time'' $=\Omega^n\times [0,\infty)$.
Recall that each atom $c$ comes with its own transition matrix $P_c$.
Unless specified otherwise, a fixed perturbation value of $\delta$ is assumed
once and for all.
Think of $U_v$ and $V_v$ as the end-sections at times
$0$ and $t_v$ of a 
phase tube containing all the orbits originating from $U_v$.
At time $t_v$, the {\em SP} discontinuities might split the tube. This
happens only if $V_v$ intersects the margin, which
is the ``branching condition'' in the boxed algorithm.
That intersection indicates the vanishing
time of some points in $U_v$, so we place a leaf as an indicator, and call it 
a {\em vanishing node}. Whereas $U_v$ is an open $n$-cell, 
$V_v$ can be a cell of any dimension; hence so can be the connected components 
of $V_v\setminus {\mathcal R}_{\eps}$.
For each one, $c$, we attach a new child $w$ to $v$
and denote by $P_w$ the matrix of the map's restriction to $c$.
The image of $c$ at time $t_v$, ie, $P_w\, c$, forms the 
end-section $V_w$ of a new phase tube from the root,
whose starting section $U_w$ is the portion of $U_v$
mapping to $c$ at time $t_v$ (Fig.~\ref{fig-coding-tree}).\footnote{\,
Note that $U_w$ cannot be defined as the portion of $U_v$
mapping to $V_w$ at time $t_w$: the orbits must pass through $c$.}

\bigskip\bigskip
{\small
\par
\renewcommand{\sboxsep}{0.7cm}
\renewcommand{\sdim}{0.8\fboxsep}
\hspace{.0cm}
\shabox{\parbox{11.9cm}{
\vspace{-0.6cm}
\begin{center}{\bf Building ${\mathcal T}$}\end{center}
\smallskip
\begin{itemize}
\item[\bf{[1]$\,\,$}]
The root $v$ has depth $t_v=0$;
set $U_v\leftarrow V_v\leftarrow \Omega^n$.
\item[\bf{[2]$\,\,$}]
Repeat forever:
\begin{itemize}
\item[\bf{[2.1]}]
    For each newly created node $v$:
    \begin{itemize}
    \item[$\bullet$]
        If $V_v\cap {\mathcal R}_{\eps}\neq \emptyset$ 
                    \hspace{1.4cm}\,\, [ {\tt branching condition} ] \\
        then create a leaf and make it a child of $v$.
    \item[$\bullet$]
                   For each cell $c$ of $V_v\setminus 
                                 {\mathcal R}_{\eps}$,
           create a child $w$ of $v$ and \\
      set $P_w\leftarrow P_c$ ; \, $V_w\leftarrow  P_w \, c$ ; \,
      $U_w\leftarrow  U_v\cap f^{-t_v}(c)$.
    \end{itemize}
\end{itemize}
\end{itemize}
}}
\par\bigskip
}

\bigskip\bigskip

\noindent
Let $ww'w''\cdots$ denote the upward, $t_w$-node path 
from $w$ to the root (but excluding the root). 
Using the notation $P_{\leq w}= P_wP_{w'}P_{w''}\cdots$,
we have the identity $V_w= P_{\leq w}\, U_w$.
No point in $U_w$ vanishes before time $t_w$, and, 
in fact, ${\mathcal S}_k= \bigcup_w \{\,U_w\,|\, t_w= k \,\}$.
The points of ${\mathcal S}_\infty$ are precisely
those whose orbits follow an infinite path 
$v_\infty= v_0,v_1,v_2,\ldots$ down the  
coding tree. Each such path has its own limit cell 
$U_{v_\infty}= \bigcap_{\, t\geq 0} U_{v_t}$: collectively,
these form the cells of ${\mathcal S}_\infty$.
Example~\ref{ex1} features two infinite paths each of whose nodes has
two children, one vanishing and one not.
\begin{itemize}
\item
The {\em nesting time} $\nu= \nu({\mathcal T})$ is the minimum depth
at which any node has at most one nonvanishing child 
(Lemma~\ref{nesting-charact}); visually, 
below depth $\nu$, the tree consists of
single paths, some finite, others infinite,
with vanishing leaves hanging off some of them.
A node $v$ is {\em deep} if $t_v > \nu$
and {\em shallow} otherwise.
\item
The {\em word-entropy} $h({\mathcal T})$ is
the logarithm of the number of shallow nodes.\footnote{\,
The trajectories form a language $L({\mathcal T})$ over
the alphabet of atom labels. Its growth rate
plays a key role in the analysis and is bounded via
the word-entropy.} As we observed, 
${\mathcal S}_\nu= \bigcup_v \{\,U_v\,|\, t_v= \nu \,\}$;
therefore $\# {\mathcal S}_\nu\leq 2^{h({\mathcal T})}$.
\item
The {\em period} $p({\mathcal T})$ is the maximum value
of $p(c)$ for all cells $c= U_v$, with $t_v=\nu$.
The attraction rate $\theta_\alpha({\mathcal T})$
is the maximum value of the attraction rate for any such $c$.
\end{itemize}

\smallskip
\paragraph{The global coding tree.}

Let ${\mathbb I}$ denote the interval $[-1,1]$.
Since not all perturbations~$\delta$ 
are equally good, we must understand
how the coding tree ${\mathcal T}$ varies
as a function of~$\delta$. To do that, a global
approach is necessary: given $\Delta\subseteq {\mathbb I}$,
we encode the coding trees for all $\delta\in \Delta$ into a single one, 
${\mathcal T}^\Delta$, which can be viewed
as the standard coding tree for the augmented $(n+1)$-dimensional 
system $({\mathbf x}, \delta)\mapsto (f({\mathbf x}), \delta)$,
with the phase space $\Omega^n\times \Delta$.
The sets $U_v$ and $V_v$ are now cells in ${\mathbb R}^{n+1}$.
In the branching condition, one should replace 
the margin ${\mathcal R}_{\eps}$, as defined in~(\ref{Rteps-def}),
by the {\em global margin}:
\begin{equation}\label{globalmargin-defn}
\bigcup_{\text{\em SP}} \, \Bigl\{\, 
({\mathbf x}, \delta)\in {\mathbb R}^{n+1} \,:\, 
     | a_0+a_1x_1+\cdots + a_nx_n + \delta | \leq \eps \,\Bigr\}.
\end{equation}
The degree of any node is bounded
by $n^{O(n)}$, which is the maximum number of cells in
an arrangement of $n^{O(1)}$ hyperplanes in ${\mathbb R}^{n+1}$.
The definition of nesting can be extended, unchanged, to this lifted system.
Since a standard coding tree is just a ``cross-section'' of the global one,
nesting in ${\mathcal T}$ even for all $\delta$ 
does not imply nesting
in ${\mathcal T}^\Delta$.\footnote{\, Just as 
a region in the $(X,Y)$-plane need not be connected 
simply because all of its horizontal cross-sections are.}
The global word-entropy $h({\mathcal T}^\Delta)$ is defined 
in the obvious way.

\smallskip
\subsection{The arborator}\label{sec:arborator}

This algorithm assembles the coding tree by glueing smaller pieces 
together. It relies on a few primitives that we now describe. 
The direct sum and direct product are tensor-like
operations used to attach coding trees together. 
The primitives {\tt absorb} and {\tt renorm} respectively 
prune and compress trees. We present these operations
and assemble the dictionary that allows us to
bound the coding tree's parameters as we parse the arborator.

\smallskip
\paragraph{Direct sum.}

The coding tree ${\mathcal T} = {\mathcal T}_1 \oplus {\mathcal T}_2$
models two {\em independent} systems of size $n_1$ and $n_2$.
Independence means that the systems are {\em decoupled}
(no edge joins agents from distinct groups) and
{\em oblivious} (no {\em SP} discontinuity has nonzero coefficients
from both groups): this implies that the two systems
can be analyzed separately; decoupling alone is not sufficient.
The phase space of the direct sum is of dimension $n=n_1+n_2$.
A path $w_0,w_1,\ldots$ of ${\mathcal T}$ 
is a pairing of paths in the constituent trees:
the node $w_t$ is of the form $(u_t, v_t)$, 
where $u_t$ (resp. $v_t$) is a node of ${\mathcal T}_1$ 
(resp. ${\mathcal T}_2$) at depth $t$;  it is a leaf if and only if
$u_t$ or $v_t$ is one---the vanishing of one group implies
the vanishing of the whole. If $w=(u,v)$ is not a leaf, then 
$U_w = U_u\times U_v$, and $V_w = V_u\times V_v$.
The direct sum is commutative and associative.
The name comes from the fact that $P_w$ is
the direct matrix sum of $P_u$ and $P_v$:
\begin{equation*}
P_w= P_u \oplus P_v=
\begin{pmatrix}
P_u & 0 \\
0 &  P_v
\end{pmatrix} .
\end{equation*}

\begin{itemize}
\item {\em Nesting time, period, and attraction rate.}

\begin{equation}\label{DS-all}
\nu({\mathcal T})
    \leq \, \max_{i=1,2}\,\, \nu({\mathcal T}_i) 
\hspace{.4cm}\text{and}\hspace{.4cm}
p({\mathcal T})\leq \prod_{i=1,2}\,\, p({\mathcal T}_i)
\hspace{.4cm}\text{and}\hspace{.4cm}
\theta_\alpha({\mathcal T})
\leq\, \max_{i=1,2}\,\, \theta_\alpha({\mathcal T}_i).
\end{equation}
The first two inequalities are obvious, so we focus on the last one.
Consider a cell $c=c_1\times c_2$ of ${\mathcal S}_\nu$ and 
follow the path of $F$ emanating from it: this navigation
corresponds to the parallel traversal of a path in $F_i$
from $c_i$---we use the subscript $i=1,2$ to refer to either
one of the subsystems. Assume without loss of generality that $q_1\geq q_2$.
By definition, to revisit an earlier node
means doing likewise in each traversal; hence $q\geq q_1$. 
At time $q_1$, however,
both parallel traversals are already engaged
in their own respective cycles, so the node pair at time $q_1$ will 
be revisited lcm$\,(p_1,p_2)$ steps later,
the time span that constitutes the period $p$
of the direct sum; it also follows that $q=q_1$.
If $q_1>q_2$, the traversals do not
enter their cycles at the same time, so in general,
referring to~(\ref{x(t)M}), the matrix
$Q$ is not the direct sum of $Q_1^{p/p_1}$ and $Q_2^{p/p_2}$ but, rather,
of a shifted version $ Q= Q_1^{p/p_1} \oplus (BA)^{p/p_2}$,
where $Q_2=AB$. We easily verify that
\begin{equation*}
\widetilde Q= \lim_{k\rightarrow \infty} Q^k=
\widetilde Q_1 \oplus (B \widetilde Q_2 A),
\end{equation*}
where, as before, $\widetilde Q_i= \lim_{k\rightarrow \infty} Q_i^k$.
The use of the $\ell_\infty$ norm allows us
to verify the bound on the attraction rate of the direct sum by 
checking the accuracy of the approximation for each subsystem separately.
It suffices to focus on the case of ${\mathcal T}_2$, which 
presents the added difficulty that the approximation scheme
delays the cycle entrance until $q_1$. The other difference with
the approximation scheme in the original system ${\mathcal T}_2$
is that, since the period can be much longer, so can
the sequence $(M_2)_k$. In all cases, however, the approximation
scheme in ${\mathcal T}$ as it applies to ${\mathcal T}_2$
differs from the scheme in ${\mathcal T}_2$ in only one substantive way.
Consider the language consisting of the words 
$(AB)^*$, $(AB)^*A$, $(BA)^*$, and $(BA)^*B$. One approximation 
scheme involves replacing any number of ``$AB$''s by $\widetilde Q_2$,
while the other scheme replaces any number of ``$BA$''s by $B \widetilde Q_2 A$.
Because $AB \widetilde Q_2= \widetilde Q_2 AB = \widetilde Q_2$,
any application of one scheme or the other 
produces the same matrix.

\item {\em Word-entropy.}
We prove (quasi) subadditivity. 
Assume without loss of generality that 
$\nu({\mathcal T}_1) \geq \nu({\mathcal T}_2)$.
The word-entropy counts the number of shallow nodes $w=(u,v)$.
This implies that $t_u\leq \nu({\mathcal T}_1)$, which limits
the number of such nodes $u$ to $2^{h({\mathcal T}_1)}$.
If all the nodes $v$ were shallow in ${\mathcal T}_2$, the subadditivity
of word-entropy would be immediate; but it need not be the case.
If $v$ is deep, let $s(v)$ be its deepest shallow ancestor.
The function $s$ may not be injective 
but it is at most two-to-one.
Thus,
\begin{equation}\label{DS:h}
h({\mathcal T})
\leq  h({\mathcal T}_1 ) +  h({\mathcal T}_2 )+1.
\end{equation}
\end{itemize}
All of the relations in~(\ref{DS-all}, \ref{DS:h}) still hold
when the superscript $\Delta$ is added to the coding trees.
We discuss~(\ref{DS:h}) to illustrate the underlying principle.
First, we provide an independent
perturbation variable $\delta_i\in \Delta$ to ${\mathcal T}_i$
($i=1,2$) and add it as an extra coordinate to the state vector,
thus lifting system $i$ to dimension $n_i+1$.
By~(\ref{DS:h}), the word-entropy of the joint system 
${\mathcal T}^*$ in dimension $n+2$ is at most 
$h({\mathcal T}_1^{\Delta} ) +  h({\mathcal T}_2^{\Delta} )+1$.
Second, we restrict the system ${\mathcal T}^*$ 
to the invariant hyperplane $\delta_1=\delta_2$, which cannot increase
the word-entropy, hence $h({\mathcal T}^\Delta)\leq h({\mathcal T}^*)$,
as claimed.

\smallskip
\paragraph{Absorption.}

\begin{figure}[htb]
\begin{center}
\hspace{.2cm}
\includegraphics[width=5.5cm]{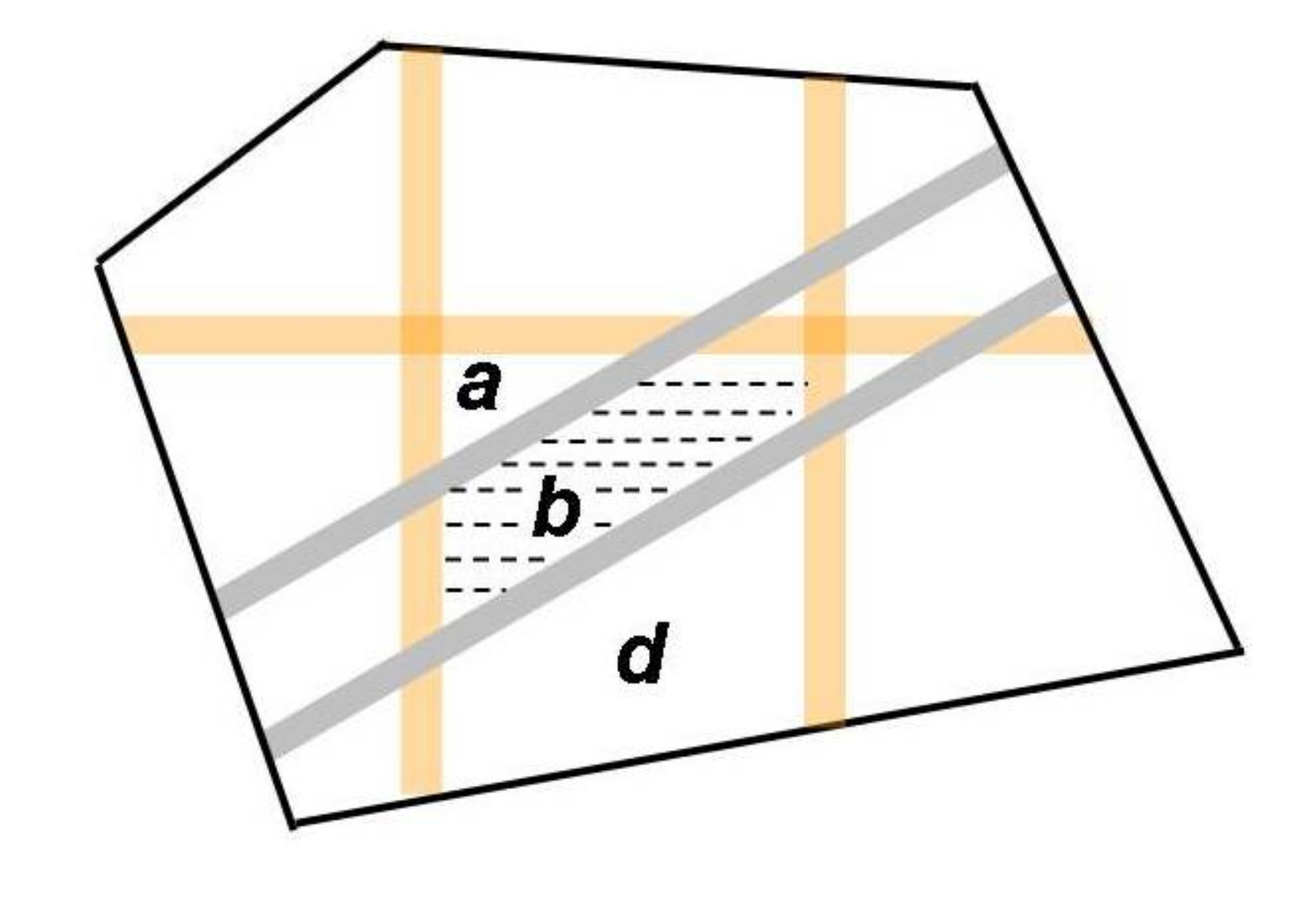}
\end{center}
\vspace{-0.5cm}
\caption{\small 
The original cell $c$ of $V_v$ 
(bottom-center trapezoid) splits
up into the wet cell $b$ and the dry cells $a$ and $d$,
both of which inherit the matrix $P_c$.}
\label{fig-absorb}
\end{figure}

The direct product, which we define below, requires an 
intermediate construction. The goal is to allow the selection
of nodes for removal, with an eye toward replacing
the subtrees they root by coding trees with
different characteristics.
The selection is carried out by an operation called
${\tt absorb}({\mathcal T})$, which replaces 
any deleted node by a leaf. For reasons that
the {\em flow tracker} will soon clarify, 
such leaves are designated {\em wet}.
An orbit that lands into one of these 
wet leaves is suddenly governed by a different dynamics, 
modeled by a different coding tree, so from the perspective of 
${\mathcal T}$ alone, wet leaves are where orbits come to a halt.
While vanishing leaves signal the termination
of an orbit (at least from the perspective
of the analysis), the wet variety merely indicates a change
of dynamics. Here is a simple illustration:

\medskip
{\small
\begin{quote}
\label{ex5}
{\sc Example \ref{ex5}}:\ 
The system consists of two independent
subsystems. Suppose we add a union of slabs,
denoted by ${\mathcal R}_{\eps}'$,
to the original margin ${\mathcal R}_{\eps}^o$, thus breaking the
direct-sum nature of the coding tree.
In Fig.~\ref{fig-absorb}, ${\mathcal R}_{\eps}'$
would consist of the two infinite strips bordering $b$. 
We keep the transition matrices unchanged everywhere
except in cell $b$, which we call {\em wet}:
all transition matrices are still direct (matrix) sums,
with the possible exception of $P_b$.
Suppose we had available the coding tree prior to the margin's
augmentation. Let $V_v$ denote the pentagon in the figure
and $w$ be the node associated
with the trapezoid $c$ that holds $a,b,d$.
We need to replace $w$ by three nodes: two of them
for $a,d$ and one, a {\em wet leaf}, for $b$.
The transition matrices for $a,d$ are both
equal to the direct sum $P_c$, while $P_b$ can be
arbitrary. The idea is that $b$ can then be made
the region $U_{\text{root}}$ of a new coding tree.

\hspace{.4cm} Minor technicality: usually, $U_{\text{root}}=\Omega^n$,
so the coding tree must be {\em cropped} by substituting
$b$ for $\Omega^n$; note that $b$ need not be
an invariant set. Cropping might involve pruning the tree
but it cannot increase any of the key parameters, such as 
the nesting time, the attraction rate, and the period.
Absorption appeals to the fact that we can ignore
$b$ and its wet leaf until we have fully analyzed
the direct sum. This separation is very useful, especially
since absorption does not require a direct sum---we never used 
the fact that the old slabs were horizontal or vertical---and is
therefore extremely general. 
\end{quote}
}

\vspace{0.5cm}
\vspace{.5cm}

\noindent
A crucial observation is that the nodes $z$ created
for the subcells $c'$ of a given $c$ 
(subcells $a$ and $d$ in Fig.~\ref{fig-absorb})
have the same matrix $P_c$. As a result of all the absorptions,
the tube $(U_v,V_v)$ is split up by up to $t_v$ linearly
transformed copies of the $n^{O(1)}$ margin slabs, 
hence into at most $t_v^n n^{O(n)}$ subcells.
This compares favorably with the naive upper bound
of $n^{O(n t_v)}$ based on the sole fact that absorption
at each ancestor of $v$ produces $n^{O(n)}$ children.

\bigskip\bigskip
{\small
\par
\renewcommand{\sboxsep}{0.7cm}
\renewcommand{\sdim}{0.8\fboxsep}
\hspace{-.3cm}
\shabox{\parbox{12.4cm}{
\vspace{-0.6cm}
\begin{center}{\bf Absorption surgery} \end{center}
\smallskip
\begin{itemize}
\item[\bf{[1]$\,\,$}]
If $v$ has no leaf,
create a vanishing leaf and make it a child of $v$. 
\item[\bf{[2]$\,\,$}]
For each cell $c$ of
$V_v\setminus {\mathcal R}_{\eps}^o$,
let $w$ be the child of $v$ for $c$ (ie, such that
$f(c)=V_w$) and let $T$ be the tree rooted at $w$.
If $c\cap {\mathcal R}_{\eps}'$, then remove $T$ and,
for each cell $c'$ of $c\setminus {\mathcal R}_{\eps}'$,
create a node $z$ and make it a child of $v$.
\begin{itemize}
\item[$\bullet$]
If $c'$ is wet, make $z$ a wet leaf. 
\item[$\bullet$]
If $c'$ is dry, reattach to $z$ a suitably cropped 
copy of $T$. \\
Set $P_{z} \leftarrow  P_c$,
$V_{z}\leftarrow  P_{z} \, c$, and
$U_{z}\leftarrow  U_v\cap f^{-t_v}(c')$.
\end{itemize}
\end{itemize}
}}
\par\bigskip
}

\smallskip
\paragraph{Direct product.}

The tree ${\mathcal T}= {\mathcal T}_1 \otimes {\mathcal T}_2$ 
models the concatenation of two systems.
The direct product is associative but not commutative.
It is always preceded by a round of
absorptions at one or several nodes of ${\mathcal T}_1$.
We begin with a few words of intuition.
Consider two systems $S_1$ and $S_2$,
governed by different dynamics yet evolving 
in the same phase space $\Omega^n$.
Given an arbitrary region $\Lambda\subset \Omega^n$, 
we define the hybrid system $S$ with the dynamics of $S_2$ over $\Lambda$
and $S_1$ elsewhere. Suppose we had complete knowledge of the 
coding tree ${\mathcal T}_i$ for each $S_i$ ($i=1,2$).
Could we then combine them in some ways to assemble the coding tree 
${\mathcal T}$ of $S$?
To answer this question, we follow a three-step approach: 
\begin{itemize}
\item{(i)}
we absorb the tree ${\mathcal T}_1$
by creating wet leaves $w$ for all the nodes 
$v$ with $V_v\cap \Lambda\neq \emptyset$; 
\item{(ii)}
we attach the roots of cropped copies of ${\mathcal T}_2$ 
at the wet leaves; and 
\item{(iii)}
we iterate and glue ${\mathcal T}_1$ and ${\mathcal T}_2$ in alternation,
as orbits move back and forth in and out~of~$\Lambda$.
\end{itemize}
Absorption, direct products, and the arborator
address (i, ii, iii) in that order.
The root of ${\mathcal T}_2$ is attached to $w$, but not until
that tree itself has been properly cropped so that 
$U_{\text{root}}= V_{\text{root}}({\mathcal T}_2)
=  V_w({\mathcal T}_1) = P_w\, c$, with $P_w$
given by ${\mathcal T}_2$ and not ${\mathcal T}_1$.
To be fully rigorous, we should write a direct product
as ${\mathcal T}_1 \otimes \{ {\mathcal T}_2 \}$ since
the trees ${\mathcal T}_2$ 
we attach to the wet nodes might not all be the same:
the cropping might vary, as might the wet regions.

\begin{itemize}

\item {\em Nesting time and attraction rate.}
Bounding the nesting time of
a direct product is not merely a combinatorial matter,
as was the case for direct sums: the geometry 
of attraction plays a role.
Even the case of ${\tt absorb}({\mathcal T}_1)$ demands
some attention and this is where we begin.
Adding ${\mathcal R}_{\eps}'$ to the margin 
cannot create arbitrarily deep wet nodes: 
specifically, no $v\in {\tt absorb}({\mathcal T}_1)$ 
of depth at least 
$\max\{\, \nu({\mathcal T}_1), \theta_{\alpha_a}({\mathcal T}_1) + p\,\}$
can have a wet child, where $p= p(U_v)$ and $\alpha_a= \eps n^{-a}$ 
for a large enough constant $a$.
Indeed, suppose there is such a node $v$.
Pick ${\mathbf x}\in U_v$
such that $f^{t_v}({\mathbf x})$ lies
in a wet cell $c$ within $V_v$ and observe that
$$
\|f^{t_v}({\mathbf x})- f^{t_v-p}({\mathbf x})\|_\infty
\leq 
\|f^{t_v}({\mathbf x})- 
\Pi_{\,t_v (\text{\scriptsize mod}\,p)}\,{\mathbf x} \|_\infty
+
\|f^{t_v-p}({\mathbf x})- 
\Pi_{\,t_v (\text{\scriptsize mod}\,p)}\,{\mathbf x} \|_\infty
\leq 2\alpha_a.
$$

\hspace{.4cm}
By our choice of $\alpha_a$, this implies
that $f^{t_v}({\mathbf x})$ and $f^{t_v-p}({\mathbf x})$ are 
at a distance apart
less than the width of the margin's slabs; therefore, 
$f^{t_v-p}({\mathbf x})$ lies in a wet cell or
in a slab. It follows that the orbit of ${\mathbf x}$
either vanishes or comes to a (wet) halt at a time 
earlier than $t_v$, so 
${\mathbf x}\not\in U_v$ and we have a contradiction.
It follows that all deep nodes of ${\mathcal T}_1$
deeper than $\theta_{\alpha_a}({\mathcal T}_1) + p({\mathcal T}_1)$
are also deep in ${\tt absorb}({\mathcal T}_1)$.
With ${\mathcal T}= {\mathcal T}_1 \otimes {\mathcal T}_2$, therefore,
\begin{equation}\label{nest-dp}
\begin{cases}
\,\, \nu ({\tt absorb}({\mathcal T}_1))
            \,\leq \max\{\, \nu({\mathcal T}_1),\,
\theta_{\alpha_a}({\mathcal T}_1) +  p({\mathcal T}_1)\,\}\, ,
\hspace{.3cm} \text{for some $\alpha_a\geq \eps n^{-O(1)}$} \\
\,\,
\nu({\mathcal T})
\,\leq \,  \nu ({\tt absorb}({\mathcal T}_1))
           + \nu({\mathcal T}_2)
  \\
\,\, \theta_\alpha({\mathcal T})
\, \leq \max\{\, 
             \theta_\alpha({\mathcal T}_1), \,
             \nu ({\tt absorb}({\mathcal T}_1))
                  + \theta_\alpha({\mathcal T}_2) \,\} 
\end{cases}
\end{equation}

\item {\em Word-entropy.}
Absorption can occur only at nodes $v$ of depth
$t_v\leq \,\nu ({\tt absorb}({\mathcal T}_1))$. This means that 
the number of nodes where wet cells can emerge is at most 
$2^{h({\mathcal T}_1)} 
(\theta_{\alpha_a}({\mathcal T}_1) 
+ p({\mathcal T}_1) )$.
As we argued earlier, each such node $v$ can 
give birth to at most $t_v^n n^{O(n)}$ new nodes, so the
number of shallow nodes in ${\mathcal T}$ is (conservatively) at most
\begin{equation*}
\overset{\# v\,\,{\tt with\,\, wet\,\, child}}{\overbrace{\, 
2^{h({\mathcal T}_1)} (\theta_{\alpha_a}({\mathcal T}_1) 
                                      + p({\mathcal T}_1) ) \,}} 
\times
\overset{\# {\tt splits}/v}{\overbrace{\, 
       ( \nu({\mathcal T}_1)  +
       \theta_{\alpha_a}({\mathcal T}_1) + p({\mathcal T}_1) )^n
       n^{O(n)} \,}} 
\times
\overset{\# {\mathcal T}_2 \,\, {\tt nodes}}
            {\overbrace{\, 2^{h({\mathcal T}_2)} \,}} .
\end{equation*}
We use the fact that cropping cannot increase the word-entropy.
Taking logarithms, we find that
\begin{equation}\label{entr-dpp}
h({\mathcal T})
\leq h({\mathcal T}_1) + h({\mathcal T}_2) +
(n+1)\log ( \, \nu({\mathcal T}_1) + \theta_{\alpha_a}({\mathcal T}_1) 
+ p({\mathcal T}_1) \, ) + O(n\log n).
\end{equation}
Since both $\nu({\mathcal T}_1)$ and $p({\mathcal T}_1)$
are no greater than $2^{ h({\mathcal T}_1) }$, we can simplify the bound:
\begin{equation}\label{entr-dp}
h({\mathcal T})
\leq (n+2)h({\mathcal T}_1) + h({\mathcal T}_2) + 
(n+1)\log \theta_{\alpha_a}({\mathcal T}_1) + O(n\log n).
\end{equation}
\end{itemize}

\noindent
We repeat our earlier observation that,
by viewing the perturbation variable $\delta$
as an extra coordinate of the state vector,
the relations above still hold for global
coding trees with $n$ incremented by one.

\smallskip
\paragraph{Renormalization.}

This operation is both the simplest and the most
powerful in the arborator's toolkit: the simplest
because all it does is compress time by folding together
consecutive levels of ${\mathcal T}$;
the most powerful because it reaches beyond lego-like assembly 
to bring in the full power of algorithmic recursion 
into the analysis. 
The primitive {\tt renorm} takes disjoint subtrees of
${\mathcal T}$ and regards them as nodes of the renormalized tree.
This is done in the obvious way: if $u$ is any node
in ${\mathcal T}$ with two children $v_1,v_2$, each one
with two children, $v_{11}, v_{12}$ and $v_{21}, v_{22}$, then  
compressing the subtree $u,v_1,v_2$ means replacing it by
a node $z$ with the same parent as $u$'s (if any) and the four
children $v_{ij}$. We discuss this process in more detail below.
Although inspired by the renormalization
group of statistical physics, our approach is more general.
For one thing, the compressed subtrees may differ in size,
resulting in nonuniform rescaling across
${\mathcal T}$. This lack of uniformity rules out
closed-form composition formulae for the nesting time, attraction rate, 
and word-entropy of renormalized coding trees,
which must then be resolved algorithmically.

\smallskip
\subsection{The flow tracker}\label{flowtracker}

We approach periodicity through the study
of an important family, the {\em block-directional} influence
systems, whose agents can be ordered so that 
\begin{equation}\label{bd-defn}
{\mathcal G}= 
\begin{pmatrix}
{\mathcal G}_A & {\mathcal G}_C \\ 0 & {\mathcal G}_B
\end{pmatrix},
\end{equation}
where $0$ denotes the $(n-m)$-by-$m$ matrix 
whose entries are the constant function ${\mathbf x} \mapsto 0$;
in other words,
in a block-directional system, no $B$-agent ever
links to an $A$-agent. Suppose that $m<n$.
Wet the $B$-agents with water while keeping 
all the $A$-agents dry. 
Whenever an edge of the communication graph links a dry agent
to a wet one, the former gets wet. Note how
water flows in the {\em reverse} direction of the edges.
As soon as all agents become wet (if ever),
dry them but leave the $B$-agents wet,
and repeat forever. The case $m=n$ is similar, with one agent
designated wet once and for all.
The sequence of times at which water
spreads or drying occurs plays a key role in
building the arborator.

\smallskip
\paragraph{Coupling times and renormalization scales.}

Let ${\mathcal T}_{m\,\rightarrow\, n-m}$ denote the coding tree of
a block-directional system consisting of $m$ 
(resp. $n-m$) $A$-agents (resp. $B$-agents). The arrow
indicates that no $B$-agent can ever link to an $A$-agent:
${\mathcal G}_{ij}$ is identically zero for any
$B$-agent $i$ and $A$-agent~$j$. 
We use the notation ${\mathcal T}_{m\,\|\, n-m}$ 
for the decoupled case: no edge ever joins the two groups
in either direction,
but the discontinuities may still mix variables from both groups.
Note that the metrical case implies full independence
(\S\ref{introduction}), so that
$$
{\mathcal T}_{m\,\|\,n-m}
= {\mathcal T}_{m}\,\oplus\, {\mathcal T}_{n-m}.
$$
Assume that $n>1$ and $0<m\leq n$.
We write ${\mathcal T}_{m\,\rightarrow\, 0}$ as ${\mathcal T}_{m}$.
Likewise, we can always express
${\mathcal T}_{m\,\rightarrow\, n-m}$ as ${\mathcal T}_{m}$,
but doing so is less informative.
When the initial state ${\mathbf x}$ is undersood,
we use the shorthand 
$G_t= {\mathcal G}(f^t({\mathbf x}))$ to designate
the communication graph at time $t$
and we denote by $W_t$ the set of wet agents 
at that time. The flow tracker is not concerned with
information exchanges among the $B$-agents:
these are permanently wet and, should they not exist
($m=m$), agent 1 is kept wet at all times~[2.1]. 
Thus the set $W_t$ of wet agents is never empty.
The assignments of $t_0$ in step~[2.3] divide 
the timeline into {\em epochs}, time intervals during which
either all agents become wet or, failing that, the flow
tracker comes to a halt (breaking out of
the repeat loop at ``stop''). Each epoch is itself divided 
into subintervals by the {\em coupling times} $t_1<\cdots<t_\ell$,
with $W_{t_k}\subset W_{t_k+1}$.
The last coupling time $t_\ell$ marks either the
end of the flow tracking (if not all $A$-agents become get)
or one less than the next value of $t_0$ in the loop.

The notion of coupling is purely {\em syntactical},
being only a matter of information transfer.
Our interest in it is {\em semantic}, 
however: as befits a dissipative system, a certain
quantity, to which we shall soon return, can be bounded by
a decreasing function of time. To get a handle on 
that quantity is the main purpose of the flow tracker.

\smallskip
\paragraph{Flow tracking in action.}

Suppose that, for a long period of time,
the wet agents fail to interact with any dry one.
The two groups can then be handled recursively. 
While this alone will not tell us whether dry-wet
interaction is to occur ever again, 
it will reveal enough fine-grained information about
the groups' behavior to help us resolve that very question.
Suppose that such interaction takes place, 
to be followed by another long period of interaction.
Renormalization squeezes these ``non-interactive'' periods into
single time units, thus providing virtual time
scales over which information flows at a steady rate across the system.
Thus, besides analyzing subsystems recursively, 
renormalization brings uniformity to the information transfer rate.

\bigskip\bigskip
{\small
\par
\renewcommand{\sboxsep}{0.7cm}
\renewcommand{\sdim}{0.8\fboxsep}
\hspace{.4cm}
\shabox{\parbox{11cm}{
\vspace{-0.6cm}
\begin{center}{\bf Flow tracker} \end{center}
\smallskip
\begin{itemize}
\item[\bf{[1]$\,\,$}]
$t_0\leftarrow 0$.
\item[\bf{[2]$\,\,$}]
Repeat forever:
\begin{itemize}
 \item[\bf{[2.1]$\,\,$}]
 If $m<n$  then $W_{t_0}\leftarrow \{m+1,\ldots, n\}$ 
         else $W_{t_0}\leftarrow \{1\}$.
 \item[\bf{[2.2]$\,\,$}]
 For $t= t_0,t_0+1,\ldots, \infty$
     \begin{itemize}
     \item[]
 $W_{t+1}\leftarrow W_t \cup \{\,i\,\,|\,\, \exists\, (i,j)\in G_t \,\,\,\&\,\,\,
                                       j\in W_t\,\}$.
     \end{itemize}
 \item[\bf{[2.3]$\,\,$}]
 If $|W_\infty|= n$  then $t_0\leftarrow \min\{\, t>t_0 \,:\, |W_t|= n\,\}$
 else stop.
 \end{itemize}
\end{itemize}
}}
\par\bigskip
}

\bigskip

{\small
\begin{quote}
\label{ex2}
{\sc Example \ref{ex2}}:\ 
The third column below lists a
graph sequence $G_0,\ldots, G_{11}$ in chronological order,
with the superscript $w$ indicating the edges through
which water propagates to dry nodes.
The system is block-directional with three $A$-agents
labeled $a,b,c$ and one $B$-agent labeled $d$.
For clarity, we spell out
the agents by writing the corresponding
coding tree ${\mathcal T}_{3\,\rightarrow \,1}$ as
${\mathcal T}_{abc\,\rightarrow \,d}$, instead, thus
indicating that no edge may link $d$ to any of $a,b,c$.

\begin{center}
\begin{tabular}{|l|l|l|l|}
\hline
\multicolumn{4}{|c|}{Flow tracking} \\
\hline
\multirow{3}{*}{renorm} & $\stackrel{}{W_0=\{d\,\}}$  &
   $d\,\,\,\,\,\,\,\,\, a\rightarrow b \rightarrow c$ 
& \\
  & $W_1=\{d\,\}$ & $d\,\,\,\,\,\,\,\,\, a \leftarrow b \rightarrow c$ 
& ${\mathcal T}_{d\,\|\,abc}$ \\
& $W_2=\{d\,\}$ & $d\,\,\,\,\,\,\,\,\, a \rightarrow b \leftarrow c$ 
& \\ 
\hline
$t_1=3$ & $W_3=\{d\,\}$ & $d 
                    \stackrel{\stackrel{}{w}}{\leftarrow} 
                     a \leftarrow b \leftarrow c$
& $\stackrel{}{{\mathcal T}_{abcd}}$ \\ 
\hline
\multirow{2}{*}{renorm} & $\stackrel{}{W_4=\{a,d\,\}}$  &
   $d\leftarrow a \rightarrow b \rightarrow c$ 
& $\stackrel{}{{\mathcal T}_{a\,\rightarrow \,bcd}}$ \\
  & $W_5=\{a,d\,\}$ & $d\,\,\,\,\,\,\,\,\, a \rightarrow b \rightarrow c$ 
&  \\
\hline
$t_2=6$ & $W_6=\{a,d\,\}$ & $d\leftarrow a 
        \stackrel{\stackrel{}{w}}{\leftarrow} 
              b \leftarrow c$ 
& $\stackrel{}{{\mathcal T}_{abcd}}$ \\ 
\hline
\multirow{3}{*}{renorm} & $\stackrel{}{W_7=\{a,b,d\,\}}$ &
   $d\leftarrow a \rightarrow b \rightarrow c$ 
& \\
  & $W_8=\{a,b,d\,\}$ & $d\leftarrow a \leftarrow b \,\,\,\,\,\,\,\,\, c$
& ${\mathcal T}_{ab\,\rightarrow \,cd}$  \\
  & $W_9=\{a,b,d\,\}$ & $d\leftarrow a \rightarrow b \rightarrow c$
&  \\
\hline
$t_3=10$ & $W_{10}=\{a,b,d\,\}$ & $d\leftarrow a \rightarrow b 
           \stackrel{\stackrel{}{w}}{\leftarrow} 
                         c$ 
& $\stackrel{}{{\mathcal T}_{abcd}}$ \\
\hline
   & $W_{11}=\{a,b,c,d\,\}$ & 
       $d\,\,\,\,\,\,\,\,\, a \leftarrow b \,\,\,\,\,\,\,\,\, c$ 
            & ${\mathcal T}_{d\,\|\,abc}$ \\
\hline
\end{tabular}
\end{center}
\smallskip\smallskip

\noindent
In the first renormalized 3-step phase, the system ``waits''
for an edge from $\{a,b,c\}$ to $d$, and so can be modeled
as ${\mathcal T}_{d\,\|\,abc}$.
In the metrical case, this is further reducible to
${\mathcal T}_{d}\oplus {\mathcal T}_{abc}$.
The times $t_1,t_2,t_3$ coincide with the growth
of the wet set: these are one-step event, which 
are treated trivially as height-one absorbed trees.
They entail no recursion, so inductive soundness
is irrelevant and writing the uninformative ${\mathcal T}_{abcd}$ is harmless.
The other renormalized phases are counterintuitive
and should be discussed.
Take the last one: it might be tempting to renormalize
it as ${\mathcal T}_{abd\,\rightarrow \,c}$ to indicate that
the phase awaits the wetting of $c$ (with $a,b,d$ already wet).
This strategy is inductively unsound, however,
as it attempts to resolve a system ${\mathcal T}_{abc\,\rightarrow \,d}$
by means of another one, ${\mathcal T}_{abd\,\rightarrow \,c}$,
of the same combinatorial type.
Instead, we use the fact that not only no edge can link 
$c$ to $\{a,b\}$ (by definition of the current phase) 
but no edge can link $d$ to $\{a,b\}$ either
(by block-directionality). This allows us to use
${\mathcal T}_{ab\,\rightarrow \,cd}$, instead,
which is inductively sound.

\hspace{.4cm} Renormalization, which is denotated by underlining,
compresses into single time units
all the time intervals during which
wetness does not spread to dry agents.
With the subscripts (resp. superscript)
indicating the time compression rates (resp. tree height),
the 11-node path of ${\mathcal T}_{abc\,\rightarrow \,d}$
matching the graph sequence above can be expressed as
$$
\underline{   
     {\mathcal T}_{d\,\| \,abc}  }_{\,|3} 
\,\otimes\,
{\mathcal T}_{abcd}^{|1}\,\otimes\,
\underline{  {\mathcal T}_{a\,\rightarrow \,bcd} }_{\,|2} 
\,\otimes\,
{\mathcal T}_{abcd}^{|1} \, \otimes\,
\underline{  {\mathcal T}_{ab\,\rightarrow \,cd} }_{\,|3}
\,\otimes\,
{\mathcal T}_{abcd}^{|1} \, . 
$$
\end{quote}
}

\bigskip 

\noindent
As the example above illustrates, 
the coupling time $t_k$ is
immediately followed by a renormalization phase
of the form ${\mathcal T}_{w_k\,\rightarrow \,n-w_k}$, where
$w_k= |W_{t_k+1}|-n+m$ is the renormalization scale ($k=1,\ldots, \ell-1$).
Thus, any path of the coding tree can be renormalized as
\begin{equation}\label{renorm-expr}
{\mathcal T}_{m \,\rightarrow \,n-m} \,\Longrightarrow \,\, 
\underline{  
 {\mathcal T}_{m\,\|\, n-m}
          }_{\,|\, t_1}  
\,\otimes\,
{\mathcal T}_n^{|1} \, \otimes\,
\Bigl\{\, \bigotimes_{k=1}^{\ell-1} \,
\Bigl(\, 
\underline{  {\mathcal T}_{w_k \,\rightarrow  \, n-w_k} 
          }_{\,|\, t_{k+1}-t_k-1}
\, \otimes\, {\mathcal T}_n^{|1} \,\Bigr)
\, \Bigr\} 
\otimes {\mathcal T}_{m \,\rightarrow \, n-m}\, .
\end{equation}
The recursion comes in two forms: 
as calls to inductively smaller subsystems
${\mathcal T}_{w_k\, \rightarrow \,n-w_k}$;
and as a rewriting rule, ${\mathcal T}_{m \,\rightarrow \,n-m} \Rightarrow 
\cdots \{\,\}  \otimes {\mathcal T}_{m \,\rightarrow \,n-m}$.
It is the latter that makes the arborator, if expanded in full,
an infinitely long expression.
We note that all these derivations easily extend to
the global coding trees.

\bigskip
\section{Bidirectional Systems}\label{sec:bidirect}

We begin our proof of the bidirectional case of 
Theorem~\ref{general-case}
by establishing a weaker result for 
metrical systems: recall that these 
make the presence of an edge
between two agents a sole function of their distance.
The proof is almost automatic and a good
illustration of the algorithmic machinery we 
have put in place.  By appealing to known
results on the total $s$-energy, we are able to
improve the bounds and extend them to the nonmetrical case.

\smallskip
\subsection{The metrical case}

It is worth noting that, even for this special case, 
perturbations are required for any uniform
convergence rate to hold.

\bigskip

{\small
\begin{quote}
\label{ex3}
{\sc Example \ref{ex3}}:\ 
Consider the 3-agent system:
\begin{equation*}
\begin{pmatrix}
x_1 \\ x_2 
\end{pmatrix}
\,\longmapsto \,
 \hbox{$\frac{1}{3}$}
\begin{pmatrix}
2 & 1 \\ 1 & 2 
\end{pmatrix}
\begin{pmatrix}
x_1 \\ x_2 
\end{pmatrix},
\end{equation*}
with $x_3\mapsto \frac{1}{2}(x_2+x_3)$ if $x_3-x_2\geq 1$ and
$x_3\mapsto x_3$ else.
Initialize the system with $x_2=-x_1=1$ and $x_3$
slightly bigger than $x_2$.
The edge joining agents 2 and 3  
will then appear only after on the order of $|\! \log(x_3-x_2)|$ steps,
which implies that the convergence time cannot be bounded
uniformly without perturbation.
\end{quote}
}

\medskip

Fix $\delta$ in $\Delta= (n^{-b}\, {\mathbb I})\setminus
(n^{-2b}\, {\mathbb I})$, where ${\mathbb I}= [-1,1]$ and 
$b$ is a suitably large constant.\footnote{\,
Recall that ideally $\Delta$ should be $\{0\}$ so
the more confined around $0$ we can make it the better;
thus a higher value of $b$ is an asset, not a drawback.}
The margin slabs of a metrical system are of the form
$|a_0+ x_i-x_j + \delta | \leq \eps$. Because $a_0$ is
an $O(\log n)$-bit rational, as long as $\eps< n^{-3b}$,
${\mathbf x}$ cannot lie in that slab if $|x_i-x_j|\leq n^{-3b}$.
Let {\em diam}$\,(s)$ be the diameter of the
system after the $s$-th epoch. From~(14) in~\cite{chazelle-total},
we conclude that water propagation to all the agents
entails the shrinking of the system's diameter
by at least a factor of $1- n^{-O(n)}$. 
Since an epoch witnesses the wetting of all the agents,
repeated applications of this principle yields
\begin{equation}\label{diam-shrink}
\text{\em diam}\,(s)\leq e^{-s n^{-O(n)}} .
\end{equation}
After $n^{c n}$ epochs have elapsed (if ever), for a 
large enough constant $c$, 
the diameter of the system falls beneath $n^{-3b}$
and, by convexity, never rises again. By our previous observation,
the orbit can never hit a margin subsequently.
The maximum time it takes for
$n^{cn}$ epochs to elapse, over all ${\mathbf x}\in \Omega^n$
and $\delta\in \Delta$, is an upper bound
on the nesting time of the global coding tree.
Furthermore, past that time, the communication graph is {\em frozen}, 
meaning that it can never change again.

\begin{lemma}\label{all-mobile:Pconverge}
$\!\!\! .\,\,$
If $P$ is the transition matrix associated with the undirected
communication graph $G$, 
there is a matrix $\Pi$ such that
$\|P^k- \Pi\|_\text{max}= e^{-k n^{-O(n)}}$, for any $k\geq 0$.
\end{lemma}
\proof
By repeating the following argument for each connected 
component if needed, we can assume that $G$
is connected. The positive diagonal 
ensures that $P$ is primitive (being the stochastic
matrix of an irreducible, aperiodic Markov chain),
hence $P^n$, which we denote by $M$, is positive.
Since each nonzero entry of $P$ is at least $n^{-O(1)}$,
the {\em coefficient of ergodicity} of $M$, defined as 
$$\beta= \, \hbox{$\frac{1}{2}$}\, \max_{i,j}\, \sum_l |M_{il}-M_{jl}|= \,
1- \min_{i,j}\, \sum_l \min\{M_{il}, M_{jl}\}$$
satisfies $\beta\leq 1- n^{-O(n)}$.
Two classic results from
the theory of nonnegative matrices~\cite{seneta06}
hold that $\beta$ is an upper bound on the 
second largest eigenvalue of $M$ (in absolute value)
and that $\beta$ is submultiplicative.\footnote{\, 
The stochastic matrix $P$ may not correspond to a reversible Markov chain
and might not be diagonalizable. It is primitive, however;
therefore, by Perron-Frobenius,
it has unique left and right unit eigenvectors associated
with the dominant eigenvalue $1$.}
Given any probability distribution 
${\mathbf x}$, if ${\mathbf y}= M^l {\mathbf x}$, then
\begin{equation}\label{beta-xij}
\max_{i,j} |y_i-y_j|\leq 
 \beta^l \max_{i,j} |x_i-x_j|\leq e^{-l n^{-O(n)}}.
\end{equation}
By Perron-Frobenius and the ergodicity of $P$, its powers
tend to the rank-one matrix ${\mathbf 1}v^T$,
where $v$ is the dominant left-eigenvector of $P$
with unit $\ell_1$-norm; furthermore,
$$\|P^k- v{\mathbf 1}^T\|_\text{max}= 
e^{-k n^{-O(n)}}.$$
Indeed, setting ${\mathbf x}$
to the $j$-th basis vector $(0,\ldots,1,\ldots, 0)^T$ 
in~(\ref{beta-xij}) shows that the $j$-th column of 
$M^{l} = P^{ln}$, for $l= \lfloor k/n \rfloor$,
consists of identical entries plus or minus
a term in $e^{-l n^{-O(n)}}$.
By convexity, these near-identical
entries cannot themselves oscillate as $l$ grows.
Indeed, besides~(\ref{beta-xij}), it is also true that 
$[\min y_i, \max y_j]\subseteq [\min x_i, \max x_j]$.
\hfill $\Box$
\proofend

The next step in deriving the coding tree's parameters 
is to specialize the arborator's expression~(\ref{renorm-expr})
to the metrical case. The outer product enumerates
the first $n^{O(n)}$ epochs leading to the combinatorial
(but not physical) ``freezing'' of the system. 
The coupling times and renormalization scales 
might vary from one epoch to the next; to satisfy
the rewriting rule below, we set $w_0=1$ and $t_0=-1$.
The cropped coding tree ${\mathcal T}^*_{n}$ 
models the post-freezing phase. 
\begin{equation}\label{renorm-expr-mobile}
{\mathcal T}_{n} \,\Longrightarrow \,\, 
\Bigl\{\,\,\bigotimes_{s=1}^{n^{O(n)}}\,
\,\,\bigotimes_{k=0}^{\ell_s} \,
\Bigl(\, 
{\underline{{\mathcal T}_{w_k}\, \oplus 
     \, {\mathcal T}_{n-w_k}} }_{\,|\, t_{k+1}-t_{k}-1}\,
\otimes\, {\mathcal T}_{n}^{|1} \,\Bigr)
\,\Bigr\}\otimes {\mathcal T}^*_{n}\, .
\end{equation}
The following derivations entail little more than
looking up the dictionary compiled in~\S\ref{sec:arborator}.

\begin{itemize}

\item {\em Nesting time and attraction rate.}
It is convenient to define 
$$\mu_\alpha({\mathcal T})= 
    \max\{\,  \nu({\mathcal T}), \theta_\alpha({\mathcal T})\,\}.$$
If the coding trees $T_1,\ldots, T_k$ have period one
then, by~(\ref{nest-dp}),
\begin{equation}\label{mu-alphaUB}
\mu_\alpha\Bigl(\,\bigotimes_{i=1}^{k} T_i\,\Bigr)
\leq \, k + \sum_{i} \mu_{\alpha_a}( T_i )
+ \max_{i} \mu_{\alpha}( T_i ).
\end{equation}
The coding tree ${\mathcal T}^*_{n}$ 
involves a single matrix whose powers
converge to a fixed matrix $\Pi$ and,
by Lemma~\ref{all-mobile:Pconverge},
$\mu_{\alpha}({\mathcal T}^*_{n}) 
\leq n^{O(n)}\log \frac{1}{\alpha}$.
The following bounds derive from monotonicity and successive
applications of~(\ref{DS-all}, \ref{nest-dp}).
For some suitable $\alpha_a= \eps n^{-O(1)}$
and any $\alpha\leq \alpha_a$,
\begin{equation}\label{all-mobile:thetaUB}
\begin{split}
\mu_\alpha({\mathcal T}_{n}) 
&\leq \,
\mu_{\alpha_a}({\mathcal T}^*_{n}) + n^{O(n)} 
 \sum_{k=1}^{n-1} \,\,
   \max\Bigl\{\, 
          \mu_{\alpha_a}({\mathcal T}_{k}),
 \mu_{\alpha_a}({\mathcal T}_{n-k})\Bigr\} 
+ \max_{k} \{\, \mu_{\alpha}({\mathcal T}_{k}), 
                     \mu_{\alpha}({\mathcal T}^*_{n})\,\}
\\
&\leq \,
n^{O(n)}\, \mu_{\alpha_a}({\mathcal T}_{n-1})
+ \mu_{\alpha}({\mathcal T}_{n-1})  
+ n^{O(n)}\log \hbox{$\frac{1}{\alpha}$} \\
&\leq 
n^{O(n^2)}\log \hbox{$\frac{1}{\eps}$} +
n^{O(n)}\log \hbox{$\frac{1}{\alpha}$} 
\, .
\end{split}
\end{equation}
In view of this last upper bound, the condition $\alpha\leq \alpha_a$
can be relaxed to $\alpha<1$. Thus,
\begin{equation}\label{all-mobile:nestingUB}
\nu({\mathcal T}_{n})
\leq n^{O(n^2)}\log \hbox{$\frac{1}{\eps}$} 
\hspace{.4cm}\text{and}\hspace{.4cm}
\theta_\alpha({\mathcal T}_n) \leq 
n^{O(n^2)}\log \hbox{$\frac{1}{\eps}$} +
n^{O(n)}\log \hbox{$\frac{1}{\alpha}$} 
\, .
\end{equation}

\item {\em Word-entropy.}
By~(\ref{entr-dpp}) and the attraction rate bound above, for $0<\eps<1/2$,
\begin{equation}\label{hT1T1ub}
\begin{split}
h( T_1 \otimes T_2 )
&\leq h( T_1 ) + h( T_2 ) +
(n+1)\log ( 2\mu_{\alpha_a}(T_1) +1)
+ O(n\log n) \\
&\leq h(T_1) + h(T_2) +
(n+1)\log\log \hbox{$\frac{1}{\eps}$} + O(n^3\log n).
\end{split}
\end{equation}
By~(\ref{DS:h}) and $h({\mathcal T}^*_{n}) =0$,
it follows that
\begin{equation*}
\begin{split}
h({\mathcal T}_{n})
&\leq \,
\sum_{s=1}^{n^{O(n)}} \,\,\, \sum_{k=1}^{n-1} \,\,
\Bigl\{\,
h( \, ({\mathcal T}_{k}\, \oplus 
     \, {\mathcal T}_{n-k})\otimes
        {\mathcal T}_{n}^{|1}\,) \,
\, \Bigr\} + h({\mathcal T}^*_{n}) 
\\
& \hspace{3cm} 
+ (n+1)\log ( 2\mu_{\alpha_a}( {\mathcal T}^*_{n}) +1 )
+ n^{O(n)}\log\log \hbox{$\frac{1}{\eps}$} \\
&\leq \, n^{O(n)} h({\mathcal T}_{n-1}) 
+ n^{O(n)}\log\log \hbox{$\frac{1}{\eps}$} 
\leq n^{O(n^2)} \log\log \hbox{$\frac{1}{\eps}$} \, .
\end{split}
\end{equation*}
Our earlier observation that such derivations
apply to the global coding trees tells us that
\begin{equation}\label{all-mobile:entropy}
h({\mathcal T}^\Delta_{n})
\leq \,
n^{O(n^2)} \log\log \hbox{$\frac{1}{\eps}$}
\, .
\end{equation}
\end{itemize}

\medskip\noindent
Note the crucial fact that, from the vantage point
of~(\ref{all-mobile:nestingUB}, \ref{all-mobile:entropy}),
the global word-entropy
is lower than the nesting time, which
shows that the coding tree's average node is less than 2.
By Lemmas~\ref{nesting-periodic} and~\ref{infinite-path},
any vanishing point 
${\mathbf x}$ hits an enlarged margin fairly early:
$f^{t_a}({\mathbf x})\in {\mathcal R}_{2\eps}$
for $t_a\leq \theta_{\alpha_o}+ \#{\mathcal S}_\nu$
and some $\alpha_o\geq  \eps n^{-O(1)}$; therefore,
\begin{equation}\label{toUBeps}
t_a\leq \theta_{\alpha_o} + 2^{h({\mathcal T}^\Delta_{n})} 
\leq |\! \log \eps|^{ n^{O(n^2)} }.
\end{equation}
For random $\delta\in \Delta$, a fixed point lies
in a given slab of ${\mathcal R}_{2\eps}$ with 
probability at most $4\eps/(2 n^{-b}- 2n^{-2b})$; by a union
bound over the margin slabs, the probability of being in 
${\mathcal R}_{2\eps}$ does not exceed $\eps n^{O(1)}$.
Therefore, the probability that a fixed ${\mathbf x}$ ever
vanishes is at most 
$\eps n^{O(1)}$ times the number of paths of
depth at most $t_a$ in the global
coding tree ${\mathcal T}^\Delta_{n}$, which, by~(\ref{toUBeps}),
is  $$|\!\log \eps|^{ n^{O(n^2)} }
       2^{h({\mathcal T}^\Delta_{n})} .$$
By~(\ref{all-mobile:entropy}),
this puts the vanishing probability~at
$$
\eps (\log \hbox{$\frac{1}{\eps}$})^{ n^{O(n^2)} } < \sqrt{\eps},
$$
for $\eps$ small enough, which means that
it can be set arbitrarily low. Removing a small interval
in the middle of $n^{-b}\, {\mathbb I}$ to form $\Delta$
was only useful for the analysis: in practice, 
we might as well pick the random perturbation
uniformly in $n^{-b}\, {\mathbb I}$ since it would add only 
$2n^{-b}$ to the error probability.
The merit of the proof is that it is 
a straightforward, automatic application of
the arborator's dictionary. It illustrates
the power of renormalization, which can be seen
in the fact that no explicit bound on $t_{k+1}-t_{k}$ is ever needed.
By appealing to
known results about the {\em total $s$-energy}~\cite{chazelle-total}
we can both extend and improve the bound on the convergence rate.

\smallskip
\subsection{The bidirectional case}\label{subsec:bidirect-case}

To give up the metrical assumption means that 
the presence of an edge in the communication graph 
no longer depends on its two agents alone
but possibly on all of them. In such a system, for example,
two agents might be joined by an edge if and only if 
fewer than ten percent of them lie in between.
We revisit the previous argument and show how to extend
it to general bidirectional systems. 
We retain the ability of the communication graph
to freeze when the agents' diameter becomes negligible by
enforcing the {\em agreement rule}:
${\mathcal G}_{ij}$ is constant over the slab $|x_i-x_j| \leq n^{-bn}$,
for some suitably large constant $b$. 
The difficulty with nonmetrical dynamics is that, though decoupled,
subsystems are no longer independent, so in~(\ref{renorm-expr-mobile})
the direct sum ${\mathcal T}_{w_k} \oplus {\mathcal T}_{n-w_k}$ 
is no longer operative.

We set $\Delta= n^{-b}\, {\mathbb I}$ and fix
${\mathbf x}\in \Omega^n$ for the time being.
This induces a length on each edge of any communication graph 
${\mathcal G}(f^t({\mathbf x}))$, so we can call
a node $v$ of ${\mathcal T}_n^\Delta$ 
{\em heavy} if its communication graph contains
one or more edges of length at least $n^{-2bn}$.
The number of times the communication graph
has at least one edge of length $\lambda$ or more is called
the {\em communication count} $C_\lambda$: it has been 
shown, using the {\em total $s$-energy}~\cite{chazelle-total},
that $C_\lambda\leq \lambda^{-1} \rho^{-O(n)}$,
where $\rho$ is the smallest nonzero entry
in the stochastic matrices; here $\rho\geq n^{-O(1)}$.
It follows that, along any path of ${\mathcal T}_n$,
the number of heavy nodes is $n^{O(n)}$. 
Let us follow one such path and let $G^k$ 
denote the communication graph
common to the subpath between the $k$-th
and $(k+1)$-st heavy nodes. To see why that graph
is unique, suppose two consecutive light-node graphs 
are different. Then some $(i,j)$ is 
an edge of one but not the other. But, since the first graph
only has edges of length less than $n^{-2bn}$,
the locations of both $i$ and $j$ cannot vary
by more than $n^{-2bn}$ between the two graphs. It means
that in both graphs their distance cannot exceed
$3n^{-2bn}<n^{-bn}$; therefore, by the agreement rule, $(i,j)$ is an edge 
in both graphs, which is a contradiction. 
We rewrite~(\ref{renorm-expr}), for fixed ${\mathbf x}$, as
\begin{equation*}
{\mathcal T}^\Delta_n
\,\Longrightarrow \,\, 
\Bigl\{\,\,\bigotimes_{k=1}^{n^{O(n)}}\,\,
(\, {\mathcal T}^\Delta_{\text{$|G^k$}}
\,\otimes\,
   {\mathcal T}^{\Delta|1}_n
\,)
\,\Bigr\}\otimes 
 {\mathcal T}^\Delta_{\text{$|G^\infty$}} \, ,
\end{equation*}
where $G^\infty$ is the final graph, which forms 
an infinite suffix of the graph sequence
${\mathcal G}(f^t({\mathbf x}))_{t\geq 0}$.
We reduce unnecessary branching as follows:
whenever $V_v$ (which, with ${\mathbf x}$ fixed, 
is an interval along the $\delta$-axis) is split into 
two or more cells by the
switching partition, we give it two or more 
children (besides vanishing leaves) only if
at least one of these cells corresponds to a heavy node.
The reasoning is that, in the absence of heavy
nodes, splitting $U_v$ into subcells is pointless
since the communication graphs of all the children
are the same; so we might as well give $v$
a single child and, if need be, a vanishing leaf.
This ensures that the nesting time of
${\mathcal T}^\Delta_{\text{$|G^k$}}$ is 0.
By Lemma~\ref{all-mobile:Pconverge},
$\theta_{\alpha}( {\mathcal T}^\Delta_{\text{$|G^k$}} )
\leq n^{O(n)}\log \frac{1}{\alpha}$, and, by~(\ref{mu-alphaUB}),
\begin{equation*}
\mu_\alpha( {\mathcal T}^\Delta_n )
\leq \, n^{O(n)}+ \mu_{\alpha_a}( {\mathcal T}^\Delta_{\text{$|G^\infty$}} )
+ \sum_{k=1}^{n^{O(n)}}\,
\mu_{\alpha_a}( {\mathcal T}^\Delta_{\text{$| G^k$}} )
+ \max_k\{\, \mu_{\alpha}( {\mathcal T}^\Delta_{\text{$| G^k$}} ),
\mu_{\alpha}( {\mathcal T}^\Delta_{\text{$|G^\infty$}} )\,\}
\leq n^{O(n)}\log \hbox{$\frac{1}{\eps\alpha}$} \, .
\end{equation*}
Since $\theta_\alpha( {\mathcal T}^\Delta_n )
\leq n^{O(n)}\log \frac{1}{\eps\alpha}$ and
$h( {\mathcal T}^\Delta_{\text{$|G^k$}} ) = 0$,
by~(\ref{entr-dpp}), inequality~(\ref{hT1T1ub}) becomes
$$
h( T_1 \otimes T_2 ) \leq h(T_1) + h(T_2) +
(n+1)\log\log \hbox{$\frac{1}{\eps}$} + O(n^2\log n);$$
therefore, $h( {\mathcal T}^\Delta_n )
\leq  n^{O(n)} \log\log \frac{1}{\eps}$.
Repeating the argument we used for the metrical case
implies that the vanishing
probability of ${\mathbf x}$ is at most 
$$
\eps (\log \hbox{$\frac{1}{\eps}$})^{ n^{O(n)} } < \sqrt{\eps},
$$
for $\eps< 2^{-n^{c}}$ and constant $c$ large enough.
The attraction rate is at most $n^{O(n)}\log \frac{1}{\alpha}$,
for any $\alpha<\eps$, and the proof of the bidirectional case of
Theorem~\ref{general-case} is complete. 
\hfill $\Box$
\proofend

\bigskip
\section{General Influence Systems}

We prove Theorem~\ref{general-case}.
The centerpiece of our proof is the
bifurcation analysis of a certain non-Markovian extension
of an influence system. We focus on that extension first
and then show how it relates to the original system.
We impose a timeout mechanism to prevent any edge from reappearing
after an absence of $t_o$ consecutive steps, for arbitrarily large $t_o$.
Fix a directed graph $H$
with $n$ nodes labeled $1$ through $n$.
Given any ${\mathbf x}\in \Omega^n$,
as soon as either the communication 
graph ${\mathcal G}(f^t({\mathbf x}))$ 
contains an edge not in $H$
or some edge of $H$ fails to appear in at least one
of ${\mathcal G}(f^{t-t_o+1}({\mathbf x})),\ldots, 
                                 {\mathcal G}(f^{t}({\mathbf x}))$
for some $t\geq t_o$, 
set all future communication graphs to be the trivial
graph consisting of $n$ self-loops.
This creates a new coding tree, still denoted ${\mathcal T}_n$
for convenience, which has special 
{\em switching} leaves associated with the
trivial communication graph.
We show that, almost surely, the orbit
of any point is attracted to a limit cycle
or its path in the coding tree reaches a switching leaf.\footnote{\, 
Vanishing leaves and switching leaves are distinct: the former 
``cover'' the chaotic regions of the system
and are the places perturbations help us avoid; 
the switching leaves, on the other hand, represent
a change in dynamics type and plug into the roots
of other coding trees.}
As in the bidirectional case,
we assume the agreement rule, which sets
${\mathcal G}_{ij}$ to a constant function
over the thin slab $|x_i-x_j| \leq n^{-bn}$.

What is $H$? Any infinite graph sequence such as 
${\mathcal G}({\mathbf x}), {\mathcal G}(f({\mathbf x})), 
                       {\mathcal G}(f^2( {\mathbf x}))$, etc,
defines a unique {\em persistent graph}, which 
consists of all the edges that appear infinitely often.
The timeout mechanism allows an equivalent
characterization, which includes the edges appearing at least once
every $t_o$ steps. The persistent graph depends on
the initial state and is unknown ahead of time,
so our analysis must handle all possible such graphs.
While it plays a key role in the analysis, it would
be wrong to think of the persistent graph as determining the dynamics:
influence systems can be chaotic and nontrivially
periodic, two behaviors that can never be found
in systems based on a single graph.

Consider the directed graph derived from $H$ by identifying
each strongly connected component with a single node.
Let $B_1,\ldots, B_r$ be the
components whose corresponding nodes are sinks
and let $n_i$ denote the number of agents in the group $B_i$;
write $n= m+ n_1+\cdots +n_r$.
(In Markov chain terminology, $B_i$ is 
a closed communicating class.) The linear subspace
spanned by the agents of each $B_i$ is forward-invariant and, as we shall see,
the phase space evolves toward a subspace of rank $r$.
We reserve the indices $1,\ldots, m$ to denote the agents
outside of the $B_i$'s. Unless they hit a vanishing
or switching node, the agents indexed $m+1,\ldots, n$ 
are expected to settle eventually, while the other
agents orbit around them, being attracted to a limit cycle.
We shall see that nontrivial periodicity is possible only if $r>1$.
We are left with a block-directional system 
with $m$ (resp. $n-m$) $A$-agents (resp. $B$-agents),
and the former exercising no influence 
on the latter~(\S\ref{flowtracker}).
It follows from~(\ref{bd-defn}) that, for each node $v$ 
of the global coding tree~${\mathcal T}_{m\,\rightarrow\, n-m}$,
\begin{equation}\label{bd-tree-v}
P_{\leq v} =
\begin{pmatrix}
A_{\leq v} & C_v \\ 0  & B_{\leq v}
\end{pmatrix}\, .
\end{equation}
To resolve the system requires a fairly subtle 
bifurcation analysis which, for convenience,
we break down into four stages: 
in~\S\ref{subsec:thinrate}
we bound the thinning rate;
in~\S\ref{subsec:genericity}
we argue that, deep enough
in the coding tree, perturbations keep the coding tree's
expected (mean) degree below 1;
in~\S\ref{subsec:structure},
we show how perturbed phase tubes avoid being split 
by {\em SP} discontinuities at large depths;
finally, in~\S\ref{subsec:persistence},
we show to remove the switching leaves and do away
with the persistent graph assumption.
We also explain why it is legitimate to ignore the non-Markovian
aspect of the system in most of the discussion.

\smallskip
\subsection{The thinning rate}\label{subsec:thinrate}

We prove that, as the depth of a node $v$ of the global coding tree
grows, $A_{\leq v}$ and $B_{\leq v}$ tend 
to matrices of rank $0$ and rank $r$, respectively,
with the thinning rates $\gamma$ and $\gamma'$ 
telling us how quickly.

\begin{lemma}\label{thinrate-ub}
$\!\!\! .\,\,$
Given a node $v$ of ${\mathcal T}_{m\,\rightarrow\, n-m}$,
there exist vectors ${\mathbf z}_i\in {\mathbb R}^{n_i}$
$(i=1,\ldots, r)$, such that, for any 
$t_v\geq t_c = n^{cnt_o}$ and a large enough constant $c$,
\begin{equation*}
\text{\rm (i)}\, \, \,
\| A_{\leq v} {\mathbf 1}_m\|_\infty \leq e^{-\gamma t_v}
\hspace{.4cm}\text{and}\hspace{.4cm}
 \text{\rm (ii)}\, \, \,
\Bigl\|\,B_{\leq v} - \text{\rm diag}\, 
      ( {\mathbf 1}_{n_1}{\mathbf z}_1^T,\ldots,
{\mathbf 1}_{n_r}{\mathbf z}_r^T) \,\Bigr\|_\text{\rm max} 
\leq e^{-\gamma' t_v},
\end{equation*}
where $\gamma = 1/t_c$ and $\gamma' = n^{-cn}$.
\end{lemma}

\proof
We begin with (i). Consider the initial state 
${\mathbf x}= ({\mathbf 1}_m, {\mathbf 0}_{n-m})$,
with all the $A$-agents at 1 and the $B$-agents at $0$,
and let ${\mathbf y}= P_{\leq v} {\mathbf x}$;
obviously, $\| A_{\leq v} {\mathbf 1}_m\|_\infty=
\|{\mathbf y}\|_\infty$.
To bound the $\ell_\infty$-norm of ${\mathbf y}$, we
apply to ${\mathbf x}$ the sequence of maps specified along
the path of ${\mathcal T}_{m\,\rightarrow\, n-m}$
from the root to $v$.\footnote{\,
The path need not track the orbit of ${\mathbf x}$.}
Referring to the arborator~(\ref{renorm-expr}),
let's analyze the factor
$$
\underline{  {\mathcal T}_{w_k \,\rightarrow  \, n-w_k} 
          }_{\,|\, t_{k+1}-t_k-1}
\, \otimes\, {\mathcal T}_n^{|1} \, .
$$
The wait period $t_{k+1}-t_k$ before wetness
propagates again at time $t_{k+1}$ is at most $t_o$: 
indeed, by definition, any $A$-agent can reach some $B$-agent
in $H$ via a directed path, so all of them will  
eventually get wet. It follows that
the set $W_k$ cannot fail to grow
in $t_0$ steps unless it already contains all $n$ nodes or
the trajectory reaches a switching leaf.
Assume that the agents of $W_{t_k+1}$, the wet agents 
at time $t_k+1$ lie in $(0,1-\sigma]$.
Because their distance to $1$ can decrease by
at most a polynomial factor at each step,
they all lie in $(0, 1-\sigma n^{-O(t_o)}]$ between
times $t_k$ and $t_{k+1}$. The agents newly wet at 
time $t_{k+1}+1$, ie,
those in $W_{t_{k+1}+1}\setminus W_{t_{k+1}}$,
move to a weighted average of up to $n$ 
numbers in $(0,1)$, at least one of which is in 
$(0, 1-\sigma n^{-O(t_o)}]$. This implies
that the agents of $W_{t_{k+1}+1}$ lie in
$(0, 1-\sigma n^{-O(t_o)}]$.
Since $\sigma\leq 1$,
when all the $A$-agents are wet, which happens
within $nt_o$ steps, their positions are confined within
$(0,1- n^{-O(n t_o)}]$. It follows that
$$
\|{\mathbf y}\|_\infty \leq 
e^{-\lfloor t_v/(nt_o)\rfloor n^{-O(nt_o)}},
$$
which proves~(i). We establish~(ii) along similar lines.
Although $B_i$ and $B_j$ ($i\neq j$)
are decoupled, they are not independent; so 
their joint coding tree cannot be expressed
as a direct sum.
The subgraph $H_{|B_i}$ of $H$ induced by the agents of 
any given $B_i$ is strongly connected, so viewed
as a separate subsystem, the $B$-agents
are newly wetted at least once every $nt_o$ steps. 
By repeating the following argument for each $B_i$,
we can assume, for the purposes of this proof,
that $B=B_1$, $n_1=n-m$ and $r=1$.

Initially, place $B$-agent $j$ at 1 and all the others at $0$;
then apply to it the sequence of maps leading to
$B_{\leq v}$ (this may not be the actual trajectory
of that initial state).
The previous argument shows that the entries of
the $j$-th column of $B_{\leq v}$, which denote the locations
of the agents at time $t_v$, are confined to an interval of length 
$e^{-\lfloor t_v/(nt_o)\rfloor n^{-O(nt_o)}}$.
By the agreement rule, this implies that
the communication subgraph among the $B$-agents 
must freeze at some time $t_c= n^{c nt_o}$ for
a constant $c$ large enough, hence become 
$H_{|B}$.\footnote{\, 
We emphasize that we are making no heuristic {\em assumption} about
the repeated occurrence of the edges of $H$: switching
leaves are there precisely to allow violations of the rule.}
Let $\{u_i\}$ be the $n^{O(nt_c)}$ nodes of the coding tree
at depth $t_c$. Any deeper node $v$ is such that
$B_{\leq v} = Q^{t_v-t_{u_i}} B_{\leq u_i}$
for some $i$, where $Q$ is the stochastic matrix
associated with $H_{|B}$. Since that graph is strongly connected,
the previous argument shows that the entries in column $j$ of 
$Q^k$ lies in an interval of length $e^{-k n^{-O(n)}}$.
Since $Q^{k+1}$ is derived from $Q^k$ by taking 
convex combinations of the rows of $Q^k$,
as $k$ grows, these intervals are nested downwards and 
hence converge to a number $z_j$.
It follows that $Q^k$ tends to ${\mathbf 1}_{n_1}{\mathbf z}^T$,
with $\|Q^k - {\mathbf 1}_{n_1} {\mathbf z}^T\|_\text{\rm max} \leq 
e^{-k n^{-O(n)}}$. Doubling the value of $t_c$ yields 
part~(ii) of the lemma.
\hfill $\Box$
\proofend

\noindent
The proof suggests that, for any node $v$ 
deep enough in the coding tree, the matrix
$A_{\leq v}$ becomes an error term while
$B_{\leq v}$ tends to a matrix that depends only
on the ancestor of $v$ of depth~$t_c$. 
The bifurcation analysis requires a deeper understanding
of the error term and calls for more sophisticated arguments.
We state the thinning bound in terms of
the global coding tree for the perturbation
interval ${\mathbb I}= [-1,1]$.

\smallskip

\begin{lemma}\label{thinMatrixStruct}
$\!\!\! .\,\,$
Any node $v$ of ${\mathcal T}^{\,\mathbb I}_{m\,\rightarrow\, n-m}$
of depth $t_v\geq t_c$ has an ancestor $u$ of depth $t_c$
such that
\begin{equation*}
\Bigl\|\,
P_{\leq v} -
\begin{pmatrix}
0 & C_v \\ 0  & D_u
\end{pmatrix}
\,\Bigr\|_\text{max} \leq e^{-\gamma t_v},
\end{equation*}
where $D_u$ is a stochastic matrix of the form
$D_u=  \text{\rm diag}\,( \,{\mathbf 1}_{n_1}{\mathbf z}_1(u)^T,\ldots,
{\mathbf 1}_{n_r}{\mathbf z}_r(u)^T\,)$.
\end{lemma}

\smallskip
\subsection{Sparse branching}\label{subsec:genericity}

If we look deep enough in the coding tree
for the thinning rate to ``kick in,'' we 
observe that, under random margin perturbation,
the average branching factor is less than two.
Bruin and Deane observed a similar phenomenon 
in single-agent contractive systems~\cite{bruinD09}.
Their elegant dimensionality argument
does not seem applicable in our case, so
we follow a different approach, based on geometric considerations.
We begin with some terminology:
$\text{Lin}\,[x_1,\ldots,x_n]$ refers to a real linear form 
over $x_1,\ldots, x_n$, with $\text{Aff}\,[x_1,\ldots,x_n]$
designating the affine version; in neither case may the coefficients 
depend on $\delta$ or on the agent positions.\footnote{\, For example,
we can express $y= \delta + x_1-2x_2$ 
as $y= \delta + \text{Lin}\,[x_1,x_2]$
and $y= \delta + x_1 - 2x_2 +5$ as $y= \delta + \text{Aff}\,[x_1,x_2]$.}
With $y_1,\ldots, y_r$ understood,  
a {\em gap} of type $\omega$ denotes an interval 
of the form $a+ \omega\, {\mathbb I}$, where 
$a= \text{Aff}\,[y_1,\ldots, y_r]$.
We define the set
\begin{equation*}
{\mathbb C}[y_1,\ldots,y_r]
= \Bigl\{\, (\, \xi\, ,\,\, \overset{n_1} {\overbrace{\,y_1,\ldots,y_1\,}}
,\ldots, \overset{n_r} {\overbrace{\,y_r,\ldots,y_r\,}}\,\,)\,\,|\,\,
\xi\in \Omega^m\, \Bigr\}.
\end{equation*}
The variables $y_1,\ldots, y_r$ denote the limit positions
of the $B$-agents: they are linear combinations
of their initial positions $x_{m+1},\ldots, x_n$
(but functions of the full initial state ${\mathbf x}$).
Let $v$ be a node of the global 
coding tree ${\mathcal T}^{\,\mathbb I}_{m\,\rightarrow\, n-m}$.
The matrix $P_{\leq v}$
is a product of the form $P_{t_v}\cdots P_0$, with
$P_0= \text{\rm Id}$ and
$P_0,\ldots, P_{t_v}$ forming what we call a {\em valid matrix sequence}.
Fix a parameter $\rho>0$ and a point ${\mathbf x}$ in ${\mathbb R}^n$.
The phase tube formed by the cube
${\mathbf B}= {\mathbf x}+ \rho\, {\mathbb I}^n$ and
the matrix sequence $P_0,\ldots, P_{t_v}$ consists of the cells 
$P_0 \, {\mathbf B}, \ldots, 
(P_{t_v}\cdots P_0) {\mathbf B}$.
It might not track any orbit from ${\mathbf B}$ and hence have
little relation with a phase tube of the actual system.
The phase tube {\em splits} at node $v$  
if the global margin ${\mathcal R}_\eps$
defined in~(\ref{globalmargin-defn}) makes  
$(P_{k}\cdots P_0\, {\mathbf B}) \setminus {\mathcal R}_\eps$
disconnected. The following result is the key to sparse branching:

\smallskip

\begin{lemma}\label{low-branching}
$\!\!\! .\,\,$
Fix $\eps,\rho>0$, $D_0\geq 2^{(1/\gamma)^{n+1}}$, and
$(y_1,\ldots,y_r)\in {\mathbb R}^r$, where $\gamma= n^{-cnt_o}$.
There exists a union $W$ of $n^{O(nD_0)}$ gaps of type
$(\eps+\rho) n^{O( n^5 D_0 )}$ such that,
for any interval $\Delta\subseteq {\mathbb I}\setminus W$
of length $\rho$ 
and any ${\mathbf x}\in {\mathbb C}[y_1,\ldots,y_r]$,
the phase tube formed by the box 
$\,{\mathbf x}+ \rho\, {\mathbb I}^n$
along any path of ${\mathcal T}^\Delta_{m\,\rightarrow\, n-m}$
of length at most~$D_0$ cannot split at more
than $D_0^{1-\gamma^{n+1}}$ nodes.\footnote{\,
The crux of the lemma is the {\em uniformity} over ${\mathbf x}$: only
$(y_1,\ldots, y_r)$ needs to be fixed.}
\end{lemma}

\smallskip

\proof
We begin with a technical lemma which we prove later.
For $k=0,\ldots, D$, let $a_k$ be a row vector in 
${\mathbb R}^m$ with $O(\log n)$-bit rational coordinates
and $A_k$ be an $m$-by-$m$
nonnegative matrix whose entries are rationals
over $O(\log N)$ bits, for $N>n$.\footnote{\, The 
coefficients $a_k$ express the discontinuities.
Being extracted from the product of several
transition matrices, $A_k$ requires more bits.}
Write $v_k= a_k A_k\cdots A_0$, with 
$A_0= \text{\rm Id}$, and 
assume that the maximum row-sum 
$\alpha= \max_{\, k>0} \|A_k {\mathbf 1}\|_\infty$ satisfies
$0<\alpha<1$. 
Given $I\subseteq \{0,\ldots,D\}$,
denote by $V_{|I}$ the matrix whose rows are, 
from top to bottom, the row vectors $v_k$ with the 
indices $k\in I$ sorted in increasing order.
The following result is an elimination device
meant to factor out the role of the $A$-agents.
It is a type of matrix rigidity statement.

\smallskip

\begin{lemma}\label{rigidity}
$\!\!\! .\,\,$
Given any integer $D\geq 2^{(1/\beta)^{m+1}}$ and 
$I\subseteq \{0,\ldots,D\}$ of size 
$|I|\geq D^{1-\beta^{m+1}}$,
where
$\beta= |\! \log \alpha|/(c m^3\log N)$ for a
constant $c$ large enough,
there exists a unit vector $u$ such that 
$$u^T V_{|I} = {\mathbf 0} \hspace{.4cm} \text{and} \hspace{.4cm}
u^T {\mathbf 1} \geq N^{-c m^3D}.$$
\end{lemma}

\smallskip

\noindent
Although $c$ is unrelated
to its namesake in Lemma~\ref{thinrate-ub}, we 
use the same constant by picking the larger of the two;
in general, such constants are implied by the bit complexity
of the transition matrices and the {\em SP} discontinuities.
Note also that $\alpha\geq N^{-O(1)}$, so $\beta$ can be assumed
to be much less than 1. 
To prove Lemma~\ref{low-branching}, we first consider the case where
the splitting nodes are well separated, which
allows for Lemma~\ref{thinrate-ub} to be used; then we extend
this result to all cases.
Given a valid matrix sequence $P_0,\ldots, P_{D_0}$,
choose $D\geq 2^{(1/\beta)^{m+1}}$ and pick   
a sequence of $D+1$ integers  
$0=s_0 < \cdots < s_D \leq D_0$ such that 
\begin{equation}\label{Ds_k-gamma}
D\geq 2^{(1/\beta)^{m+1}}
\hspace{.8cm}
\text{and}
\hspace{1cm}
1/\gamma\leq s_k-s_{k-1}\leq 3/\gamma,
\end{equation}
for $k=1,\ldots, D$: we identify the matrix $A_k$ of 
Lemma~\ref{rigidity} with the $m$-by-$m$ upper left
principal submatrix of $P_{s_k}P_{s_k-1}\cdots P_{s_{k-1}+1}$;
using the notation of~(\ref{bd-tree-v}), 
$A_k= A_{\leq w}$, for some node $w$ (not necessarily an ancestor of $v$)
of depth $t_w= s_k-s_{k-1}\geq 1/\gamma$. Thus,
by Lemma~\ref{thinrate-ub}, for $k>0$,
the maximum row-sum of any $A_k$ satisfies $\alpha \leq 1/e$:
each $A_k$ is a submatrix of a product of at most $3/\gamma$
transition matrices, so each entry is an 
$O(\log N)$-bit rational, with $N= n^{n^2/\gamma}$.
What is the row vector $a_k$?
For $k=0,\ldots, D$, pick any one of the $n^{O(1)}$ margin slabs and 
denote by $a_k$ the $m$-dimensional 
vector of $O(\log n)$-bit rational coefficients
indexed by the $A$-agents.\footnote{\, With $m=3$,
$x_1-x_3+\delta=0.2$ gives $a_k=(1,0,-1)$ 
and $x_2-x_4+\delta= 0.7$ produces $a_k=(0,1,0)$.}
Fix $\delta\in {\mathbb I}$ and pick $I$ in Lemma~\ref{rigidity} to be 
of size $\lceil D^{1-\beta^{m+1}} \rceil$. Assume 
that, given ${\mathbf x}\in {\mathbb C}[y_1,\ldots,y_r]$,
the phase tube formed by 
the box ${\mathbf x}+ \rho\, {\mathbb I}^n$
and $P_0,\ldots, P_{s_D}$ splits at every index of $I$
along the chosen slabs.\footnote{\,
It is immaterial that
${\mathbf x}+ \rho\, {\mathbb I}^n$ might slightly
bulge out of the phase space $\Omega^n$.}
In other words, for each $k\in I$, there exist 
a node $z_k$ of depth $t_{z_k}= s_k$ and
$\rho_i = \rho_i(k)$, for $i=1,\ldots, n$,
such that $|\rho_i|\leq \rho$ and
\begin{equation*}
\Bigl|\, c_k+ (a_k,b_k)
\begin{pmatrix}
A_{\leq z_k} & C_{z_k} \\ 0  &  B_{\leq z_k}
\end{pmatrix}
(x_1+\rho_1,\ldots, x_n+ \rho_n)^T
+\delta \, \Bigr| \leq \eps,
\end{equation*}
where the chosen slab is of the form 
$|c_k + a_k(x_1,\ldots, x_m)^T+ b_k(x_{m+1},\ldots, x_n)^T
+\delta|\leq \eps$, with 
$b_k\in {\mathbb R}^{n-m}$ and $c_k\in {\mathbb R}$.
Since $v_k= a_k A_{\leq z_k}$
and ${\mathbf x}\in {\mathbb C}[y_1,\ldots,y_r]$,
it follows that
\begin{equation}\label{Aff-delta-eps}
\Bigl|\, v_k(x_1+\rho_1,\ldots,x_m+\rho_m)^T +
\, \text{Aff}\,
[\,y_1,\ldots, y_r, \rho_{m+1},\ldots, \rho_n\,]
+ \delta \,\Bigr| \leq \eps,
\end{equation}
where the coefficients in the affine form are of 
magnitude $n^{O(1)}$.

\vspace{.5cm}
\begin{figure}[htb]
\begin{center}
\hspace{.2cm}
\includegraphics[width=8cm]{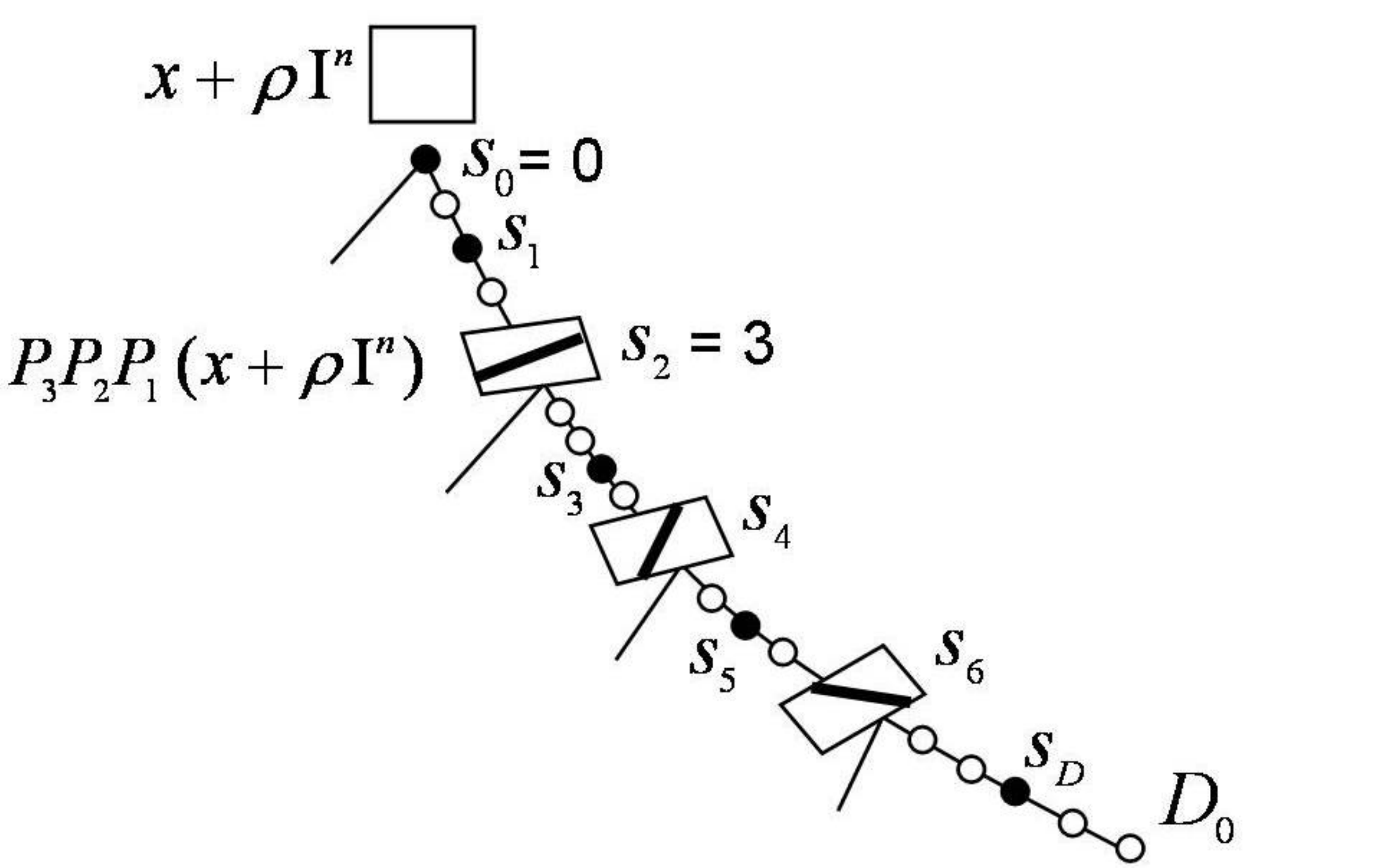}
\end{center}
\caption{\small 
The choice of slabs at the nodes causes
the phase tube to split at the nodes indexed by $I=\{2,4,6\}$.
The nodes of depth $s_k$ for $k\not\in I$ are represented 
as black dots: $s_0, s_1, s_3, s_5, s_7$ ($D=7$).
The other nodes in the paths are the white dots.}
\label{fig-codingtreesplit}
\end{figure}
\vspace{.5cm}

Lemma~\ref{rigidity} allows us to eliminate the variables
$x_1,\ldots, x_m$: we premultiply $V_{|I}$ 
by the unit vector $u$ to find that
\begin{equation}\label{Aff-rhoepsScalar}
|\,\text{Aff}\, [\,y_1,\ldots, y_r\,] + \delta\,|
\leq  (\eps+\rho) N^{O(c m^3 D)},
\end{equation}
where the coefficients of the new affine form
are bounded by $N^{O(c m^3 D)}$. 
(We leave the constant $c$ in the exponent 
to highlight its influence.) The remarkable
fact is that the variable $\delta$ is assured not to vanish
during the elimination.
Thus, as long as $\delta$ remains outside 
a gap of type $(\eps+\rho) N^{O(c m^3 D)}$,
the phase tube formed by ${\mathbf x}+ \rho\, {\mathbb I}^n$
and $P_0,\ldots, P_{D}$ cannot split at every index of $I$.
Counting the number of possible choices of
slabs per node raises the number of gaps to $n^{O(|I|)}$.
The argument assumes that $\delta$ has the same value
in each of $|I|$ inequalities.
It need not be so: each $\delta$ in~(\ref{Aff-delta-eps})
can be replaced by $\delta +\nu_k$ ($k\in I$), for
$|\nu_k|\leq \rho$,
and the new system of inequalities will still
imply~(\ref{Aff-rhoepsScalar}).\footnote{\, This
observation is crucial for the degree structure analysis 
to come next and the need to randomize~$\delta$.} 
A combinatorial argument shows how adding more gaps
to the ``exclusion zone'' keeps branching low. 
Before proceeding with that final part of the proof, we
summarize our results, using the bound
$|\! \log \alpha|\geq \log e>1$.
\smallskip

\begin{lemma}\label{lowbranch-singleI}
$\!\!\! .\,\,$
Let $N= n^{n^2/\gamma}$ and $\beta= 1/(c m^3\log N)$,
where $c$ is the constant of Lemma~\ref{rigidity}.
Fix a path in
${\mathcal T}^{\,\mathbb I}_{m\,\rightarrow\, n-m}$
from the root and pick $D+1$ nodes on it 
of depth $0=s_0< \cdots <s_D$
satisfying~(\ref{Ds_k-gamma}); out of these
nodes, choose a subset $I$ of size $\lceil D^{1-\beta^{m+1}} \rceil$.
There exists an exclusion zone $W$ consisting
of the union of at most $n^{O(|I|)}$ gaps 
of type $(\eps+\rho) N^{O( cm^3 D )}$,
such that, for any interval $\Delta\subseteq {\mathbb I}\setminus W$
of length $\rho$ and
any ${\mathbf x}\in {\mathbb C}[y_1,\ldots,y_r]$,
the phase tube formed by 
${\mathbf x}+ \rho\, {\mathbb I}^n$
cannot split at all the nodes of $I$ in
${\mathcal T}^{\Delta}_{m\,\rightarrow\, n-m}$
(assuming they exist).
\end{lemma}

\smallskip

The crux of the lemma is that it holds uniformly for
all ${\mathbf x}$.
To prove Lemma~\ref{low-branching},
we need to extend the previous lemma to all the paths of
the coding tree of the prescribed length and remove
from~(\ref{Ds_k-gamma})
the lower bound of $1/\gamma$ on the distance between
consecutive splitting nodes.
Fix $D_0 \geq 2^{(1/\gamma)^{n+1}}$, and let $v$ be a node 
of ${\mathcal T}^{\,\mathbb I}_{m\,\rightarrow\, n-m}$
of depth $t_v=D_0$. Since the path is fixed, we
can uniquely identify the node $v$ and its
ancestors by their depths and denote by $P_t$ the
transition matrix of the node at depth $t$.
Define the node set $J=\{1/\gamma,2/\gamma,\ldots, D_0\}$,
with $|J|= \lceil \gamma D_0 \rceil$; recall that
$1/\gamma= t_c$ is an integer. 
Let $K$ be the set of ancestors of $v$
at which the phase tube formed by 
${\mathbf x}+ \rho\, {\mathbb I}^n$
and $P_0,\ldots, P_{D_0}$ splits
(with respect to ${\mathcal T}^{\,\mathbb I}_{m\,\rightarrow\, n-m}$);
assume that 
\begin{equation}\label{KlbD0}
|K|\geq D_0^{1-\gamma^{n+1}}.
\end{equation}
We define $I$ to be the largest subset of $K$ 
with no two elements of $I\cup\{0\}$ 
at a distance less than $1/\gamma$;
obviously, $|I|\geq \lfloor \gamma |K|\rfloor -1$. 
To define $s_1,\ldots,s_D$,
we add all of $J$ to $I$ (to keep distances between consecutive
nodes small enough) and then clean up the set to avoid
distances lower than allowed: we define
$J'$ to be the smallest subset of $J$ such that
$L= I\cup (J\setminus J')$ contains
no two elements at a distance less than $1/\gamma$.
Each element of $I$ can cause the disappearance of
at most two elements in $J$ for the addition of one into $L$,
hence $|J|/2\leq |L|\leq \gamma D_0 +1$. 
By construction, consecutive elements of $L$ are
at most $3/\gamma$ away from each other, so we can
identify $L$ with the sequence $s_1< \cdots < s_D$.
By $m<n$ and the specifications of $\gamma$ in Lemma~\ref{thinrate-ub}
and $N, \beta$ in Lemma~\ref{lowbranch-singleI},
we can verify that
\begin{equation}\label{DbetaD0bounds}
\text{\rm (i)}\, \, \,
D_0\geq 2^{(1/\gamma)^{n+1}}\geq 
\gamma^{-1} 2^{(1/\beta)^{m+1}+1}
\hspace{.4cm}\text{and}\hspace{.4cm}
\text{\rm (ii)}\, \, \,
D_0^{1-\gamma^{n+1}}\geq \hbox{$\frac{2}{\gamma}$} 
(\gamma D_0 + 1)^{1-\beta^{m+1}}.
\end{equation}
Part (i) ensures~(\ref{Ds_k-gamma}).
By Lemma~\ref{lowbranch-singleI}, keeping $\delta$ 
outside the union $W$
of at most $n^{O(|I|)}$ gaps of type $(\eps+\rho) N^{O( m^3 D )}$
prevents $I$ from witnessing
a phase tube split at each of its nodes,
and hence keeps $K\supseteq I$ from being, as claimed,
made entirely of ``splitting'' nodes.
For this, we need to ensure that $|I|\geq D^{1-\beta^{m+1}}$,
which follows from:~(\ref{KlbD0});
$|I|\geq \lfloor \gamma |K|\rfloor -1$;
$D=|L| \leq \gamma D_0 +1$; 
and part~(ii) of~(\ref{DbetaD0bounds}).

We conclude that, as long as we choose an interval
$\Delta\subseteq {\mathbb I}\setminus W$ of length $\rho$,
the coding tree ${\mathcal T}^{\Delta}_{m\,\rightarrow\, n-m}$
cannot witness splits at all of the nodes of $K$
(if they exist---their existence is ensured
only in ${\mathcal T}^{\,\mathbb I}_{m\,\rightarrow\, n-m}$) 
for the phase tube
formed by any box ${\mathbf x}+ \rho\, {\mathbb I}^n$,
where $y_1,\ldots,y_r$ are fixed
and ${\mathbf x}\in {\mathbb C}[y_1,\ldots,y_r]$.
Note the order of the quantifiers:
first, we fix the coordinates $y_k$ and the target length $D_0$,
and we pick a large enough candidate splitting node set $K$ in
${\mathcal T}^{\,\mathbb I}_{m\,\rightarrow\, n-m}$;
these choices determine the exclusion zone $W$;
next, we pick a suitable $\Delta$ and then claim
an impossibility result for {\em any} ${\mathbf x}$ in 
${\mathbb C}[y_1,\ldots,y_r]$.
To complete the proof of Lemma~\ref{low-branching},
we bound, by $2^{D_0}$ and $n^{O(nD_0)}$
respectively, the number of ways of choosing $K$ 
(hence $I$, $L$) and the number of nodes $v$ 
in ${\mathcal T}^{\,\mathbb I}_{m\,\rightarrow\, n-m}$
of depth $t_v= D_0$.
\hfill $\Box$
\proofend

\smallskip
\paragraph{Proof of Lemma~\ref{rigidity}.}

We can make the assumption
that $I$ includes $0$, since all cases easily reduce to it.
Indeed, let $l$ be the smallest index in $I$. If $l>0$,
subtract $l$ from the indices of $I$ to define $I'\supseteq \{0\}$.
Form the matrix $V'_{|I'}$ of vectors $v'_k$, where
$v_{k+l}= v'_k A_l\cdots A_0$. Rewriting 
$V_{|I}$ as $V'_{|I'}A_l\cdots A_0$ takes us to the desired case:
we (cosmetically) duplicate the last matrix, $P_D$, $l$ times to match
the lemma's assumptions and observe that, if $u^T V'_{|I'}={\mathbf 0}$,
then so does $u^T V_{|I}$. 
We may also assume that all $v_k$ are nonzero since
the lemma is trivial otherwise.
All the coordinates of $v_k$ can be expressed as 
$O(m^2(k+1)\log N)$-bit rationals sharing
a common denominator; therefore, 
\begin{equation}\label{vk-bounds}
N^{-O((k+1) m^2)}\leq \|v_k\|_1\leq 2^{-k |\! \log\alpha| + O(\log n)}.
\end{equation}
The {\em affine hull} of $V_{|I}$ is
the flat defined by $\{\, z^T V_{|I} \,:\, z^T {\mathbf 1} = 1 \,\}$:
its dimension is called the {\em affine rank} of $V_{|I}$.
Let $g(D,r)$ be the maximum value of $|I|$, for
$\{0\}\subseteq I\subseteq \{0,\ldots,D\}$, such that
$V_{|I}$ has affine rank at most $r$ and its affine hull
does not contain the origin.
Lemma~\ref{rigidity} follows from this inequality, whose proof 
we postpone: for $r=0,\ldots, m-1$,
\begin{equation}\label{g(D,r)-ub}
g(D,r)< D^{1-\beta^{m+1}}\, ,
        \hspace{.5cm} \text{for any}\,\,\, D\geq 2^{(1/\beta)^{m+1}},
\end{equation}
where $\beta= |\! \log \alpha|/(c m^3\log N)$,
for constant $c$ large enough.
Indeed, given any $\{0\}\subseteq I\subseteq \{0,\ldots,D\}$
of size at least $D^{1-\beta^{m+1}}$, we have
$|I|> g(D,m-1)$, so the affine hull of $V_{|I}$
contains the origin. If $r$ is its affine rank,
then there exists $J\subseteq I$ of size
$r+1$ such that the affine rank of $V_{|J}$ 
is $r$ and its affine hull contains the origin,
hence coincides with the row space of $V_{|J}$,\footnote{\,
Because any $y^T V_{|J}$ can be written as
$(y+ (1-y^T{\mathbf 1})z)^T V_{|J}$, where 
$z^T V_{|J}= {\mathbf 0}$ and $z^T{\mathbf 1}= 1$.}
which is therefore of
dimension~$r$. This implies the existence of 
$r$ independent columns in $V_{|J}$ spanning
its column space: add a column of $r+1$ ones
to the right of them to form the $(r+1)$-by-$(r+1)$ matrix $M$.
Since the affine hull of $V_{|J}$ contains the origin,
there exists $z$ such that $z^TV_{|J}= {\mathbf 0}$
and $z^T{\mathbf 1}=1$, which in turn shows that
${\mathbf 1}_{r+1}$ lies outside the column space of $V_{|J}$;\footnote{\,
Otherwise, $1= z^T{\mathbf 1}= z^T V_{|J}\, y= 0$.}
therefore $M$ is nonsingular. Since each one of its rows
consists of $O(m^2 D\log N)$-bit rationals 
with a common denominator,
\begin{equation}\label{detVI1}
|\! \det M| \geq  N^{-O(m^3D)}.
\end{equation}
Let $\xi$ be the $(r+1)$-dimensional vector whose $k$-th coordinate
is the cofactor of the $k$-th entry in the last column of ones in $M$.
Simple determinant cofactor expansions show that
\begin{equation*}
\xi^T M= (\, \overset{r}{\overbrace{\,0,\ldots,0\,}}, \det M).
\end{equation*}
Since the first $r$ columns of $M$ span the column
space of $V_{|J}$, it follows that 
\begin{equation*}
\xi^T (\,V_{|J}, {\mathbf 1}_{r+1}\,) 
       = (\, \overset{m}{\overbrace{\,0,\ldots,0\,}}, \det M).
\end{equation*}
By Hadamard's inequality,
each coordinate of $\xi$ is at most $n^{O(m)}$ in absolute value,
so straightforward rescaling and padding with zeroes
turns $\xi$ into a suitable vector $u$ 
such that $u^T V_{|I} = {\mathbf 0}$ and
$u^T {\mathbf 1} \geq N^{-c_1 m^3D}$,
for an absolute constant $c_1$ that does not depend on $c$.
Replacing $c$ by $\max\{c,c_1\}$ establishes Lemma~\ref{rigidity}.

It suffices now to prove~(\ref{g(D,r)-ub}), which
we do by induction on $r$. 
If $V_{|I}$ has affine rank $r=0$ and its affine hull
does not contain the origin, then all the rows
of $V_{|I}$ are equal and nonzero.
Since $V_{|I}$ has the row $v_0$,
it follows from~(\ref{vk-bounds})
that $|I|\leq 1+ \max\{ k\in I \} 
= O(|\! \log\alpha|^{-1}m^2\log N)$,
hence
\begin{equation}\label{g(D,0)-ub}
g(D,0)\leq \beta^{-1}.
\end{equation}

\vspace{.5cm}
\begin{figure}[htb]
\begin{center}
\hspace{.0cm}
\includegraphics[width=8cm]{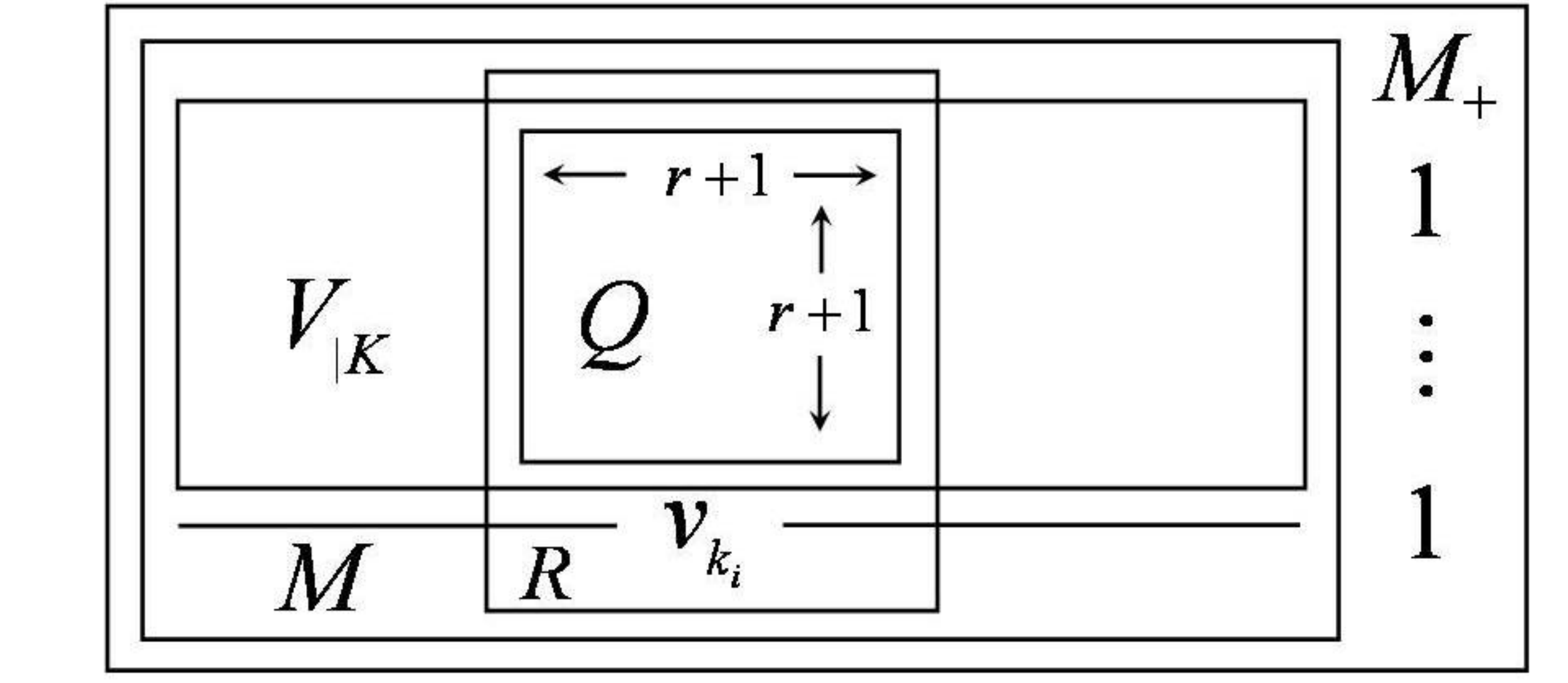}
\end{center}
\caption{\small 
Why a large value of $k_i$ implies that the affine hull of
$V_{|I}$, hence of $M$, contains the origin. }
\label{fig-A1}
\end{figure}
\vspace{.5cm}

Assume now that $r>0$ and that 
$V_{|I}$ has affine rank exactly $r$.
Put $I=\{k_0,k_1,\ldots, k_i\}$, with $k_0=0$,
and consider the smallest $j$
such that $V_{|J}$ has affine rank $r$, where 
$J=\{k_0,k_1,\ldots, k_j\}\subseteq I$.
Since the origin is not in the affine hull of 
$V_{|I}$ hence of $V_{|J}$, we can always 
pick a subset $K\subseteq J$ consisting 
of $r+1$ independent rows: let $M= V_{|K\cup \{k_i\}}$ denote
the $(r+2)$-by-$m$ matrix formed by adding
the row $v_{k_i}$ at the bottom of $V_{|K}$.\footnote{\,
It may be the case that $i=j$ or $k_i\in K$.
Since $r>0$, we have $k_i\geq k_j\geq 1$ and $j>0$.}
Since $V_{|I}$ has affine rank $r$, its 
rank is $r+1$ (using once again the noninclusion of $O$ in
the affine hull of $V_{|I}$), hence so is the rank of $M$.
We show that if $k_i$ is large enough,
the system below is feasible in $\xi\in {\mathbb R}^{r+2}$:
\begin{equation}\label{vM+}
\xi^T M_+= (\, \overset{m}{\overbrace{\,0,\ldots,0\,}}, 1),
\end{equation}
where $M_+$ is the $(r+2)$-by-$(m+1)$ matrix $(M,{\mathbf 1}_{r+2})$,
which leads to a contradiction.
This is the crux of the argument and makes essential use
of the rapid decay of the vectors $v_k$.
Assume that $k_i > c k_j |\! \log \alpha|^{-1} m^3\log N$,
for a large enough constant $c$. 
We first show that $M_+$ is of rank $r+2$. 
Pick $r+1$ independent columns 
of $V_{|K}$, which is possible since the latter has rank $r+1$,
to form the full-ranked $(r+1)$-by-$(r+1)$ matrix $Q$.
Add a new row to it by fitting the relevant part of $v_{k_i}$
(the last row of $M$)
and call $R$ the resulting $(r+2)$-by-$(r+1)$ matrix
(Fig.~\ref{fig-A1});
consistent with our notation, 
$R_+$ will denote the matrix $(R,{\mathbf 1})$.
A cofactor expansion of the determinant of $R_+$ along
the bottom row shows that
$$
|\! \det R_+| \geq |\! \det Q| - \Delta \|v_{k_i}\|_1 ,
$$
where $\Delta$ is an upper bound on the 
absolute values of the cofactors other than $\det Q$.
In view of~(\ref{vk-bounds}),
the matrix entries involved in these cofactors are all
in $n^{O(1)}$; by Hadamard's inequality, this 
shows that we can set $\Delta= n^{O(m)}$.
Likewise, we find that 
$$\|v_{k_i}\|_1 \leq 2^{-k_i |\! \log\alpha| + O(\log n)}.$$
Since $Q$ is nonsingular, we can 
adapt~(\ref{detVI1}) to derive $|\! \det Q|\geq  N^{-O(m^3k_j)}$, 
hence $|\! \det R_+|>0$. It follows that the linear 
system~(\ref{vM+}) is feasible if we replace $M_+$ by $R_+$.
As it happens, there is no need to do so since every column
of $M$ missing from $R$ lies in the column space of the latter:
thus the missing homogeneous equalities are automatically
satisfied by the solution $\xi$.
The feasibility of~(\ref{vM+}) contradicts our assumption
that the origin is outside the affine hull of 
$V_{|I}$; therefore 
\begin{equation}\label{kikjub}
k_j\geq \beta k_i >0,
\end{equation}
where $\beta = |\! \log \alpha|/(c m^3\log N)$.
The affine rank of $V_{|\{k_0, \ldots, k_{j-1}\}}$ is $r-1$
and its affine hull does not contain the origin, so
$j \leq g(k_{j-1},r-1)$, with $g(0,r-1)=1$.
Let $w_0= a_{k_j}$ and, for $k>0$,
$w_k = a_{k_j+k} A_{k_j+k}\cdots A_{k_j+1}$,
thus ensuring that $v_{k_j+k}= w_k A_{k_j}\cdots A_0$.
Since the affine hull of $V_{|I}$ does not contain the origin,
neither does that of the matrix $W$ with rows
$w_0,w_{k_{j+1}-k_j},\ldots, w_{k_i-k_j}$. 
It follows that the affine rank of $W$ is less than~$m$,
so $i-j+1\leq g(k_i-k_j,m-1)$, hence\footnote{\, 
It would be nice to bound the affine rank 
as a function of~$r$, but since we never perturb 
the transition matrices it is unclear how to do that.}
$i\leq g(k_{j-1},r-1) + g(k_i-k_j,m-1) -1$.
By~(\ref{kikjub}) and $i=|I|-1$, we derive,
by monotonicity,
$$
|I|\leq g(k,r-1)+ g(D-k, m-1),
$$
where $\beta D\leq k\leq D$; hence, by~(\ref{g(D,0)-ub}), 
for $m>0$ and $D\geq 0$:
\begin{equation*}
g(D,r)\leq 
\begin{cases}
\, 1 & \text{ if $D=0$ } \\
\, \beta^{-1}  & \text{ if $r=0$}\,  \\
\, g(n_1,m-1)+\cdots + g(n_r,m-1)+ \beta^{-1} &\text{ if $0<r<m$,}
\end{cases}
\end{equation*}
where $n_1+\cdots + n_r \leq (1-\beta^s)D$, 
with $s= |\{\,i\,|\,n_i>0\,\}|$.
Setting $\eta= \beta^m$, we check that, for all $D>0$ and $m>0$,
\begin{equation}\label{g(D,r)-ub2}
g(D,m-1)\leq \beta^{-2}(2 D^{1-\eta} -1).
\end{equation}
The case $m=1$ follows from $g(D,0)\leq \beta^{-1}$.
For $m>1$, we begin with the case $s=0$, where
$$g(D,m-1)\leq m-1 + \beta^{-1} \leq 
\beta^{-2}(2D^{1-\eta}-1).
$$
This follows from $\alpha\geq N^{-O(1)}$, which implies that
$\beta m^3$ can be made arbitrarily small by increasing~$c$.
For $s=1$,
\begin{equation*}
\begin{split}
g(D,m-1)&
\leq \beta^{-2}(2 (1-\beta)^{1-\eta} D^{1-\eta} -1)
+ m-2 +\beta^{-1} \\
&\leq 2\beta^{-2} D^{1-\eta} 
      - (2 \beta^{-1}(1-\eta)- O(1)) D^{1-\eta} 
      - \beta^{-2} + \beta^{-1} + m - 2  \\
 &  \leq \beta^{-2}(2D^{1-\eta}-1).
\end{split}
\end{equation*}
Assume that $s>1$. Being concave, the function 
$x\mapsto x^{1-\eta}$ is subadditive for $x\geq 0$; therefore,
$$ n_1^{1-\eta}+\cdots + n_r^{1-\eta} \leq (1-\beta^s)^{1-\eta}D^{1-\eta}.$$
Setting $r=m-1$, relation~(\ref{g(D,r)-ub2}) follows from
the inequality,
\begin{equation*}
\begin{split}
g(D,m-1)&\leq 
   \beta^{-2}( 2(1-\beta^{s})^{1-\eta} D^{1-\eta} -s)
     +  m-s-1+\beta^{-1} \\
&\leq
  2\beta^{-2}(1-\beta^{m-1})^{1-\eta} D^{1-\eta}
     -  \hbox{$\frac{3}{2}$}\beta^{-2} \leq 
  2\beta^{-2} D^{1-\eta} - \beta^{-2},
\end{split}
\end{equation*}
which proves~(\ref{g(D,r)-ub2}), hence~(\ref{g(D,r)-ub})
and Lemma~\ref{rigidity}.
\hfill $\Box$
\proofend

\smallskip
\subsection{The degree structure}\label{subsec:structure}

We decompose the global coding tree 
into three layers: the top one has
no degree constraints; the second has mean degree less than two;
and the third has no branching.
Consider a placement of the $B$-agents,
such that the diameter of each $B_i$ is less than $n^{-bn}$.
By the agreement rule, the communication subgraph
induced by the $B$-agents is frozen and its
transition matrix $Q$ is fixed and independent of
the particular placement of the $B$-agents.\footnote{\,
The system under consideration is the non-Markovian extension
defined by the persistent graph~$H$.}
By Perron-Frobenius, or simply by repeating the
proof of Lemma~\ref{thinrate-ub}, we derive the existence
of a rank-$r$ stochastic matrix 
$${\widetilde Q} = \text{\rm diag}\, ( {\mathbf 1}_{n_1}\chi_1^T,\ldots,
{\mathbf 1}_{n_r}\chi_r^T)
$$
such that $\chi_i\in {\mathbf R}^{n_i}$
and $\|Q^k- {\widetilde Q}\|_\text{\rm max} \leq e^{-kn^{-O(n)}}$.
The $B$-agents find themselves attracted
to the fixed point ${\mathbf y}= {\widetilde Q} \xi$, where
$\xi\in {\mathbb R}^{n-m}$ is their initial state vector and
$$
{\mathbf y}= ( \, \overset{n_1} {\overbrace{y_1,\ldots,y_1}}
,\ldots,
\overset{n_r} {\overbrace{y_r,\ldots,y_r}}\,).
$$
Define 
$\Upsilon= \Omega^m\times (\Omega^{n-m}\cap \Upsilon_B)$,
where
$$\Upsilon_B= {\mathbf y} +
(n^{-2bn}\,{\mathbb I}^{n-m})\cap \ker {\widetilde Q}.
$$
If ${\mathbf x}\in \Upsilon$, the diameter of any group
$B_i$ is at most $2n^{-2bn}< n^{-bn}$ so the communication graph
induced by their agents is frozen and remains so.
The $B$-agents are attracted to ${\mathbf y}$.\footnote{\,
Although the $B$-agents in $\Upsilon_B$ have 
been essentially immobilized around ${\mathbf y}$,
they are not decoupled from the rest.
Indeed, while the increasingly microscopic movement
of the $B$-agents
can no longer affect their own communication graph,
it can still influence the communication among
the $A$-agents: furthermore, this may still be true
even if no edge is ever to join an $A$-agent to a $B$-agent.}
This follows easily, as does the next lemma, whose proof we omit,
from the stochasticity of $Q$ and the identities: 
${\widetilde Q}Q=Q{\widetilde Q}={\widetilde Q}^2= {\widetilde Q}$.

\begin{lemma}\label{upsilon-facts}
$\!\!\! .\,\,$
The set $\Upsilon$ is forward-invariant. Furthermore,
any $\xi\in {\mathbf y}+ n^{-2bn}\,{\mathbb I}^{n-m}$ belongs to
$\Upsilon_B$ if and only if ${\widetilde Q} \xi= {\mathbf y}$.
\end{lemma}

We set $\eps,\rho,D_0$ as in Lemma~\ref{low-branching} and call 
an interval $\Delta$ {\em free} if
it does not intersect the exclusion zone $W= W({\mathbf y})$.
As usual, we choose the perturbation sample space $n^{-b}\, {\mathbb I}$ 
to make perturbations inconsequential in practice.
For counting purposes, it is convenient to partition 
$n^{-b}\, {\mathbb I}$ into {\em canonical} intervals of 
length $\rho$ (with possibly a single smaller one). 
A gap of $W$ can keep only $(1+\eps/\rho) n^{O( n^5 D_0 )}$
canonical intervals from being free, so the Lebesgue measure of 
the free ones satisfies:
\begin{equation}\label{Wbound}
\text{Leb}\, \Bigl\{\, \bigcup\,\text{free canonical intervals}\, 
     \Bigr\}\, \geq  2n^{-b} - (\eps+\rho) n^{O( n^5 D_0 )}.
\end{equation}

\smallskip
\paragraph{Fixing the $B$-agent attractor.}

With ${\mathbf y}$ fixed, we pick a free canonical interval $\Delta$
and focus on the global coding tree
${\mathcal T}^{\Delta|\Upsilon}_{m\,\rightarrow\, n-m}$,
with the superscripts indicating the perturbation
and phase spaces, respectively.\footnote{\,
The reason we do not fix the perturbation $\delta$ is that
it needs to be randomized and it is easier to avoid
randomizing the coding tree itself.}
For any node $v$ of depth $t_v\geq t_c$, the limit matrix $D_u$
in Lemma~\ref{thinMatrixStruct} is the same for all nodes $u$
of depth $t_c$. Indeed,
\begin{equation*}
\Bigl\|\,
P_{\leq v} -
\begin{pmatrix}
0 & C_v \\ 0  & {\widetilde Q}
\end{pmatrix}
\,\Bigr\|_\text{max} \leq e^{-\gamma t_v}.
\end{equation*}
Pick $v$ of depth $t_v\geq 3t_c$ and let $w$ be
its ancestor at depth $t_w= \lfloor t_v/2\rfloor$.
Given ${\mathbf x}\in U_v\subseteq \Upsilon$, 
\begin{equation*}
\begin{split}
{\mathbf x}'
&= f^{t_w}({\mathbf x}) = P_{\leq w} \, {\mathbf x} \in
\begin{pmatrix}
C_w 
\\ {\widetilde Q}
\end{pmatrix} (x_{m+1},\ldots, x_n)^T
+ n e^{-\gamma t_w}\, {\mathbb I}^n \\
&\in 
\begin{pmatrix}
C_w (x_{m+1},\ldots, x_n)^T
\\ {\mathbf y}
\end{pmatrix} 
+ n e^{-\gamma t_w}\, {\mathbb I}^n.
\end{split}
\end{equation*}
By Lemma~\ref{upsilon-facts},
${\mathbf x}'\in \Upsilon$, so
there exists a node $v'$ of
depth $t_{v'}= t_v-t_w \geq t_c$ such that,
\begin{equation*}
f^{t_v}({\mathbf x})
= f^{t_{v'}}({\mathbf x}')
= P_{\leq v'} \, {\mathbf x}' 
\in
\begin{pmatrix}
C_{v'}
\\ {\widetilde Q}
\end{pmatrix} ({\mathbf y} + n e^{-\gamma t_w}\, {\mathbb I}^n )
+ n e^{-\gamma t_{v'}}\, {\mathbb I}^n 
\subseteq
\begin{pmatrix}
C_{v'} {\mathbf y}\\ 
{\mathbf y}
\end{pmatrix}
+ 2 n e^{-\gamma t_v/3}\, {\mathbb I}^n.
\end{equation*}
It is important to note that $v'$ depends only on $v$
and not on ${\mathbf x}\in U_v$: indeed,
the phase tube from $U_v$ between time $t_w$ and $t_v$
does not split; therefore $f^{t_w}(U_v)\subseteq U_{v'}$.
It follows that, for 
$t_v\geq 3t_c$ and $v'=v'(v)$,
\begin{equation}\label{Pv-Upsilon-CDI}
V_v \subseteq 
\begin{pmatrix}
C_{v'}{\mathbf y}   \\ {\mathbf y}
\end{pmatrix} 
+ 2n e^{-\gamma t_v/3}\, {\mathbb I}^n.
\end{equation}
The $A$-agents evolve toward
convex combinations of the $B$-agents, which
themselves become static. The weights of
these combinations (ie, the barycentric
coordinates of the $A$-agents), however, might change at every node, so 
there is no assurance that the orbit is always attracted to a limit cycle.
The layer decomposition of the coding tree, which we describe next,
allows us to bound the nesting time 
while exhibiting weak yet sufficient conditions for periodicity.

\vspace{0cm}
\begin{figure}[htb]
\begin{center}
\hspace{.2cm}
\includegraphics[width=6cm]{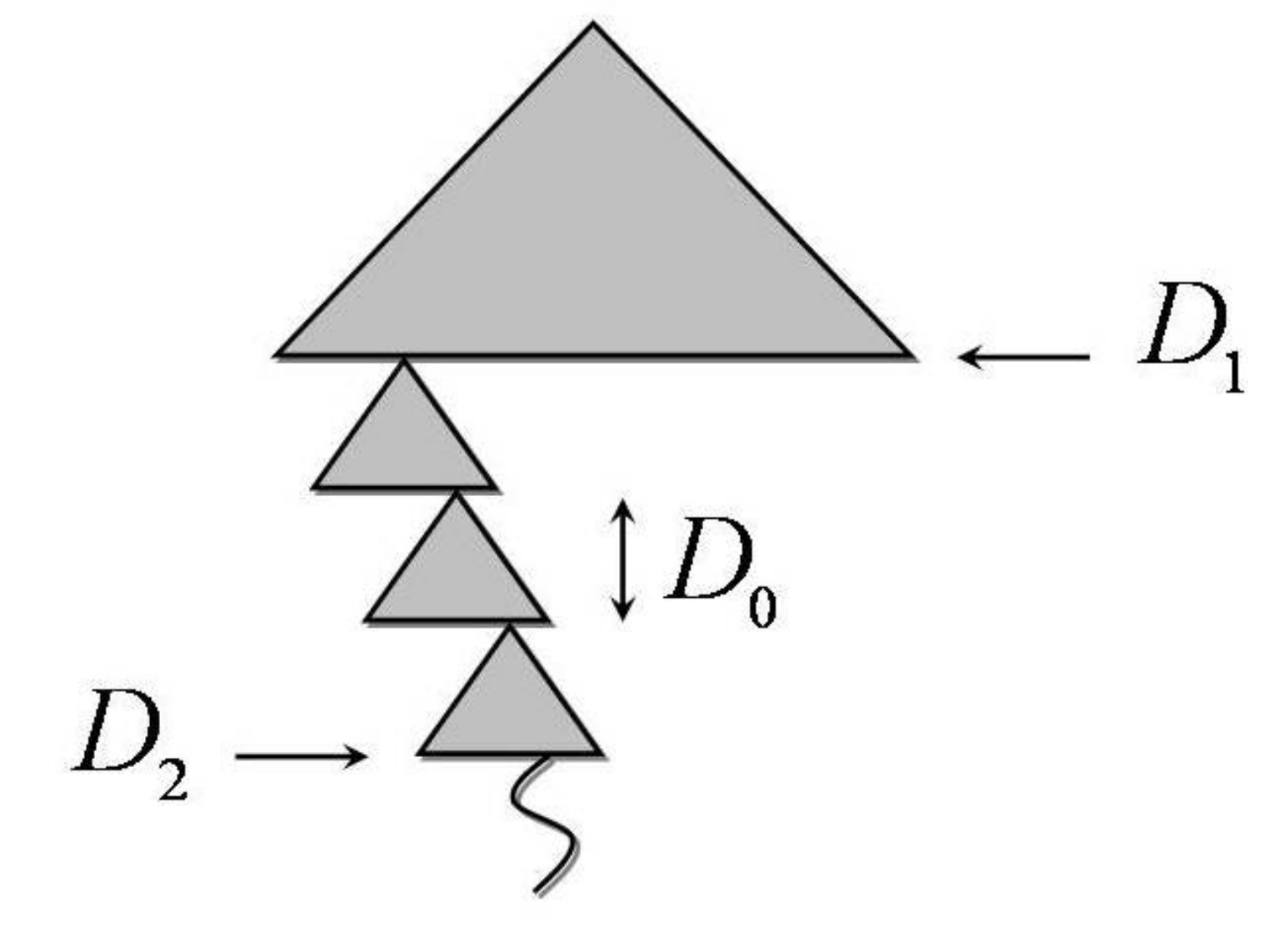}
\end{center}
\vspace{-.5cm}
\caption{\small The global coding tree 
is stratified into three layers, with decreasing branching rates.
\label{fig-strattree}}
\end{figure}
\vspace{.4cm}

\noindent
To stratify the coding tree 
${\mathcal T}^{\Delta|\Upsilon}_{m\,\rightarrow\, n-m}$
into layers, we set up three parameters $D_0$, $D_1$, and $D_2$:
the first targets the topological entropy;
the second specifies the height of the first layer;
the third indicates the nesting time.
We examine each one in turn and indicate their
purposes and requirements.

\smallskip
\paragraph{\sc First layer.}

By~(\ref{Pv-Upsilon-CDI}), the
phase tubes get thinner over time at a rate of roughly
$e^{-\gamma/3}$, while the tree is branching at a rate of
$n^{O(n)}$. To ensure that the topological
entropy is zero, the product of these two rates
should be less than 1: with $\gamma<1$, 
this is far from being 
the case, so we need a sparsification mechanism.
This is where Lemma~\ref{low-branching} comes in.
Indeed, deep enough in 
${\mathcal T}^{\Delta|\Upsilon}_{m\,\rightarrow\, n-m}$,
the size of a subtree of height $D_0$ is at most\footnote{\,
This assumes a thinness condition we discuss below. 
The factor $D_0$ comes from the nonbranching paths
in the subtree spanned by the phase tubes from $\Upsilon$.}
$$ D_0(n^{O(n)})^{D_0^{1-\gamma^{n+1}}},$$
while the tubes get thinner at a rate of $2n e^{-\gamma D_0/3}$
for every consecutive $D_0$ nodes: the choice 
of $D_0$ below ensures that the product
is less than 1, as desired. 
We justify this choice formally below.
\begin{align}
\hspace{1cm}
D_0 \geq  2^{(1/\gamma)^{n+2}}  && 
   \text{\small  [ {\em $D_0$ big enough for thinning to outpace
                                     branching} ].}\label{D0-lb}
\end{align}

\smallskip
\paragraph{\sc Second layer.}

Technically, Lemma~\ref{low-branching}
addresses only the branching of the phase tube
formed by a small box ${\mathbf x}+ \rho\, {\mathbb I}^n$,
for ${\mathbf x}\in {\mathbb C}[y_1,\ldots,y_r]$,
whereas we are concerned here with phase tubes
originating at some cell $V_v$ of 
${\mathcal T}^{\Delta|\Upsilon}_{m\,\rightarrow\, n-m}$.
To make $V_v$ thin enough, we choose a node $v$ deep in the tree.\footnote{\,
Factoring out the $B$-agents gives us the sort of fixed-point
attraction that is required by Lemma~\ref{low-branching}: it
is a dimension reduction device in attractor space.}
By~(\ref{Pv-Upsilon-CDI}),
$V_v\subseteq {\mathbf x}+ \rho\, {\mathbb I}^n$, for
${\mathbf x}\in {\mathbb C}[y_1,\ldots,y_r]$, provided that
$t_v\geq D_1$ and
\begin{align}
\hspace{1cm}
D_1 \geq  \frac{3}{\gamma} \log \frac{2n}{\rho}  && 
   \text{\small  [ {\em $D_1$ big enough for tree branches 
                     to be thinner than $\rho$} ].}\label{D1-lb}
\end{align}
Note that the requirement in~(\ref{Pv-Upsilon-CDI}) that
$t_v\geq 3t_c= 3/\gamma$ is implied 
by $t_v\geq D_1$. In view of Lemma~\ref{low-branching},
the number of nodes in
${\mathcal T}^{\Delta|\Upsilon}_{m\,\rightarrow\, n-m}$
of depth no greater than $t\geq D_1$, 
is bounded by 
\begin{equation*}
\underset{{\tt depth \,\, D_1}}{\underbrace{\, 
n^{O(nD_1)} \,}} 
\times
\underset{{\tt from \,\, D_1\,\, to \,\, t \,\, in\,\, chunks\,\, 
         of\,\, D_0}} {\underbrace{\, 
n^{O( n D_0^{1-\gamma^{n+1}} \lfloor (t-D_1)/D_0 \rfloor )} \,}}
\times
\underset{{\tt truncated \,\, chunk}}
  {\underbrace{\, 
n^{O( n D_0 )} \,}}
\times
\underset{{\tt single\,\, paths}}
  {\underbrace{\, 
D_0 \,}} ;
\end{equation*}
hence, for any $t\geq D_1$,
\begin{equation}\label{widthUB}
\Bigl|\, \{\,v\in {\mathcal T}^{\Delta|\Upsilon}_{m\,\rightarrow\, n-m}
\,|\, t_v\leq t\,\} \, \Bigr|
\leq  
n^{O( nD_0 + nD_1+ n t D_0^{-\gamma^{n+1}} )}  \, .
\end{equation}

\smallskip
\paragraph{\sc Third layer.}

The bottom layer of the stratified global coding tree begins
at a depth $D_2\geq D_0+D_1$. If the node $v$ of depth  $t_v\geq D_2$
has nontrivial branching,\footnote{\,
A node is branching nontrivially if it has at least two children 
neither of which is switching or vanishing.}
then, by continuity, $V_v$ contains a point right on
the boundary of the global margin.
By~(\ref{Pv-Upsilon-CDI}), this implies
the existence of $\zeta\in {\mathbb R}^n$
such that $\|\zeta\|_\infty\leq 2n e^{-\gamma \, D_2/3}$ and 
$\text{Aff}\,[{\mathbf y}+ \zeta]= \delta$, where
the coefficients of the affine form are of magnitude $n^{O(1)}$
and depend only on the node $v$.
It then follows from~(\ref{widthUB}) that 
${\mathcal T}^{\Delta'|\Upsilon}_{m\,\rightarrow\, n-m}$
has no nontrivial branching at depth $D_2$,
provided that $\Delta'= \Delta\setminus W'$, 
where $W'$ consists of gaps of type 
$n^{O(1)} e^{ -\gamma\, D_2/3}$ numbering at most 
\begin{equation*}
\underset{{\tt \#\, nodes\,\, at\,\, depth\,\, D_2}}
    {\underbrace{\, n^{O( nD_0+ nD_1 + n D_2 D_0^{-\gamma^{n+1}} )}
\,}} 
\times
\underset{{\tt \#\, margin \,\, slabs}}
    {\underbrace{\, n^{O(1)} \,}} .
\end{equation*}
This calculation, in which $\eps$ played no role,
puts a bound of $D_2$ on the nesting time.
It follows that
\begin{equation}\label{lengthW'ub}
\text{Leb}\,(W')\leq e^{ -\gamma\, D_2/3}
n^{O( nD_0+ nD_1 + n D_2 D_0^{-\gamma^{n+1}} )}
\end{equation}
Pick an arbitrarily small $\eps_o>0$ and
a large enough constant $d= d(b,c)$;
recall that $\gamma= n^{-cnt_o}$.
We set the parameters $\rho= \eps_o^2 \, n^{-dn^5D_0}$ and
$\eps\leq \min\, \{\, \rho, e^{-\gamma D_2 \,}\}$, where,
rounding up to the nearest integer,
\begin{equation}\label{D0D1=defn}
\begin{cases}
\,  D_0= \, 2^{d(1/\gamma)^{n+2}}  \\
\,  D_1 =  \, \frac{d^2}{\gamma}( n^6D_0 
                  + |\! \log \eps_o| )  \\
\,  D_2 = \, \frac{d}{\gamma}( n^2D_1
                        + |\! \log \eps_o| ).
\end{cases}
\end{equation}

\medskip
\noindent
We verify that conditions~(\ref{D0-lb}, \ref{D1-lb}) 
are both satisfied and that
\begin{equation}\label{D1geqD2D0}
D_1\geq D_2D_0^{-\gamma^{n+1}}.
\end{equation}
Thus the measure bound~(\ref{lengthW'ub}) implies that
$\text{Leb}\,(W')\leq \rho 2^{-D_0}$.
By making $\eps$ tend to $0$,
the point ${\mathbf x}$ vanishes with arbitrarily
small probability for random $\delta\in \Delta'$.
By Lemma~\ref{nesting-periodic}, this 
implies that, with probability at least $1- 2^{-D_0}$,
subjecting the system's margin to 
a perturbation $\delta$ chosen randomly in $\Delta$
makes the orbit of any ${\mathbf x}\in \Upsilon$
attracted to a limit cycle (or a switching leaf):\footnote{\,
The regions $W$ and $W'$, which make perturbation
a requirement, depend only ${\mathbf y}$. But perturbation
is also needed to avoid vanishing, which
depends on the initial state ${\mathbf x}$.}
we call this {\em success.}
The sum of the period and preperiod is bounded by
the number of nodes of depth at most $D_2$ (the nesting time),
which, by~(\ref{widthUB}, \ref{D1geqD2D0}), is no
greater, conservatively, than 
\begin{equation}\label{p-bound-D0}
\bar p= n^{O(nD_1)}\leq 
(1/\eps_o)^{O(\gamma^{-2})}\, 2^{D_0\gamma^{-1}n^{O(1)}}.
\end{equation}
We bound the attraction rate by appealing to 
Lemmas~\ref{matrixproduct} and \ref{thinMatrixStruct}.
Note that if $p$ is the period then so is
$p\lceil (\log 2n)/\gamma\rceil$. This choice of $p$
still satisfies the upper bound~(\ref{p-bound-D0}) while
ensuring that, at every period, the error bound in 
Lemma~\ref{thinMatrixStruct} is at most $\frac{1}{2n}$.
The row-sums in $A$ in Lemma~\ref{matrixproduct} 
are at most $1/2$, so we can set $\mu=e^{-\gamma}$.
Since $\nu\leq D_2$
and $(\#{\mathcal S}_\nu)\lceil (\log 2n)/\gamma\rceil
 \leq \bar p$, it follows that
\begin{equation}\label{theta-upsilon-alpha}
\theta_\alpha \leq D_0^{O(D_0)} 
(1/\eps_o)^{O(\gamma^{-2})} \log \hbox{$\frac{1}{\alpha}$},
\end{equation}
for any $0<\alpha<1/2$.
The perturbation space is not $\Delta$
but $n^{-b}\,{\mathbb I}$, so we apply the 
previous result to each free canonical interval
and argue as follows. If $\Lambda$ is the measure of
the union of all the free canonical intervals, then the perturbations
that do not guarantee success have measure at most
$(2n^{-b}- \Lambda) +  2^{-D_0}\Lambda$. 
Dividing by $2n^{-b}$ and applying~(\ref{Wbound}) shows that
\begin{equation}\label{failureProbUpsilon}
\text{Prob}\,[\,\text{failure in}\,\, 
{\mathcal T}^{\,n^{-b}\,\mathbb I
\,|\, \Upsilon}_{m\,\rightarrow\, n-m}
\,]\leq 
1- (1- 2^{-D_0})(\,1-(\eps+\rho) n^{O( n^5 D_0 )}\,)
\leq 2^{1-D_0}.
\end{equation}
The nesting time is at most $D_2$, which, 
by~(\ref{widthUB}, \ref{D1geqD2D0}), implies that
\begin{equation}\label{wordentropyTupsilon}
h({\mathcal T}^{\,n^{-b}\,\mathbb I
\,|\, \Upsilon}_{m\,\rightarrow\, n-m})
\leq O(D_1 n\log n)
\leq \hbox{$\frac{1}{\gamma}$}(D_0+ |\! \log \eps_o|)n^{O(1)}.
\end{equation}

\smallskip
\paragraph{Freeing the $B$-agents.}

Set $D_3= \lceil 3b t_c n\log n\rceil$ and
fix ${\mathbf x}$ in $\Omega^n$.
Let $\xi$ denote the projection of $f^{D_3}({\mathbf x})$
onto the last $n-m$ coordinate axes.
By Lemma~\ref{thinMatrixStruct} and $t_c=1/\gamma$
(Lemma~\ref{thinrate-ub}), the coding tree 
${\mathcal T}^{\,n^{-b}\,\mathbb I}_{m\,\rightarrow\, n-m}$
has $n^{O(nt_c)}$ nodes $u$ such that $t_u=t_c$ and
$$
\xi\in {\mathbf y} + ne^{-\gamma D_3}\,{\mathbb I}^{n-m}
\subseteq  {\mathbf y} + n^{-2bn}\,{\mathbb I}^{n-m},
$$
where ${\mathbf y}=  D_u (x_{m+1},\ldots, x_n)^T$.
The state vector for the $B$-agents is $\xi$ at time
$D_3$ and $Q^{t-D_3}\xi$ at $t>D_3$, where $Q$ is
the transition matrix of the frozen communication
subgraph joining the $B$-agents at time $D_3$. By taking $t$ 
to infinity, it follows that ${\mathbf y}= {\widetilde Q}\xi$
and, by Lemma~\ref{upsilon-facts}, $\xi\in \Upsilon_B$
hence $f^{D_3}({\mathbf x})\in \Upsilon$. 
We can then apply the previous result. Since ${\mathbf x}$ is
fixed, only the choice of random perturbation $\delta$
can change which path in 
${\mathcal T}^{\,n^{-b}\,\mathbb I}_{m\,\rightarrow\, n-m}$
the orbit will follow.
The failure probability of~(\ref{failureProbUpsilon})
needs to be multiplied by the number of nodes $u$,
which yields an upper bound of $n^{O(nt_c)} 2^{1-D_0}$; hence
\begin{equation}\label{failureProbH}
\text{Prob}\,[\,\text{failure in}\,\, 
{\mathcal T}^{\,n^{-b}\,\mathbb I}_{m\,\rightarrow\, n-m}
\,]
\leq 2^{-D_0/2}.
\end{equation}
If ${\mathcal T}^*$ denotes the part of the
global coding tree extending to depth $D_3$, then
$$
{\mathcal T}^{\,n^{-b}\,\mathbb I}_{m\,\rightarrow\, n-m}
= {\mathcal T}^* \,\otimes\, 
{\mathcal T}^{\,n^{-b}\,\mathbb I
\,|\, \Upsilon}_{m\,\rightarrow\, n-m}.
$$
The tree ${\mathcal T}^*$ has at most 
$n^{O(n t_c)} D_3$ nodes; therefore, 
by~(\ref{wordentropyTupsilon}),
\begin{equation}\label{wordentropy:T+D3}
h({\mathcal T}^{\,n^{-b}\,\mathbb I}_{m\,\rightarrow\, n-m})
\leq O(t_cn\log n +\log D_3) +
h({\mathcal T}^{\,n^{-b}\,\mathbb I
\,|\, \Upsilon}_{m\,\rightarrow\, n-m})
\leq \hbox{$\frac{1}{\gamma}$}(D_0+ |\! \log \eps_o|)n^{O(1)}.
\end{equation}

\smallskip
\subsection{Removing persistence}\label{subsec:persistence}

We use direct products to relax the condition that 
the permanent graph $H$ be fixed once and for all.
This touches on the non-Markovian nature of the system,
a feature we chose to ignore in the previous section.
We explain now why this was legitimate.
Because the switching condition is about  
time differences and not absolute times,
any subpath in the coding tree 
has an incarnation as a path from the root.
Equivalently, any interval of a trajectory appears
as the prefix of another trajectory.
This property explains why, following~(\ref{Ds_k-gamma}),
we could argue that $A_k$ was of the form $A_{\leq w}$.
Likewise, in the derivations 
leading to~(\ref{Pv-Upsilon-CDI}), we
used the fact that  
$f^{t_v}({\mathbf x}) = f^{t_v-t_w}( f^{t_w}({\mathbf x}) )$, 
an identity that might not always hold in a
non-Markovian setting, but which, in this case, did.
Finally, weren't we too quick to appeal to 
Lemma~\ref{nesting-periodic} for periodicity
since its proof relied heavily on the Markov property?
To see why the answer is no, observe that
the argument did establish the periodicity 
of the ``wrap-around'' system derived
from ${\mathcal T}^{\Delta'|\Upsilon}_{m\,\rightarrow\, n-m}$
by redirecting any trajectory that reaches the
nesting depth to the root. The only problem is that this
system, being Markovian, is not the one modeled by the coding tree.
Wrapping around resets the time to zero, which
might cause switching conditions to be missed
and trajectories to be continued when they should be stopped:
none of this stops nonvanishing orbits from being periodic, however.

We now show how to relax the permanent graph assumption.
The idea is to begin with $H$ set as the complete graph
and update it at each switching leaf by removing the edge(s)
whose missing presence causes the node to be a switching leaf.
We then append to each such leaf the coding tree, suitably cropped,
defined with respect to the new value of $H$.
We model this iteration by means of
direct products, using $m_k$ to denote
the number of $A$-agents in the block-directional
system used in the $k$-th product:
\begin{equation}\label{T-dirprod}
{\mathcal T}^{\,n^{-b}\,\mathbb I}_n
\,\Longrightarrow \,\, 
\bigotimes_{k=1}^{k_0}\,
{\mathcal T}^{\,n^{-b}\,\mathbb I}_{m_k\,\rightarrow\, n-m_k}.
\end{equation}
The upper limit $k_0$ is bounded by $n(n-1)$;
note that each decision procedure ${\mathcal G}_{ij}$ 
needs it own counter.
To keep the failure probability small
throughout the switching of dynamics,
we need to update the value of $D_0$ in~(\ref{D0D1=defn})
at each iteration, so we define $C_k$ as
its suitable value for a persistent graph
consisting of $k$ (nonloop) directed edges and  
let $\phi_k$ denote the maximum  
failure probability for such a graph:
$C_{n(n-1)}\geq D_0$ and $\phi_0=0$.
The logarithm of the number of switching leaves is at most 
the word-entropy;\footnote{\,
No two switching leaves can have the same ancestor 
at a depth equal to the nesting time.
Because we can bound the number of switching leaves,
we may dispense with~(\ref{entr-dp}) altogether.}
by~(\ref{failureProbH}, \ref{wordentropy:T+D3}), for $k>0$,
\begin{equation*}
\phi_k \leq  2^{-C_k/2}+ 
2^{ \gamma^{-1} (C_k + |\! \log \eps_o|)n^a } \phi_{k-1} ,
\end{equation*}
for some constant $a>0$. Setting 
$$C_{n(n-1)-j}= \lceil \gamma^{-j} n^{2aj}(D_0+ 3|\! \log \eps_o|)\rceil,$$
for $j=0,\ldots, n(n-1)$, 
we verify by induction that $\phi_k\leq 2^{1-C_k/2}$,
for $k=0,\ldots, n(n-1)$; hence,
\begin{equation*}
\text{Prob}\,[\,\text{failure in nonpersistent}\,\, 
{\mathcal T}^{\,n^{-b}\,\mathbb I}_{m\,\rightarrow\, n-m}
\,]
\leq \phi_{n(n-1)} \leq  2^{1-  \frac{1}{2}( D_0 +3 |\! \log \eps_o|)}
\leq \eps_o.
\end{equation*}
The attraction rate is still exponential: using~(\ref{theta-upsilon-alpha}) 
yields a geometric series summing up~to
\begin{equation}\label{theta-upsilon-alpha-final}
\theta_\alpha \leq C_0^{O(C_0)} 
(1/\eps_o)^{O(\gamma^{-2})} \log \hbox{$\frac{1}{\alpha}$}
\leq O_{n,\eps_o,t_o}( \log \hbox{$\frac{1}{\alpha}$} ),
\end{equation}
for any $0<\alpha<1/2$. 
By~(\ref{p-bound-D0}), the period and preperiod are bounded by
$$
(1/\eps_o)^{O(\gamma^{-2})}\, 2^{C_0\gamma^{-1}n^{O(1)}}
\leq (1/\eps_o)^{O_{n,\eps_o,t_o}(1)},
$$
which completes the proof of Theorem~\ref{general-case}.
\hfill $\Box$
\proofend

\subsection*{Acknowledgments}

I wish to thank Pascal Koiran and John Tsitsiklis for helpful conversations.

\end{document}